\documentclass[aps,prl,twocolumn]{revtex4}

\usepackage{graphicx,amssymb,amsfonts,amsmath,chemarr,color}

\newcommand{\beq}{\begin{equation}}
\newcommand{\eeq}{\end{equation}}
\newcommand{\beqn}{\begin{eqnarray}}
\newcommand{\eeqn}{\end{eqnarray}}
\newcommand{\elabel}[1]{\label{eq:#1}}
\newcommand{\eref}[1]{Eq.\ \ref{eq:#1}}
\newcommand{\erefs}[2]{Eqs.\ \ref{eq:#1} and \ref{eq:#2}}

\newcommand{\erefn}[2]{Eqs.\ \ref{eq:#1}-\ref{eq:#2}}
\newcommand{\flabel}[1]{\label{fig:#1}}
\newcommand{\fref}[1]{Fig.\ \ref{fig:#1}}

\newcommand{\avg}[1]{\left\langle#1\right\rangle}
\newcommand{\abs}[1]{\left|#1\right|}

\begin{document}

\title{Fundamental limits to collective concentration sensing in cell populations}

\author{Sean Fancher}
\affiliation{Department of Physics and Astronomy, Purdue University, West Lafayette, IN 47907, USA}

\author{Andrew Mugler}
\email{amugler@purdue.edu}
\affiliation{Department of Physics and Astronomy, Purdue University, West Lafayette, IN 47907, USA}

\begin{abstract}
The precision of concentration sensing is improved when cells communicate. Here we derive the physical limits to concentration sensing for cells that communicate over short distances by directly exchanging small molecules (juxtacrine signaling), or over longer distances by secreting and sensing a diffusive messenger molecule (autocrine signaling). In the latter case, we find that the optimal cell spacing can be large, due to a tradeoff between maintaining communication strength and reducing signal cross-correlations. This leads to the surprising result that sparsely packed communicating cells sense concentrations more precisely than densely packed communicating cells. We compare our results to data from a wide variety of communicating cell types.
\end{abstract}

\maketitle

Single cells sense chemical concentrations with extraordinary precision. In some cases this precision approaches the physical limits set by molecular diffusion \cite{berg1977physics, bialek2005physical}. Yet, no cell performs this sensory task in isolation. Cells exist in communities, such as colonies, biofilms, and tissues. Within these communities, cells communicate in diverse ways. Communication mechanisms include the exchange of molecules between cells in contact (juxtacrine signaling), and secretion and detection of diffusible molecules over distances comparable to the cell size or longer (autocrine signaling \cite{foot1}) \cite{alberts2002molecular, singh2005autocrine, youk2014secreting, coppey2007time}. This raises the question of whether cell-cell communication improves a cell's sensory precision, beyond what the cell achieves alone.

Experiments have shown that cells are more sensitive in groups than they are alone. Groups of neurons \cite{rosoff2004new}, lymphocytes \cite{malet2015collective}, and epithelial cells \cite{ellison2016cell} exhibit biased morphological or motile responses to chemical gradients that are too shallow for cells to detect individually. Groups of cell nuclei in fruit fly embryos detect morphogen concentrations with a higher precision than is expected for a single nucleus \cite{gregor2007probing, little2013precise, erdmann2009role}. In some cases, such as with epithelial cells \cite{ellison2016cell}, cell-cell communication has been shown to be directly responsible for the enhanced sensitivity. Yet, from a theoretical perspective, the fundamental limits to concentration sensing \cite{berg1977physics, bialek2005physical, berezhkovskii2013effect, wang2007quantifying, kaizu2014berg, lang2014thermodynamics, mora2015physical, bicknell2015limits, endres2009maximum} or gradient sensing \cite{ueda2007stochastic, endres2008accuracy, hu2010physical} have been largely limited to single receptors or single cells. Analogous limits for groups of communicating cells have been derived only for specific geometries \cite{mugler2016limits}, and are otherwise poorly understood. In particular, it remains unknown whether the limits depend on the communication mechanism (juxtacrine vs.\ autocrine), and how they scale with collective properties like communication strength and population size.

Here we derive the fundamental limits to the precision of collective sensing by one-, two-, and three-dimensional (3D) populations of cells. We focus on the basic task of sensing a uniform chemical concentration. We compare two ubiquitous communication mechanisms, juxtacrine signaling and autocrine signaling. Intuitively one expects that sensory precision is enhanced by communication, that communication is strongest when cells are close together, and therefore that small cell-to-cell distances should result in the highest sensory precision. Instead, we find that under a broad range of conditions, it is not optimal for cells to be as close as possible. Rather, an optimal cell-to-cell distance emerges due to a tradeoff between maintaining sufficient communication strength and minimizing signal cross-correlations. For sufficiently large populations, this distance can be many times the cell diameter, meaning that these populations are sparsely packed, not densely packed.
These sparsely packed populations then sense concentrations with a precision that can be many times higher than that of populations in which cells are adjacent and communicate directly.
We discuss the implications of these findings for cell populations, compare our results to data from a wide variety of communicating cell types, and make predictions for future experiments.

Consider $N$ cells with radii $a$ in the presence of a ligand that diffuses in three dimensions with coefficient $D_c$ (\fref{cartoon}). The ligand concentration $c\left(\vec{x},t\right)$ fluctuates due to the particulate nature of the ligand molecules, but on average the steady-state concentration $\bar{c}(\vec{x}) = \bar{c}$ is uniform. Ligand molecules bind and unbind to receptors on the surface of cell $i$ with rates $\alpha$ and $\mu$, respectively, leading to $r_i(t)$ bound receptors. The dynamics of $c$ and $r$ are
\begin{align}
\elabel{c1}
\dot{c} &= D_{c}\nabla^{2}c-\sum_{i=1}^N\delta\left(\vec{x}-\vec{x}_i\right)\dot{r}_i+\eta_{c}, \\
\elabel{r1}
\dot{r}_i &= \alpha c\left(\vec{x}_i,t\right)-\mu r_i+\eta_{ri}.
\end{align}
The first term on the righthand side of \eref{c1} describes the ligand diffusion, while the second term describes the binding and unbinding of ligand at cell positions $\vec{x}_i$, with the binding dynamics given by \eref{r1}. The noise terms obey $\left\langle\eta_{c}\left(\vec{x},t\right)\eta_{c}\left(\vec{x}',t'\right)\right\rangle = 2D_{c}\bar{c}\delta\left(t-t'\right)\vec{\nabla}_{x}\cdot\vec{\nabla}_{x'}\delta^{3}\left(\vec{x}-\vec{x}'\right)$ and $\left\langle\eta_{ri}\left(t\right)\eta_{rj}\left(t'\right)\right\rangle = 2\mu\bar{r}\delta_{ij}\delta\left(t-t'\right)$, and account for the spatiotemporally correlated diffusive fluctuations \cite{gardiner1985handbook} and the stochastic nature of the binding reactions \cite{gillespie2000chemical}, respectively. Here $\bar{r} = \alpha\bar{c}/\mu$ is the mean bound receptor number of each cell in steady state. \eref{r1} neglects the effects of receptor saturation.

We first consider juxtracrine signaling, in which a messenger molecule is exchanged between adjacent cells at a rate $\gamma$ (\fref{cartoon}A). Messenger molecules are produced in each cell by the bound receptors at a rate $\beta$ and degraded at a rate $\nu$, such that the messenger acts as both the mediator of communication and the sensory readout. Including the messenger molecule extends work in which receptor counts are integrated mathematically to form the readout \cite{bialek2005physical, wang2007quantifying, bicknell2015limits} because here the integration is modeled physically \cite{govern2012fundamental, govern2014energy}. The dynamics of $m_i(t)$, the number of messenger molecules in cell $i$, are
\beq
\elabel{m1}
\dot{m}_i = \beta r_{i}-\nu m_{i}+\gamma\sum_{j\in\mathcal{N}_i}\left(m_{j}-m_{i}\right)+\eta_{mi},
\eeq
where ${\cal N}_i$ denotes the neighbors of cell $i$. The noise term obeys $\left\langle\eta_{mi}\left(t\right)\eta_{mj}\left(t'\right)\right\rangle = 2\bar{m}M_{ij}\delta\left(t-t'\right)$, where the matrix $M_{ij} = \left(\nu+\left|{\cal N}_i\right|\gamma\right)\delta_{ij} -\gamma\delta_{j\in{\cal N}_i}$ accounts for the stochasticity of the reactions (first term) and the anti-correlations induced by the exchange (second term) \cite{gillespie2000chemical}. Here $\bar{m} = \beta\bar{r}/\nu$ is the mean messenger molecule number of each cell in steady state, and $\left|{\cal N}_i\right|$ is the number of neighbors of cell $i$.

The precision of concentration sensing is given by the signal-to-noise ratio of the readout in a particular cell, $\bar{m}^2/(\delta m_i)^2$. We assume that each cell integrates its messenger molecule count over a time $T$, such that $(\delta m_i)^2$ is the variance in the time average $T^{-1}\int_0^Tdt\ m_i(t)$ \cite{berg1977physics}.
Because \erefn{c1}{m1} are linear with Gaussian white noise, finding $(\delta m_i)^2$ is straightforward: we Fourier transform \erefn{c1}{m1} in space and time, calculate the power spectrum of $m_i$, and recognize that $(\delta m_i)^2$ is given by its low-frequency limit (see the supplementary material \cite{supp}). During this procedure, we cut off the wavevector integrals at the maximal value $k=2\pi/(ga)$ to regularize the divergence caused by the delta function in \eref{c1} \cite{bialek2005physical, mugler2016limits}. Here $g$ is a geometric factor of order unity that depends on the shape of the cell, as further discussed below. Using the low-frequency limit of the power spectrum assumes that $T$ is longer than any of the intrinsic signaling timescales \cite{supp}. Specifically, $T \gg \{\tau_1, \tau_2, \tau_3\}$, where $\tau_1 \equiv a^2/D_c$ is the characteristic time for a ligand molecule to diffuse across a cell; $\tau_2 \equiv \mu^{-1} + (k_DK_D)^{-1}$ is the receptor equilibration timescale including rebinding, with diffusion-limited rate $k_D = \pi gaD_c$ and dissociation constant $K_D = \mu/\alpha$ \cite{kaizu2014berg}; and $\tau_3\equiv(\nu+\gamma)^{-1}$ is the turnover timescale of the messenger molecule. Since $g=4$ corresponds to the diffusion-limited rate $k_D = 4\pi a D_c$ for a sphere of radius $a$, we take $g=4$ from here on. This procedure yields $(\delta m_i)^2$ for arbitrary parameters and cell configuration \cite{supp}. Special cases illuminate the key physics, below.

First, for either $N=1$ or $\gamma = 0$, we obtain the result for a single, isolated cell \cite{supp},
\beq
\elabel{nsr_m1}
\left(\frac{\delta m}{\bar{m}}\right)^{2} = \frac{1}{2}\frac{1}{\pi a\bar{c}D_cT}
	+\frac{2}{\mu T\bar{r}}+\frac{2}{\nu T\bar{m}}.
\eeq
\eref{nsr_m1} is the inverse of the precision, which we call the error. The first term is the well-known Berg-Purcell limit for the extrinsic noise propagated from ligand diffusion \cite{berg1977physics}. The second and third terms are the intrinsic noise arising from the finite numbers of bound receptors $\bar{r}$ and messenger molecules $\bar{m}$. In the limit of large molecule numbers $\{\bar{r},\bar{m}\}\to\infty$, these terms vanish. From here on we consider only the extrinsic noise, since extrinsic factors such as $\bar{c}$ and $D_c$ are not under direct control of the cell (see \cite{supp} for the general case).

\begin{figure}
\begin{center}
\includegraphics[width=\columnwidth]{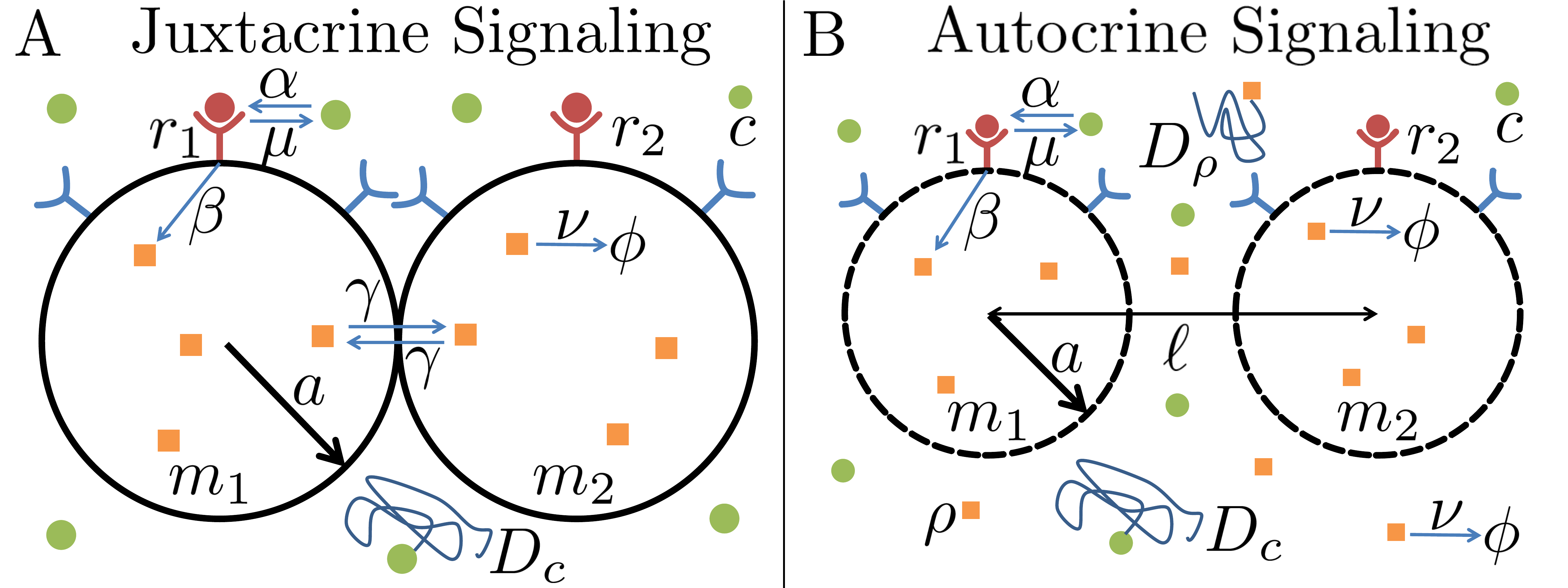}
\end{center}
\caption{Cells with (A) short-range (juxtacrine) or (B) long-range (autocrine) communication, sensing a uniform ligand concentration.}
\flabel{cartoon}
\end{figure}

Second, for $N=2$ (as in \fref{cartoon}A) the extrinsic part of the error for either cell is \cite{supp}
\begin{align}
\elabel{nsr_m2}
\left(\frac{\delta m}{\bar{m}}\right)^{2} &=
\frac{1}{2}\left[1-\frac{\hat{\lambda}^2(1+\hat{\lambda}^2)}{(1+2\hat{\lambda}^2)^2}\right]
\frac{1}{\pi a\bar{c}D_{c}T} \\
\elabel{nsr_m3}
&\to \frac{3}{8} \frac{1}{\pi a\bar{c}D_{c}T} \qquad \lambda\gg a.
\end{align}
Here $\hat{\lambda} \equiv \lambda/(2a)$, where $\lambda \equiv 2a\sqrt{\gamma/\nu}$ is the communication length: it is the lengthscale of the exponential kernel that governs the exchange of messenger molecules \cite{mugler2016limits, ellison2016cell, supp}. The prefactor in \eref{nsr_m2} is a monotonically decreasing function of $\hat{\lambda}$, which demonstrates that error decreases with increasing communication. In the limit of weak communication $\lambda \ll a$, the prefactor becomes $1/2$ as in the one-cell case (\eref{nsr_m1}). In the limit of strong communication $\lambda \gg a$, it becomes $3/8$ (\eref{nsr_m3}). The fact that $3/8$ is larger than half of $1/2$ means that two cells are less than twice as good as one cell in terms of sensory precision, even with perfect communication. The reason is that, with juxtacrine signaling, the cells are sampling adjacent regions of extracellular space, and correlations mediated by the diffusing ligand molecules prevent their measurements from being independent \cite{mugler2016limits}. Can autocrine signaling avoid this drawback?

To answer this question, we consider autocrine signaling (\fref{cartoon}B). As before, a messenger molecule is produced by each cell at rate $\beta$ and degraded at rate $\nu$, but now it diffuses within the extracellular space with coefficient $D_\rho$. Thus, the autocrine model retains \erefs{c1}{r1} but replaces \eref{m1} with
\beq
\elabel{m2}
\dot{\rho} = D_{\rho}\nabla^{2}\rho-\nu \rho+\sum_{i=1}^N\delta\left(\vec{x}-\vec{x}_{i}\right)\left(\beta r_{i}+\eta_{pi}\right)+\eta_{d},
\eeq
where $\rho\left(\vec{x},t\right)$ is the concentration of the messenger molecule. The production noise obeys \cite{gillespie2000chemical} $\left\langle\eta_{pi}\left(t\right)\eta_{pj}\left(t'\right)\right\rangle = \beta\bar{r}\delta_{ij}\delta\left(t-t'\right)$,
while the degradation and diffusion noise obeys \cite{gardiner1985handbook} $\left\langle\eta_{d}\left(\vec{x},t\right)\eta_{d}\left(\vec{x}',t'\right)\right\rangle = \nu\bar{\rho}\left(\vec{x}\right)\delta\left(t-t'\right)\delta\left(\vec{x}-\vec{x}'\right) + 2D_{\rho}\delta\left(t-t'\right)\vec{\nabla}_{x}\cdot\vec{\nabla}_{x'}\left[\bar{\rho}\left(\vec{x}\right)\delta\left(\vec{x}-\vec{x}'\right)\right]$.
Here
\beq
\elabel{rho}
\bar{\rho}\left(\vec{x}\right) = \frac{\beta\bar{r}}{4\pi D_{\rho}}\sum_{i = 1}^{N}\frac{e^{-\abs{\vec{x}-\vec{x}_{i}}/\lambda}}{\abs{\vec{x}-\vec{x}_{i}}}
\eeq
is the steady-state concentration profile of the messenger molecule, which is non-uniform due to the multiple cell sources. $\lambda \equiv \sqrt{D_\rho/\nu}$ sets the communication length in the autocrine case.
Even in the limit of strong communication, which sends to unity the exponential term in \eref{rho}, the power-law decay remains due to the inability of diffusion to fill 3D space.

We then imagine that each cell counts the number $m_i(t) = \int_{V_i}d^{3}x \rho\left(\vec{x},t\right)$ of messenger molecules within its volume $V_i$ \cite{foot2}, and we use the same procedure as above to exactly solve for the error for arbitrary parameters and cell configuration \cite{supp}. For the special case of $N=2$ cells separated by a distance $\ell > a$ (as in \fref{cartoon}B), in the limit of strong communication $\lambda \gg a$, the result for the extrinsic noise ($\{\bar{r},\beta\}\to\infty$) is
\begin{align}
\elabel{nsr_m4}
\left(\frac{\delta m}{\bar{m}}\right)^{2} &=
\frac{1+16/(9\hat{\ell}^2)}{2[1+2/(3\hat{\ell})]^2}
\frac{1}{\pi a\bar{c}D_{c}T} \\
\elabel{nsr_m5}
&\to \frac{2}{5} \frac{1}{\pi a\bar{c}D_{c}T} \qquad \ell = \ell^* = \frac{8}{3} a,
\end{align}
where $\hat{\ell} \equiv \ell/a$, $T\gg\{\tau_1,\tau_2, \tau_4\}$, and $\tau_4\equiv (\nu + D_\rho/a^2)^{-1}$ is the messenger turnover timescale.
In the limit of large separation $\ell\gg a$, the prefactor in \eref{nsr_m4} becomes $1/2$ as in the one-cell case (\eref{nsr_m1}), since here each cell only detects messenger molecules produced by itself (proper autocrine signaling). The denominator in \eref{nsr_m4} decreases with $\ell$. This is because the mean decreases with cell separation due to the decay of the messenger molecule concentration profile (\eref{rho}). The numerator in \eref{nsr_m4} also decreases with $\ell$. This is because increasing the cell separation reduces the messenger molecule variance due to two separate effects, as detailed in \cite{supp}. First, it decreases the ligand cross-correlations discussed above. Second, it allows diffusion to wash out correlations in the messenger molecules themselves, an effect that is known to reduce super-Poissonian noise during spatial averaging \cite{erdmann2009role}. The tradeoff between the decrease of the mean and of the variance with cell separation results in a minimum value of the prefactor equal to $2/5$, when $\ell = \ell^* = 8a/3$ (\eref{nsr_m5}).

\begin{figure}
\begin{center}
\includegraphics[width=\columnwidth]{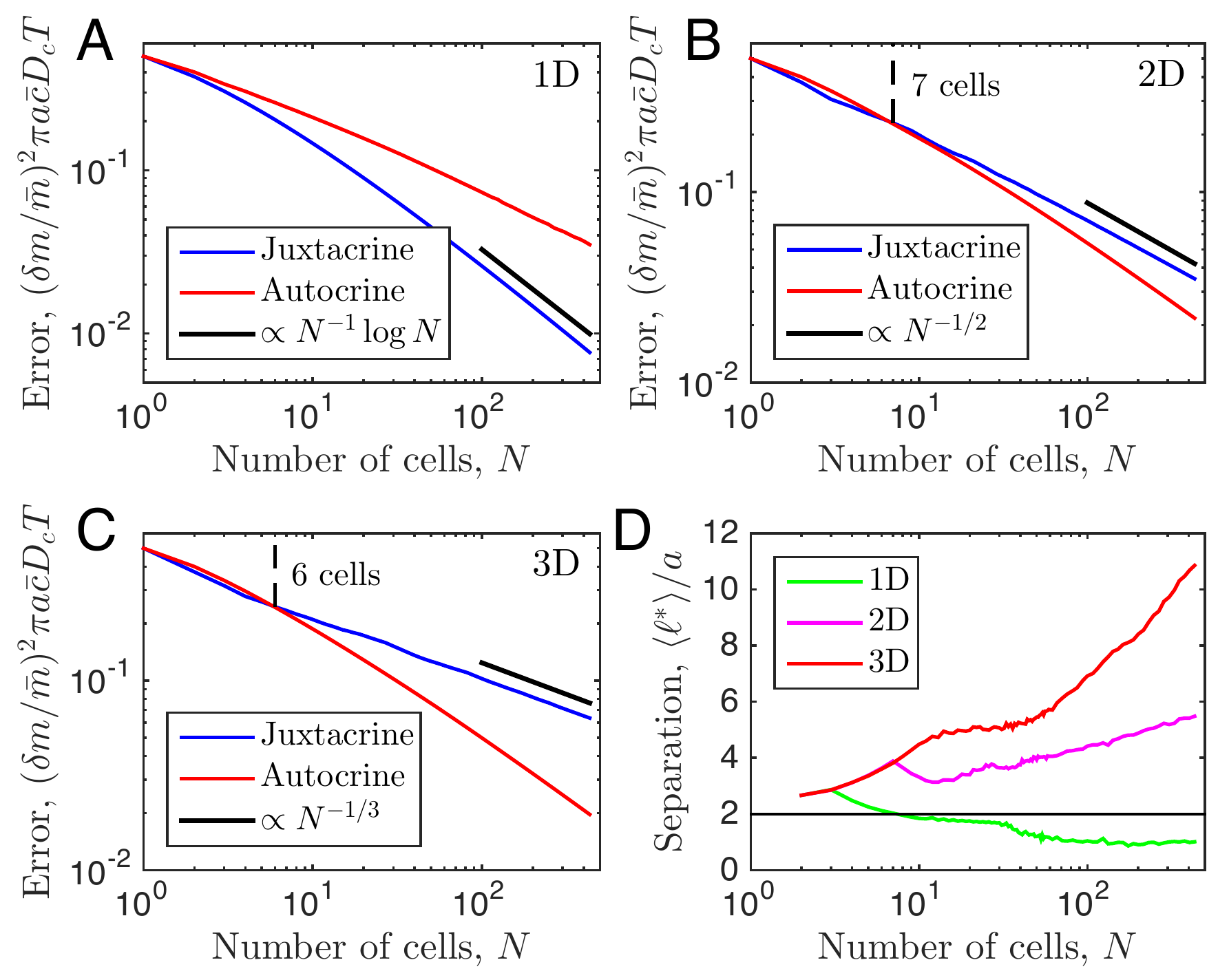}
\end{center}
\caption{(A-C) Error in concentration sensing for the two communication strategies vs.\ the size of 1D, 2D and 3D populations. Error is calculated for the center cell. (D) Average optimal nearest-neighbor separation in the autocrine case. $\avg{\ell^*}<2a$ corresponds to overlapping cells.}
\flabel{scalings}
\end{figure}

Evidently, for $N=2$ cells, the error in concentration sensing is roughly equivalent for juxtacrine signaling ($3/8 = 0.375$) and autocrine signaling ($2/5 = 0.4$). This makes sense, considering that the optimal cell separation for autocrine signaling ($\ell^* = 8a/3$) is close to that of adjacent cells ($\ell = 2a$), and therefore both signaling mechanisms correspond to close-range communication in this case.
Do the results change for larger $N$?
Because we have exact results for arbitrary $N$ and arbitrary cell positions \cite{supp}, we can answer this question immediately. \fref{scalings}A, B, and C compare as a function of $N$ the error of the two signaling mechanisms in the limit of strong communication for 1D, 2D, and 3D configurations of cells, respectively. For juxtacrine signaling, cells are arranged within a line (1D), circle (2D), or sphere (3D) on a rectangular lattice with spacing $2a$. For autocrine signaling, cells are confined to the respective dimensionality, but are otherwise allowed to adjust their positions via a Monte Carlo scheme
until the minimum error is reached. The average nearest-neighbor separation $\avg{\ell^*}$ in this case is shown in \fref{scalings}D. We see in \fref{scalings}A-C that the error always decreases with $N$, meaning that communication among an increasing number of cells monotonically improves sensory precision \cite{gregor2007probing, little2013precise, erdmann2009role}. In 1D, we see that juxtacrine signaling results in a smaller error than autocrine signaling for all $N$ (\fref{scalings}A). In fact, in the case of autocrine signaling in 1D, the optimal separation decreases with $N$, and beyond $N=7$, cells overlap, $\avg{\ell^*}<2a$ (\fref{scalings}D). However, in 2D and 3D, the two strategies are comparable for small $N$, but beyond $N = 7$ or $6$ cells, autocrine signaling clearly results in the smaller error (\fref{scalings}B and C, respectively). At the same time, the optimal separation for autocrine signaling in 2D and 3D increases with $N$ (\fref{scalings}D). By $N=400$ cells, the optimal separation in 3D becomes more than $10$ cell radii, meaning that the optimal arrangement of cells is highly sparse.

It is clear from \fref{scalings}A-C that the errors of the two communication strategies scale differently with population size. The scaling in the juxtacrine case can be understood quantitatively. In the limit of strong communication, the entire population of contiguous cells acts as one large detector. The error of a long ellipsoidal (1D), disk-shaped (2D), or spherical detector (3D) scales inversely with its longest lengthscale (with a log correction in 1D) \cite{berg1993random}. This lengthscale in turn scales with $N$, $N^{1/2}$, or $N^{1/3}$, respectively, leading to the predicted scalings in \fref{scalings}A-C, which are seen to agree excellently at large $N$. On the other hand, the scaling for autocrine signaling is different from that for juxtacrine signaling in each dimension. Evidently, diffusive communication and sparse arrangement lead to fundamentally different physics of sensing. In particular, in 2D and 3D the autocrine scaling is clearly steeper at large $N$ (\fref{scalings}B and C), meaning that not only is the autocrine strategy more precise for a sufficiently large population, but the improvement in precision will continue to grow with population size.

How do our results compare to biological systems? Arguably the most biologically unrealistic assumption that we make in \fref{scalings} is that of strong communication, $\lambda\gg a$.
Since our calculations are exact for any $\lambda$ \cite{supp}, we relax this assumption in \fref{phase}, allowing us to identify phases in the space of $\lambda$ and $N$ in which the optimal sensory precision is achieved with either densely packed or sparsely packed cells.
We now ask where biological systems fall in this phase space.
Bacteria communicate via autocrine signaling, and it has been suggested that this enables collective sensing during swarming \cite{grunbaum1998schooling, shklarsh2011smart}. Data are available from studies of bacterial quorum sensing \cite{ng2009bacterial}, which itself has been argued to also play a role in sensing environmental features \cite{hense2007does}. The quorum-sensing messenger molecule AHL has $D_\rho \sim 490$ $\mu$m$^2/$s \cite{stewart2003diffusion} and $\nu \sim 0.1$$-$$1$ day$^{-1}$ \cite{trovato2014quorum, schaefer2000detection}, yielding $\lambda \sim 5$$-$$20$ mm. Quorum or swarm sizes $N$ are typically large but can be as small as tens or hundreds of cells \cite{boedicker2009microfluidic}.
Gap junctions are a ubiquitous mediator of juxtacrine signaling \cite{goodenough2009gap}. Gap junctions extracted from mouse tissues with a range of sizes $N$ were found to propagate small molecules over approximately $1$$-$$2$ cell lengths, or $\lambda \sim 10$$-$$20$ $\mu$m \cite{elfgang1995specific}.
Mammary epithelial cells also communicate via gap junctions across $3$$-$$4$ cell lengths, or $\lambda \sim 30$$-$$40$ $\mu$m \cite{ellison2016cell}, and typical sensory units at the ends of mammary ducts contain $N\sim 10^1$$-$$10^3$ cells \cite{ellison2016cell, lu2008genetic}. We see in \fref{phase} that these systems fall within the phases where we would predict that the observed packing strategy of each leads to the larger sensory precision.

Of course, bacteria are single-cellular, whereas many mammalian cells are part of tissues, so one might argue that these packing strategies are predisposed for other functional reasons. However, other multicellular components adopt sparse cell arrangements and exhibit autocrine signaling. Recent experiments have shown that glioblastoma tumor cells in groups of $N \sim 10^3$$-$$10^4$ secrete the autocrine factors IL-6 and VEGF \cite{kravchenko2014glioblastoma}, for which $D_\rho \sim 30$ $\mu$m$^2/$s \cite{goodhill1997diffusion} and $100$ $\mu$m$^2/$s \cite{miura2009vitro}, and $\nu \sim 0.2$ hr$^{-1}$ \cite{marino2007cinacalcet} and $0.7$ hr$^{-1}$ \cite{yen2011two}, respectively, yielding $\lambda \sim 700$ $\mu$m.
Indeed, in this regime, we would predict that sparse packing provides the higher sensory precision (\fref{phase}). In fact, in these experiments, cells autonomously adopted a typical spacing of several cell diameters, which is consistent with the predicted optimal spacing $\langle\ell^*\rangle$ shown in the red contours in Fig.\ 3. The authors argued that this spacing minimized the signaling noise \cite{kravchenko2014glioblastoma}, suggesting a mechanism similar to the one we uncover here.

\begin{figure}
\begin{center}
\includegraphics[width=.9\columnwidth]{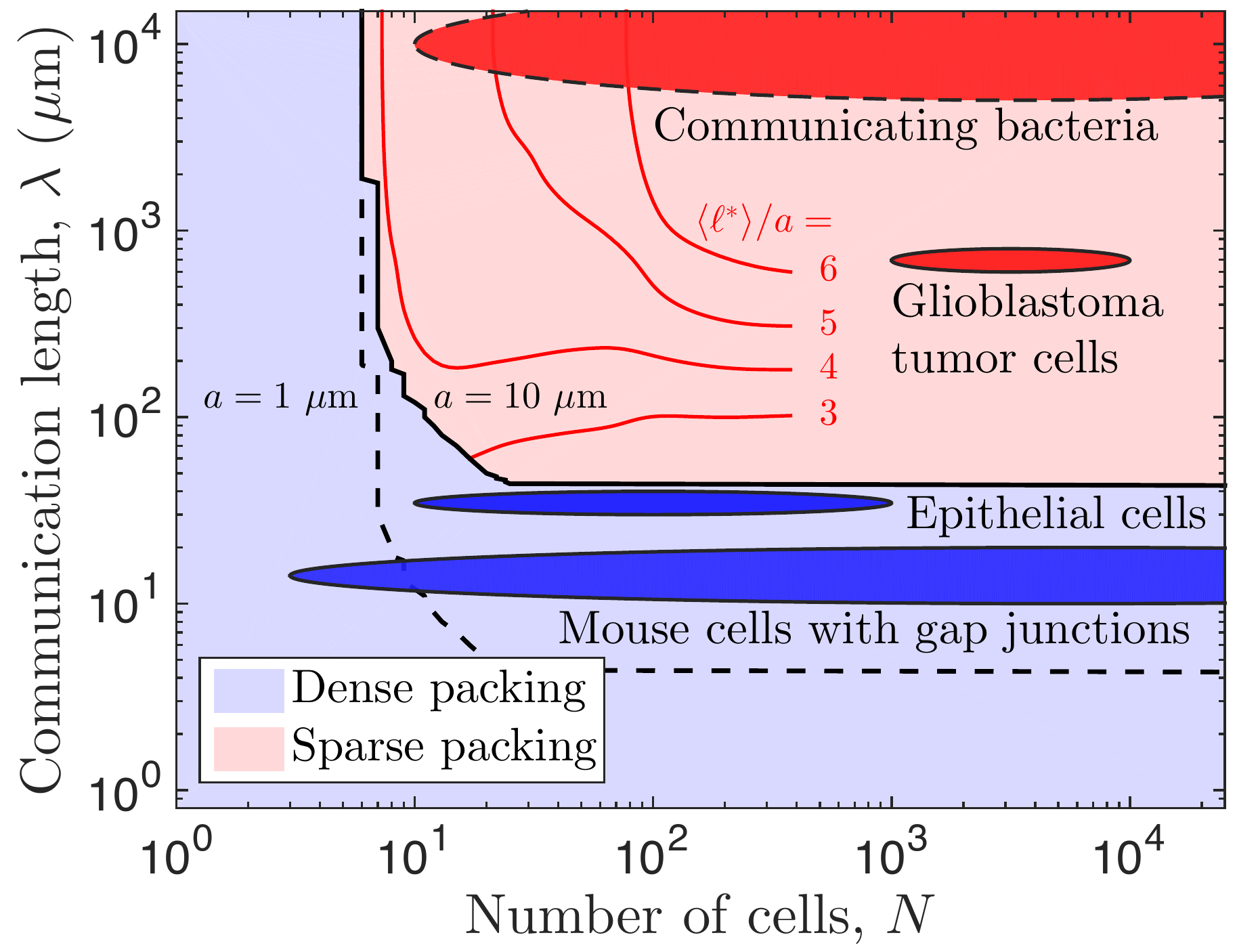}
\end{center}
\caption{Phase plot showing cell arrangement that results in lower error in 3D (results are similar for 2D). In dense packing phase, lower error results from either autocrine signaling with $\langle\ell^*\rangle = 2a$ or juxtacrine signaling. In sparse packing phase, lower error results from autocrine signaling with $\langle\ell^*\rangle > 2a$; red contour lines show $\langle\ell^*\rangle$ for $a=10$ $\mu$m up to largest numerically tractable $N=400$. Ellipses are estimates from biological systems described in the text. Solid (dashed) black phase boundary is for $a = 10$ $\mu$m ($a = 1$ $\mu$m), to be compared with mammalian (bacterial) cell ellipses. Phase boundaries are extrapolated from largest numerically tractable $N = 10^3$.}
\flabel{phase}
\end{figure}

We have shown that communicating cells maximize the precision of concentration sensing by adopting an optimal separation that can be many cell diameters. This is surprising, since separation weakens the impact of the communication. However, we have demonstrated that cells weigh this drawback against the benefit of obtaining independent measurements of their environment. We predict that the concentration detection threshold for communicating cells should decrease with the cell number, which could be tested using an intracellular fluorescent reporter. Moreover, if cell positions are controllable \cite{kravchenko2014glioblastoma}, we predict that a small concentration would be detected by modestly separated cells, but not adjacent or far-apart cells. It will be interesting to test these predictions, as well as to push the theory of collective sensing to further biological contexts.

\begin{acknowledgments}
We thank Ken Ritchie for useful discussions. This work was supported by Simons Foundation grant \#376198.
\end{acknowledgments}

\newpage
\onecolumngrid

\section{Supporting information for ``Fundamental limits to collective concentration sensing in cell populations''}

\section{Juxtacrine Signaling}

We begin with multiple cells all modeled as spheres of radius $a$ in contact with each other. The cells exist within a ligand bath of density profile $c\left(\vec{x},t\right)$ assumed to fluctuate around a spatially constant mean profile ($\bar{c}\left(\vec{x}\right)=\bar{c}$). The ligand molecules diffuse with diffusion constant $D_{c}$ and can bind and unbind to receptors on the surface of each cell at rates $\alpha$ and $\mu$ respectively. Each cell produces a messenger molecule species, $m_{j}$, at a rate $\beta$ proportional to that cell's number of bound receptors, $r_{j}$. This species degrades at rate $\nu$ and can also be exchanged between neighboring cells at rate $\gamma$.

Assume that each cell can be treated as a point particle with respect to the ligand field and that the number of receptors on each cell is large enough to neglect the effects of increased bound receptor number on the overall binding propensity (i.e. receptor saturation). Additionally, let $\mathcal{N}_{j}$ be the set of cells neighboring the $j$th cell. This system can thus be modeled via
\begin{subequations}
\begin{equation}
\frac{\partial c}{\partial t} = D_{c}\nabla^{2}c-\sum_{j}\delta^{3}\left(\vec{x}-\vec{x}_{j}\right)\frac{\partial r_{j}}{\partial t}+\eta_{c}
\label{Juxtsysdefa}
\end{equation}
\begin{equation}
\frac{\partial r_{j}}{\partial t} = \alpha c\left(\vec{x}_{j},t\right)-\mu r_{j}+\eta_{rj}
\label{Juxtsysdefb}
\end{equation}
\begin{equation}
\frac{\partial m_{j}}{\partial t} = \beta r_{j}-\nu m_{j}+\sum_{l\in\mathcal{N}_{j}}\gamma\left(m_{l}-m_{j}\right)+\eta_{mj},
\label{Juxtsysdefc}
\end{equation}
\label{Juxtsysdef}%
\end{subequations}
where $\eta_{c}$ is the noise intrinsic to the diffusion of ligand molecules, $\eta_{rj}$ is the noise intrinsic to the bound receptor number of the $j$th cell, and $\eta_{mj}$ is the noise intrinsic to the creation, degradation, and exchange of $m$ molecules in the $j$th cell. The purpose of this section is to calculate the statistics of the long-time average of the number of messenger molecules in a particular cell based off of this model.

\subsection{Power Spectrum}

First, we discuss the correlation function and power spectrum, to establish some definitions and notation. Specifically, we show that the variance in the long-time average of a variable is given by the low-frequency limit of its power spectrum. For a one dimensional function $x(t)$, the correlation function $C\left(t\right)$ takes the form
\begin{equation}
C\left(t-t'\right) = \left\langle x\left(t'\right)x\left(t\right)\right\rangle.
\label{notcorr}
\end{equation}
Since absolute time is irrelevent in any physical system with no time dependent forcing, $t'$ can be set to 0 without loss of generality. This leads to a definition for the power spectrum of $x(t)$ as
\begin{align}
S\left(\omega\right) &= \int\frac{d\omega'}{2\pi}\left\langle\tilde{x}^{*}\left(\omega'\right)\tilde{x}\left(\omega\right)\right\rangle = \frac{1}{2\pi}\int d\omega'dtdt'\left\langle x\left(t'\right)x\left(t\right)\right\rangle e^{i\omega t}e^{-i\omega't'} \nonumber\\
&= \int dtdt'C\left(t-t'\right)e^{i\omega t}\delta\left(t'\right) = \int dtC\left(t\right)e^{i\omega t}.
\label{notspec}
\end{align}
Thus, under this definition the power spectrum is seen to be the Fourier transform of the correlation function. Additionally, when $x\left(t\right)$ is averaged over a time $T$, the time averaged correlation function of $x\left(t\right)$ takes the form
\begin{align}
C_{T}\left(t-t'\right) &= \left\langle\left(\frac{1}{T}\int_{t'}^{t'+T}d\tau'x\left(\tau'\right)\right)\left(\frac{1}{T}\int_{t}^{t+T}d\tau x\left(\tau\right)\right)\right\rangle \nonumber\\
&= \frac{1}{T^{2}}\int_{t}^{t+T}d\tau\int_{t'}^{t'+T}d\tau'\left\langle x\left(\tau'\right)x\left(\tau\right)\right\rangle \nonumber\\
&= \frac{1}{T^{2}}\int_{t}^{t+T}d\tau\int_{t'}^{t'+T}d\tau'C\left(\tau-\tau'\right)
\label{notcorrT1}
\end{align}
Let $y\equiv\left(\tau-\tau'\right)-\left(t-t'\right)$ and $z\equiv\left(\tau+\tau'\right)-\left(t+t'\right)$. This transforms Eq.\ \ref{notcorrT1} into
\begin{align}
C_{T}\left(t-t'\right) &= \frac{1}{T^{2}}\int_{-T}^{T}dy\int_{\abs{y}}^{2T-\abs{y}}dz\frac{1}{2}C\left(y+t-t'\right) \nonumber\\
&= \frac{1}{T^{2}}\int_{-T}^{T}dy\left(T-\abs{y}\right)C\left(y+t-t'\right).
\label{notcorrT2}
\end{align}
By inverting the relationship found in Eq.\ \ref{notspec}, $C\left(y+t-t'\right)$ can be replaced with an inverse Fourier transform of $S\left(\omega\right)$ to produce
\begin{align}
C_{T}\left(t-t'\right) &= \frac{1}{T^{2}}\int_{-T}^{T}dy\int\frac{d\omega}{2\pi}\left(T-\abs{y}\right)S\left(\omega\right)e^{-i\omega\left(y+t-t'\right)} \nonumber\\
&= \int\frac{d\omega}{2\pi}\left(\frac{2}{\omega T}\sin\left(\frac{\omega T}{2}\right)\right)^{2}S\left(\omega\right)e^{-i\omega\left(t-t'\right)}.
\label{notcorrT3}
\end{align}
The factor of $\left(\omega T\right)^{-2}$ in the integrand of Eq.\ \ref{notcorrT3} forces only small values of $\omega$ to contribute when $T$ is large. Thus, the approximation $S\left(\omega\right)\approx S\left(0\right)$ can be made since only values of $\omega$ near 0 are contributing. This causes $C_{T}\left(0\right)$ to be exactly calculable to
\begin{equation}
C_{T}\left(0\right)\approx S\left(0\right)\int\frac{d\omega}{2\pi}\left(\frac{2}{\omega T}\sin\left(\frac{\omega T}{2}\right)\right)^{2} = \frac{S\left(0\right)}{T}.
\label{notcorr0T1}
\end{equation}

\subsection{Receptor Binding and Unbinding}

We can now begin determining the noise properties of the juxtacrine model, beginning with the receptor binding-unbinding process. Let $c\left(\vec{x},t\right) = \bar{c}+\delta c\left(\vec{x},t\right)$ and $r_{j}\left(t\right) = \bar{r}_{j}+\delta r_{j}\left(t\right)$, where $\bar{r}_{j}$ is the mean value of $r_{j}\left(t\right)$. Eq.\ \ref{Juxtsysdefb} then dictates
\begin{equation}
0 = \alpha\bar{c}-\mu\bar{r}_{j} \implies \bar{r}_{j}=\frac{\alpha\bar{c}}{\mu},
\label{crbarrel}
\end{equation}
while Eqs.\ \ref{Juxtsysdefa} and \ref{Juxtsysdefb} can be written in the form
\begin{subequations}
\begin{equation}
\frac{\partial\delta c}{\partial t} = D_{c}\nabla^{2}\delta c-\sum_{j}\delta^{3}\left(\vec{x}-\vec{x}_{j}\right)\frac{\partial\delta r_{j}}{\partial t}+\eta_{c}
\label{LBUsysdefa}
\end{equation}
\begin{equation}
\frac{\partial\delta r_{j}}{\partial t} = \alpha\delta c\left(\vec{x}_{j},t\right)-\mu\delta r_{j}+\eta_{rj}.
\label{LBUsysdefb}
\end{equation}
\label{LBUsysdef}%
\end{subequations}

Fourier transforming Eq.\ \ref{LBUsysdefa} then yields
\begin{align}
-i\omega\tilde{\delta c} &= -D_{c}k^{2}\tilde{\delta c}-\sum_{j}e^{i\vec{k}\cdot\vec{x}_{j}}\left(-i\omega\tilde{\delta r}_{j}\right)+\tilde{\eta}_{c} \nonumber\\
&\implies \tilde{\delta c} = \frac{i\omega\sum_{j}\tilde{\delta r}_{j}e^{i\vec{k}\cdot\vec{x}_{j}}+\tilde{\eta}_{c}}{D_{c}k^{2}-i\omega}.
\label{FTLBUc}
\end{align}
Similarly Fourier transforming Eq.\ \ref{LBUsysdefb} then yields
\begin{align}
-i\omega\tilde{\delta r}_{j} &= \alpha\int\frac{d^{3}k}{\left(2\pi\right)^{3}}\tilde{\delta c}\left(\vec{k},\omega\right)e^{-i\vec{k}\cdot\vec{x}_{j}}-\mu\tilde{\delta r}_{j}+\tilde{\eta}_{rj} \nonumber\\
&= \alpha\int\frac{d^{3}k}{\left(2\pi\right)^{3}}\frac{i\omega\sum_{l}\tilde{\delta r}_{l}e^{i\vec{k}\cdot\vec{x}_{l}}+\tilde{\eta}_{c}}{D_{c}k^{2}-i\omega}e^{-i\vec{k}\cdot\vec{x}_{j}}-\mu\tilde{\delta r}_{j}+\tilde{\eta}_{rj} \nonumber\\
\implies &\left(\mu-i\omega\right)\tilde{\delta r}_{j}-i\omega\sum_{l}\tilde{\delta r}_{l}\Sigma\left(\vec{x}_{l}-\vec{x}_{j},\omega\right) = \alpha\int\frac{d^{3}k}{\left(2\pi\right)^{3}}\frac{\tilde{\eta}_{c}}{D_{c}k^{2}-i\omega}e^{-i\vec{k}\cdot\vec{x}_{j}}+\tilde{\eta}_{rj},
\label{FTLBUr}
\end{align}
where 
\begin{align}
\Sigma\left(\vec{x},\omega\right)\equiv \alpha\int\frac{d^{3}k}{\left(2\pi\right)^{3}}\frac{1}{D_{c}k^{2}-i\omega}e^{i\vec{k}\cdot\vec{x}} = \frac{\alpha}{4\pi D_{c}\abs{\vec{x}}}e^{-\abs{\vec{x}}\sqrt{\frac{\omega}{2D_{c}}}}e^{i\abs{\vec{x}}\sqrt{\frac{\omega}{2D_{c}}}}.
\label{sigdef}
\end{align}
Thus, it is seen that $\Sigma\left(\vec{x},\omega\right)$ is dependent only on the magnitude of $\vec{x}$, implying $\Sigma\left(-\vec{x},\omega\right)=\Sigma\left(\vec{x},\omega\right)$.

Unfortunately, $\Sigma\left(\vec{x},\omega\right)$ diverges as $\vec{x}\to 0$. This case can be rectified by truncating the range of integration in Eq.\ \ref{sigdef} to be inside of a sphere, $S$, in $k$ space with radius $\frac{2\pi}{ga}$, where $a$ is the cell radius and $g$ is a geometric factor. This lets $\Sigma\left(0,\omega\right)$ to be evaluated as
\begin{align}
\Sigma\left(0,\omega\right) &\approx \alpha\int_{S}\frac{d^{3}k}{\left(2\pi\right)^{3}}\frac{1}{D_{c}k^{2}-i\omega} \nonumber\\
&= \frac{\alpha}{\pi gaD_{c}}\left(1+\frac{ga}{8\pi}\sqrt{\frac{\omega}{2D_{c}}}\left(\log\left(\frac{4D_{c}\pi^{2}+\omega g^{2}a^{2}-2\pi ga\sqrt{2\omega D_{c}}}{4D_{c}\pi^{2}+\omega g^{2}a^{2}+2\pi ga\sqrt{2\omega D_{c}}}\right)-2\arctan\left(\frac{2\pi ga\sqrt{2\omega D_{c}}}{\omega g^{2}a^{2}-4D_{c}\pi^{2}}\right)\right)\right) \nonumber\\
&\quad +\frac{i\alpha}{2\pi\omega}\left(\frac{\abs{\omega}}{2D_{c}}\right)^{\frac{3}{2}} \approx \frac{\alpha}{\pi gaD_{c}}+\frac{i\alpha}{2\pi\omega}\left(\frac{\abs{\omega}}{2D_{c}}\right)^{\frac{3}{2}},
\label{sig0eval}
\end{align}
where the final approximation was made assuming $\omega a^{2}\ll D_{c}$. This is equivalent to the assumption $T\gg \tau_1 = a^2/D_c$, i.e.\ that over a time $T$ the ligand can easily diffuse around the whole cell.

Now, let $R$ be a matrix defined as:
\begin{equation}
R_{jl}\left(\omega\right)\equiv\begin{cases}
\mu-i\omega\left(1+\Sigma\left(0,\omega\right)\right) & j=l \\
-i\omega\Sigma\left(\vec{x}_{j}-\vec{x}_{l},\omega\right) & j\ne l \end{cases}.
\label{Rmatdef}
\end{equation}
Since $\Sigma\left(\vec{x}_{j}-\vec{x}_{l},\omega\right)=\Sigma\left(\vec{x}_{l}-\vec{x}_{j},\omega\right)$, $R$ is seen to be a symmetric matrix. With this, Eq.\ \ref{FTLBUr} can be rewritten as
\begin{align}
\sum_{l}R_{jl}\left(\omega\right)\tilde{\delta r}_{l} &= \alpha\int\frac{d^{3}k}{\left(2\pi\right)^{3}}\frac{\tilde{\eta}_{c}}{D_{c}k^{2}-i\omega}e^{-i\vec{k}\cdot\vec{x}_{j}}+\tilde{\eta}_{rj} \nonumber\\
\implies\tilde{\delta r}_{j} &= \sum_{l}R_{jl}^{-1}\left(\omega\right)\left(\alpha\int\frac{d^{3}k}{\left(2\pi\right)^{3}}\frac{\tilde{\eta}_{c}}{D_{c}k^{2}-i\omega}e^{-i\vec{k}\cdot\vec{x}_{l}}+\tilde{\eta}_{rl}\right).
\label{FTLBUrmat}
\end{align}
Utilizing Eq.\ \ref{FTLBUrmat} yields the cross spectrum of $r_{j}\left(t\right)$ and $r_{l}\left(r\right)$ to be
\begin{align}
&\left\langle\tilde{\delta r}_{l}^{*}\left(\omega'\right)\tilde{\delta r}_{j}\left(\omega\right)\right\rangle = \left\langle\left(\sum_{u}R_{lu}^{-1}\left(\omega'\right)\left(\alpha\int\frac{d^{3}k'}{\left(2\pi\right)^{3}}\frac{\tilde{\eta}_{c}\left(\vec{k}',\omega'\right)}{D_{c}{k'}^{2}-i\omega'}e^{-i\vec{k}'\cdot\vec{x}_{n}}+\tilde{\eta}_{ru}\left(\omega'\right)\right)\right)^{*}\right. \nonumber\\
&\quad\cdot\left.\left(\sum_{s}R_{js}^{-1}\left(\omega\right)\left(\alpha\int\frac{d^{3}k}{\left(2\pi\right)^{3}}\frac{\tilde{\eta}_{c}\left(\vec{k},\omega\right)}{D_{c}k^{2}-i\omega}e^{-i\vec{k}\cdot\vec{x}_{m}}+\tilde{\eta}_{rs}\left(\omega\right)\right)\right)\right\rangle.
\label{rjspec1}
\end{align}
At this point it is necessary to determine the properties of $\eta_{c}$ and $\eta_{rj}$. $\eta_{c}$ in particular is known to have a correlation function of the form

\begin{equation}
\left\langle\eta_{c}\left(\vec{x}',t'\right)\eta_{c}\left(\vec{x},t\right)\right\rangle = 2D_{c}\delta\left(t-t'\right)\vec{\nabla}_{x}\cdot\vec{\nabla}_{x'}\left(\bar{c}\left(\vec{x}\right)\delta^{3}\left(\vec{x}-\vec{x}'\right)\right),
\label{etacorr}
\end{equation}
Performing a Fourier transformation on Eq.\ \ref{etacorr} then yields
\begin{align}
&\left\langle\tilde{\eta}_{c}^{*}\left(\vec{k}',\omega'\right)\tilde{\eta}_{c}\left(\vec{k},\omega\right)\right\rangle = \int d^{3}xd^{3}x'dtdt' \left\langle\eta_{c}\left(\vec{x}',t'\right)\eta_{c}\left(\vec{x},t\right)\right\rangle\left(e^{i\vec{k}\cdot\vec{x}}e^{i\omega t}\right)\left(e^{i\vec{k}'\cdot\vec{x}'}e^{i\omega't'}\right)^{*} \nonumber\\
&\quad = \int d^{3}xd^{3}x'dtdt' 2D_{c}\bar{c}e^{i\left(\vec{k}\cdot\vec{x}-\vec{k}'\cdot\vec{x}'\right)}e^{i\left(\omega t-\omega't'\right)}\delta\left(t-t'\right)\vec{\nabla}_{x}\cdot\vec{\nabla}_{x'}\left(\delta^{3}\left(\vec{x}-\vec{x}'\right)\right),
\label{etaspec1}
\end{align}
as in the main text. Due to the factor of $\delta\left(t-t'\right)$, the integral in $t'$ becomes trivial and leaves the only time dependent factor as $e^{it\left(\omega-\omega'\right)}$. This allows the integral in $t$ to be solved via the Fourier definition of the $d$-dimensional $\delta$ function
\begin{equation}
\delta^{d}\left(\vec{z}\right) = \int\frac{d^{d}\kappa}{\left(2\pi\right)^{d}}e^{i\vec{\kappa}\cdot\vec{z}}.
\label{ndelta}
\end{equation}
By letting $d=1$, $\vec{z} = \omega-\omega'$, and $\kappa = t$, utilizing Eq.\ \ref{ndelta} in Eq.\ \ref{etaspec1} yields
\begin{equation}
\left\langle\tilde{\eta}_{c}^{*}\left(\vec{k}',\omega'\right)\tilde{\eta}_{c}\left(\vec{k},\omega\right)\right\rangle = 2D_{c}\bar{c}\left(2\pi\delta\left(\omega-\omega'\right)\right)\int d^{3}xd^{3}x'e^{i\left(\vec{k}\cdot\vec{x}-\vec{k}'\cdot\vec{x}'\right)}\vec{\nabla}_{x}\cdot\vec{\nabla}_{x'}\left(\delta^{3}\left(\vec{x}-\vec{x}'\right)\right).
\label{etaspec2}
\end{equation}
Eq.\ \ref{ndelta} can then be put back into Eq.\ \ref{etaspec2} by letting $d=3$ and $\vec{z}=\vec{x}$ to change the form of $\delta^{3}\left(\vec{x}-\vec{x}'\right)$,
\begin{align}
&\left\langle\tilde{\eta}_{c}^{*}\left(\vec{k}',\omega'\right)\tilde{\eta}_{c}\left(\vec{k},\omega\right)\right\rangle = 2D_{c}\bar{c}\left(2\pi\delta\left(\omega-\omega'\right)\right)\int d^{3}xd^{3}x'e^{i\left(\vec{k}\cdot\vec{x}-\vec{k}'\cdot\vec{x}'\right)}\vec{\nabla}_{x}\cdot\vec{\nabla}_{x'}\int\frac{d^{3}\kappa}{\left(2\pi\right)^{3}}e^{i\vec{\kappa}\cdot\left(\vec{x}-\vec{x}'\right)} \nonumber\\
&\quad = \frac{2D_{c}\bar{c}}{\left(2\pi\right)^{3}}\left(2\pi\delta\left(\omega-\omega'\right)\right)\int d^{3}xd^{3}x'd^{3}\kappa e^{i\vec{x}\cdot\left(\vec{k}+\vec{\kappa}\right)}e^{-i\vec{x}'\cdot\left(\vec{k}'+\vec{\kappa}\right)}\kappa^{2}.
\label{etaspec3}
\end{align}
Continuing to utilize Eq.\ \ref{ndelta}, all remaining integrals in Eq.\ \ref{etaspec3} either become $\delta$ functions or are over $\delta$ functions by integrating over $x$ then $\kappa$ then $x'$. This yields for the cross spectrum of $\eta_{c}\left(\vec{x},t\right)$
\begin{align}
&\left\langle\tilde{\eta}_{c}^{*}\left(\vec{k}',\omega'\right)\tilde{\eta}_{c}\left(\vec{k},\omega\right)\right\rangle = 2D_{c}\bar{c}\left(2\pi\delta\left(\omega-\omega'\right)\right)\int d^{3}x'd^{3}\kappa\delta^{3}\left(\vec{k}+\vec{\kappa}\right)\kappa e^{-i\vec{x}'\cdot\left(\vec{k}'+\vec{\kappa}\right)}\kappa^{2} \nonumber\\
&\quad = 2D_{c}\bar{c}k^{2}\left(2\pi\delta\left(\omega-\omega'\right)\right)\int d^{3}x' e^{-i\vec{x}'\cdot\left(\vec{k}'-\vec{k}\right)} \nonumber\\
&\quad = 2D_{c}\bar{c}k^{2}\left(2\pi\delta\left(\omega-\omega'\right)\right)\left(\left(2\pi\right)^{3}\delta^{3}\left(\vec{k}-\vec{k}'\right)\right).
\label{etaspec4}
\end{align}

For $\eta_{rj}$, since the binding-unbinding process for a given cell is independent of the binding-unbinding process of any other cell and the ligand diffusion, it must be true that cross correlations between $\eta_{rj}$ and $\eta_{rl}$, for $j\ne l$, or $\eta_{c}$ vanish. Additionally, since Eq.\ \ref{LBUsysdefb} is of the form of a birth-death process, the power spectrum of $\eta_{rj}$ must be the sum of the mean propensities of its reactions. These all imply
\begin{align}
\left\langle\tilde{\eta}_{rl}^{*}\left(\omega'\right),\tilde{\eta}_{rj}\left(\omega\right)\right\rangle &= \left(\alpha\bar{c}+\mu\bar{r}_{j}\right)\delta_{jl}\left(2\pi\delta\left(\omega-\omega'\right)\right) \nonumber\\
&= 2\alpha\bar{c}\delta_{jl}\left(2\pi\delta\left(\omega-\omega'\right)\right),
\label{etarjspec}
\end{align}
as in the main text. Ignoring the vanishing cross terms between $\tilde{\eta}_{c}$ and $\tilde{\eta}_{rj}$ allows Eq.\ \ref{rjspec1} to be written in the form
\begin{align}
&\left\langle\tilde{\delta r}_{l}^{*}\left(\omega'\right)\tilde{\delta r}_{j}\left(\omega\right)\right\rangle = \sum_{s,u}R_{js}^{-1}\left(\omega\right)\left(R_{lu}^{-1}\left(\omega'\right)\right)^{*}\left(\left\langle\tilde{\eta}_{ru}^{*}\left(\omega'\right),\tilde{\eta}_{rs}\left(\omega\right)\right\rangle\vphantom{\frac{\left\langle\tilde{\eta}_{c}^{*}\left(\vec{k}',\omega'\right),\tilde{\eta}_{c}\left(\vec{k},\omega\right)\right\rangle}{\left(D_{c}k^{2}-i\omega\right)\left(D_{c}{k'}^{2}+i\omega'\right)}}\right. \nonumber\\
&\quad\quad \left.+\alpha^{2}\int\frac{d^{3}kd^{3}k'}{\left(2\pi\right)^{6}}\frac{\left\langle\tilde{\eta}_{c}^{*}\left(\vec{k}',\omega'\right),\tilde{\eta}_{c}\left(\vec{k},\omega\right)\right\rangle}{\left(D_{c}k^{2}-i\omega\right)\left(D_{c}{k'}^{2}+i\omega'\right)}e^{i\left(\vec{k}'\cdot\vec{x}_{u}-\vec{k}\cdot\vec{x}_{s}\right)}\right).
\label{rjspec2}
\end{align}
Utilizing Eqs.\ \ref{etaspec4} and \ref{etarjspec} to evaluate the noise correlation terms then yields
\begin{align}
&\left\langle\tilde{\delta r}_{l}^{*}\left(\omega'\right)\tilde{\delta r}_{j}\left(\omega\right)\right\rangle = \sum_{s,u}R_{js}^{-1}\left(\omega\right)\left(R_{lu}^{-1}\left(\omega'\right)\right)^{*}\left(2\alpha\bar{c}\delta_{su}\left(2\pi\delta\left(\omega-\omega'\right)\right) \vphantom{\frac{2D_{c}\bar{c}k^{2}\left(2\pi\delta\left(\omega-\omega'\right)\right)\left(\left(2\pi\right)^{3}\delta^{3}\left(\vec{k}-\vec{k}'\right)\right)}{\left(D_{c}k^{2}-i\omega\right)\left(D_{c}{k'}^{2}+i\omega'\right)}}\right. \nonumber\\
&\quad\quad \left.+\alpha^{2}\int\frac{d^{3}kd^{3}k'}{\left(2\pi\right)^{6}}\frac{2D_{c}\bar{c}k^{2}\left(2\pi\delta\left(\omega-\omega'\right)\right)\left(\left(2\pi\right)^{3}\delta^{3}\left(\vec{k}-\vec{k}'\right)\right)}{\left(D_{c}k^{2}-i\omega\right)\left(D_{c}{k'}^{2}+i\omega'\right)}e^{i\left(\vec{k}'\cdot\vec{x}_{u}-\vec{k}\cdot\vec{x}_{s}\right)}\right) \nonumber\\
&\quad = 2\alpha\bar{c}\left(2\pi\delta\left(\omega-\omega'\right)\right)\sum_{s,u}R_{js}^{-1}\left(\omega\right)\left(R_{lu}^{-1}\left(\omega\right)\right)^{*}\left(\delta_{su} +\alpha\int\frac{d^{3}k}{\left(2\pi\right)^{3}}\frac{D_{c}k^{2}}{\left(D_{c}k^{2}\right)^{2}+\omega^{2}}e^{i\vec{k}\cdot\left(\vec{x}_{u}-\vec{x}_{s}\right)}\right),
\label{rjspec2}
\end{align}
where all instances of $\omega'$ outside of the $\delta$ function were freely replaced with $\omega$ due to the $\delta$ function being a global factor. At this point it is useful to consider the relation

\begin{equation}
\int d\Omega_{\kappa}e^{i\vec{\kappa}\cdot\vec{z}} = 4\pi\frac{\sin\left(\kappa\abs{\vec{z}}\right)}{\kappa\abs{\vec{z}}}.
\label{solidFI}
\end{equation}

Eq.\ \ref{solidFI} thus shows that the angular portion of the integral in Eq.\ \ref{rjspec2} will cause the imaginary component to vanish. Thus, the integral is completely real and may be expressed as the real part of $\Sigma\left(\vec{x}_{u}-\vec{x}_{s},\omega\right)$
\begin{align}
&\left\langle\tilde{\delta r}_{l}^{*}\left(\omega'\right)\tilde{\delta r}_{j}\left(\omega\right)\right\rangle \nonumber\\
&\quad = 2\alpha\bar{c}\left(2\pi\delta\left(\omega-\omega'\right)\right)\sum_{s,u}R_{js}^{-1}\left(\omega\right)\left(R_{lu}^{-1}\left(\omega\right)\right)^{*}\left(\delta_{su} +\text{Re}\left(\Sigma\left(\vec{x}_{u}-\vec{x}_{s},\omega\right)\right)\right).
\label{rjspec3}
\end{align}
Comparing the last term in Eq.\ \ref{rjspec3} with Eq.\ \ref{Rmatdef} then allows Eq.\ \ref{rjspec3} to be written as
\begin{align}
\left\langle\tilde{\delta r}_{l}^{*}\left(\omega'\right)\tilde{\delta r}_{j}\left(\omega\right)\right\rangle = 2\alpha\bar{c}\left(2\pi\delta\left(\omega-\omega'\right)\right)\sum_{s,u}R_{js}^{-1}\left(\omega\right)\left(R_{lu}^{-1}\left(\omega\right)\right)^{*}\left(\frac{1}{\omega}\text{Im}\left(\left(R_{us}\left(\omega\right)\right)^{*}\right)\right).
\label{rjspec4}
\end{align}

In order to simplify Eq.\ \ref{rjspec4}, first let $a$ and $b$ be two arbitrary complex numbers. The product $a\text{Im}\left(b\right)$ can be reordered as
\begin{align}
a\text{Im}\left(b\right) &= \text{Re}\left(a\right)\text{Im}\left(b\right)+i\text{Im}\left(a\right)\text{Im}\left(b\right) \nonumber\\
&= \left(\text{Re}\left(a\right)\text{Im}\left(b\right)+\text{Im}\left(a\right)\text{Re}\left(b\right)\right)-\text{Im}\left(a\right)\left(\text{Re}\left(b\right)-i\text{Im}\left(b\right)\right) \nonumber\\
&= \text{Im}\left(ab\right)-b^{*}\text{Im}\left(a\right) = \text{Im}\left(ab\right)+b^{*}\text{Im}\left(a^{*}\right).
\label{abprodrule1}
\end{align}
Applying Eq.\ \ref{abprodrule1} to the $\left(R_{lu}^{-1}\left(\omega\right)\right)^{*}\text{Im}\left(\left(R_{us}\left(\omega\right)\right)^{*}\right)$ term in Eq.\ \ref{rjspec4} yields
\begin{align}
\left\langle\tilde{\delta r}_{l}^{*}\left(\omega'\right)\tilde{\delta r}_{j}\left(\omega\right)\right\rangle &= \frac{2\alpha\bar{c}}{\omega}\left(2\pi\delta\left(\omega-\omega'\right)\right)\sum_{s,u}R_{js}^{-1}\left(\omega\right) \nonumber\\
&\quad \cdot\left( \text{Im}\left(\left(R_{lu}^{-1}\left(\omega\right)\right)^{*}\left(R_{us}\left(\omega\right)\right)^{*}\right)+ R_{us}\left(\omega\right)\text{Im}\left(R_{lu}^{-1}\left(\omega\right)\right)\right).
\label{rjspec5}
\end{align}
Separating out the sums then yields
\begin{align}
\left\langle\tilde{\delta r}_{l}^{*}\left(\omega'\right)\tilde{\delta r}_{j}\left(\omega\right)\right\rangle &= \frac{2\alpha\bar{c}}{\omega}\left(2\pi\delta\left(\omega-\omega'\right)\right)\left(\sum_{s}R_{js}^{-1}\left(\omega\right)\sum_{u} \text{Im}\left(\left(R_{lu}^{-1}\left(\omega\right)R_{us}\left(\omega\right)\right)^{*}\right)\right. \nonumber\\
&\quad \left.+\sum_{u}\text{Im}\left(R_{lu}^{-1}\left(\omega\right)\right)\sum_{s}R_{js}^{-1}\left(\omega\right)R_{us}\left(\omega\right)\right).
\label{rjspec6}
\end{align}
In the first term of Eq.\ \ref{rjspec6} the summation over $u$ can be brought inside the $\text{Im}$ and complex conjugation operators and causes the product $R_{lu}^{-1}\left(\omega\right)R_{us}\left(\omega\right)$ to collapse to $\delta_{ls}$. However, since the Kronecker $\delta$ function is real, its imaginary component must be 0, and thus the whole first term vanishes. In the second term, the fact that $R$ is symmetric can be used to make the substitutions $R_{lu}^{-1}\left(\omega\right)\to R_{ul}^{-1}\left(\omega\right)$ and $R_{us}\left(\omega\right)\to R_{su}\left(\omega\right)$, which causes Eq.\ \ref{rjspec6} to simplify to
\begin{align}
\left\langle\tilde{\delta r}_{l}^{*}\left(\omega'\right)\tilde{\delta r}_{j}\left(\omega\right)\right\rangle &= \frac{2\alpha\bar{c}}{\omega}\left(2\pi\delta\left(\omega-\omega'\right)\right)\sum_{u}\text{Im}\left(R_{ul}^{-1}\left(\omega\right)\right) \sum_{s}R_{js}^{-1}\left(\omega\right)R_{su}\left(\omega\right) \nonumber\\
&= \frac{2\alpha\bar{c}}{\omega}\left(2\pi\delta\left(\omega-\omega'\right)\right)\sum_{u}\text{Im}\left(R_{ul}^{-1}\left(\omega\right)\right)\delta_{ju} \nonumber\\
&= \frac{2\alpha\bar{c}}{\omega}\left(2\pi\delta\left(\omega-\omega'\right)\right)\text{Im}\left(R_{jl}^{-1}\left(\omega\right)\right).
\label{rjspec7}
\end{align}
Thus, Eq.\ \ref{rjspec7} can be seen to be formed from the imaginary component of the matrix element used to connect $\tilde{\delta r}_{j}$ to the noise terms, exactly as would be predicted by the fluctuation-dissipation theorem.

Under the limit $\mu\gg\omega\left(1+\Sigma\left(0,\omega\right)\right)$, which is equivalent to $\omega\ll\left(\mu^{-1}+\left(k_{D}K_{D}\right)^{-1}\right)^{-1}$ for $k_{D}=\pi gaD_{c}$ and $K_{D} = \frac{\mu}{\alpha}$ as in the text, $R_{jl}^{-1}$ can be easily approximated. Let $A_{jl}$ be a matrix equivalent to $R$ with the $j$th row and $l$th column removed such that $\left(-1\right)^{j+l}\text{det}\left(A_{jl}\right) = C_{jl}$, where $C$ is the cofactor matrix of $R$. Under this definition,
\begin{equation}
R_{jl}^{-1}\left(\omega\right) = \left(-1\right)^{j+l}\frac{\text{det}\left(A_{lj}\right)}{\text{det}\left(R\right)}.
\label{Rinv}
\end{equation}
By the rules of determinants, any term within the determinant of $R$ that has off-diagonal elements must have at least 2 off-diagonal elements. Since all the off-diagonal terms of $R$ are of the form $-i\omega\Sigma\left(\vec{x},\omega\right)$ and $\Sigma\left(\vec{x},\omega\right)$ does not diverge as $\omega\to 0$, each of these terms must have a factor of $\omega$ that is of order 2 or higher. Thus, in the small $\omega$ limit, only the diagonal elements of $R$ contribute to its determinant. For $j\ne l$, the creation of $A_{jl}$ will cause two of the diagonal elements of $R$ to be removed and $\abs{j-l}-1$ more to be shifted to off-diagonal elements. This will in turn cause every term in the determinant of $A_{jl}$ to have at least one factor of the form $-i\omega\Sigma\left(\vec{x},\omega\right)$. However, only the term with all $N-2$ remaining factors of $\mu-i\omega\left(1+\Sigma\left(0,\omega\right)\right)$, where $N$ is the number of cells, will have an order of $\omega$ less than 2 and will thus be the only nonnegligible term. This term will also carry a prefactor of $\left(-1\right)^{\abs{j-l}-1}$ by the rules of determinants. With this, $R_{jl}^{-1}$ takes the form
\begin{align}
R_{jl}^{-1}\left(\omega\right)&\approx\left(-1\right)^{j+l+\abs{j-l}-1}\frac{-i\omega\Sigma\left(\vec{x}_{j}-\vec{x}_{l},\omega\right)\left(\mu-i\omega\left(1+\Sigma\left(0,\omega\right)\right)\right)^{N-2}}{\left(\mu-i\omega\left(1+\Sigma\left(0,\omega\right)\right)\right)^{N}} \nonumber\\
&= \frac{i\omega\Sigma\left(\vec{x}_{j}-\vec{x}_{l},\omega\right)}{\left(\mu-i\omega\left(1+\Sigma\left(0,\omega\right)\right)\right)^{2}}.
\label{Rinvoffd}
\end{align}
Thus, $R_{jl}^{-1}$ is seen to carry a dependence on the separation between the two cells. For $j=l$, $A_{jj}$ is a symmetric matrix, and just like $R$, all of its off-diagonal components are of the form $-i\omega\Sigma\left(\vec{x},\omega\right)$. By the same argument as that used for the determinant of $R$, only the diagonal terms contribute to the determinant of $A_{jj}$. This lets $R_{jj}^{-1}\left(\omega\right)$ to be approximated as
\begin{equation}
R_{jj}^{-1}\left(\omega\right)\approx\left(-1\right)^{2j}\frac{\left(\mu-i\omega\left(1+\Sigma\left(0,\omega\right)\right)\right)^{N-1}}{\left(\mu-i\omega\left(1+\Sigma\left(0,\omega\right)\right)\right)^{N}} = \frac{1}{\mu-i\omega\left(1+\Sigma\left(0,\omega\right)\right)},
\label{Rinvdiag}
\end{equation}
which is identical to $R^{-1}$ for the $N=1$ cell case. Thus, for long time averaging the presence of other cells does not affect any individual cell'��s power spectrum at the level of bound receptors. However, Eq.\ \ref{Rinvoffd} makes clear that the cross-spectrum of bound receptor numbers (the off-diagonal terms) does indeed depend on the distances between cells, as it inherits the ligand cross-correlations. We will see below that the statistics of the messenger molecules, since they are communicated among cells, depend on this cross-spectrum, and thus pick up the ligand cross-correlations.

This single cell power spectrum can be computed as
\begin{align}
\left\langle\tilde{\delta r}^{*}\left(\omega'\right)\tilde{\delta r}\left(\omega\right)\right\rangle &= \frac{2\alpha\bar{c}}{\omega}\left(2\pi\delta\left(\omega-\omega'\right)\right)\text{Im}\left(\left(\mu-i\omega\left(1+\Sigma\left(0,\omega\right)\right)\right)^{-1}\right) \nonumber\\
&= \frac{2\alpha\bar{c}}{\omega}\left(2\pi\delta\left(\omega-\omega'\right)\right)\frac{\omega\left(1+\text{Re}\left(\Sigma\left(0,\omega\right)\right)\right)}{\left(\mu+\omega\text{Im}\left(\Sigma\left(0,\omega\right)\right)\right)^{2}+\omega^{2}\left(1+\text{Re}\left(\Sigma\left(0,\omega\right)\right)\right)^{2}}
\label{singcellrjspec}
\end{align}
\begin{align}
\implies S_{r}\left(\omega\right) &= \int\frac{d\omega'}{2\pi}\left\langle\tilde{\delta r}^{*}\left(\omega'\right)\tilde{\delta r}\left(\omega\right)\right\rangle \nonumber\\
&= \frac{2\alpha\bar{c}\left(1+\text{Re}\left(\Sigma\left(0,\omega\right)\right)\right)}{\left(\mu+\omega\text{Im}\left(\Sigma\left(0,\omega\right)\right)\right)^{2}+\omega^{2}\left(1+\text{Re}\left(\Sigma\left(0,\omega\right)\right)\right)^{2}}.
\label{singcellrjpowspec}
\end{align}
To obtain the time averaged noise-to-signal ratio, Eq.\ \ref{notcorr0T1} can be utilized assuming $T\gg\tau_{2}=\mu^{-1}+\left(k_{D}K_{D}\right)^{-1}$ as was done for $\omega$ eariler. With this, the time averaged noise-to-signal ratio in the single cell case can be approximated to be
\begin{equation}
\frac{C_{r_{T}}\left(0\right)}{\bar{r}^{2}} \approx \frac{S_{r}\left(0\right)}{\bar{r}^{2}T} = \frac{1}{\bar{r}^{2}T}\frac{2\alpha\bar{c}\left(1+\text{Re}\left(\Sigma\left(0,0\right)\right)\right)}{\mu^{2}}.
\label{rjNSR1}
\end{equation}
Utilizing Eqs.\ \ref{crbarrel} and \ref{sig0eval} as well as setting $g=4$ as explained in the text, Eq.\ \ref{rjNSR1} simplifies to
\begin{align}
\frac{(\delta r)^{2}}{\bar{r}^{2}} = \frac{C_{r_{T}}\left(0\right)}{\bar{r}^{2}} &\approx \frac{1}{\bar{r}^{2}T}\frac{2\alpha\bar{c}\left(1+\frac{\alpha}{4\pi aD_{c}}\right)}{\mu^{2}} = \frac{1}{\pi aD_{c}T}\frac{\alpha^{2}\bar{c}}{2\mu^{2}\bar{r}^{2}}+\frac{2}{T}\frac{\alpha\bar{c}}{\mu^{2}\bar{r}^{2}} \nonumber\\
&= \frac{1}{2}\frac{1}{\pi a\bar{c}D_{c}T}+\frac{2}{\mu\bar{r}T}.
\label{rjNSR2}
\end{align}
Eq.\ \ref{rjNSR2} can be seen to be the sum of the noise the receptors inherit from the ligand diffusion, the $\frac{1}{2}\frac{1}{\pi a\bar{c}D_{c}T}$ term, and the noise inherent in the ligand binding-unbinding process itself, the $\frac{2}{\mu\bar{r}T}$ term.

\subsection{Messenger Molecule}

As before, let $m_{j}\left(t\right) = \bar{m}_{j}+\delta m_{j}\left(t\right)$, where $\bar{m}_{j}$ is the mean value of $m_{j}\left(t\right)$. Eq.\ \ref{Juxtsysdefc} then dictates
\begin{equation}
0 = \beta\bar{r}_{j}-\nu\bar{m}_{j}+\sum_{l\in\mathcal{N}_{j}}\gamma\left(\bar{m}_{l}-\bar{m}_{j}\right).
\label{rmbarrel1}
\end{equation}
Assuming the binding and unbinding as well as the production and degredation parameters are the same in each cell, Eq.\ \ref{crbarrel} forces $\bar{r}_{j}=\bar{r}_{l}$ for all $j$ and $l$, which by Eq.\ \ref{rmbarrel1} then forces 
\begin{equation}
\bar{m}_{j}=\bar{m}_{l}=\frac{\beta}{\nu}\bar{r}_{j}
\label{rmbarrel2}
\end{equation}
for all $j$ and $l$. Additionally, Eq.\ \ref{Juxtsysdefc} also dictates
\begin{equation}
\frac{\partial\delta m_{j}}{\partial t} = \beta\delta r_{j}-\nu\delta m_{j}+\sum_{l\in\mathcal{N}_{j}}\gamma\left(\delta m_{l}-\delta m_{j}\right)+\eta_{mj},
\label{LJuxtc}
\end{equation}
which can be Fourier transformed into
\begin{equation}
-i\omega\tilde{\delta m}_{j} = \beta\tilde{\delta r}_{j}-\nu\tilde{\delta m}_{j}+\sum_{l\in\mathcal{N}_{j}}\gamma\left(\tilde{\delta m}_{l}-\tilde{\delta m}_{j}\right)+\tilde{\eta}_{mj}.
\label{FLJuxtc1}
\end{equation}

Let $N_{j}$ be the number of cells neighboring the $j$th cell and the matrix $M$ be defined as
\begin{equation}
M_{jl}\left(\omega\right) = \begin{cases}
\nu+N_{j}\gamma-i\omega & j=l \\
-\gamma & l\in\mathcal{N}_{j} \\
0 & \text{otherwise} \end{cases}.
\label{Mdef}
\end{equation}
Thus, $M$ is seen to be a symmetric matrix. The form of $M$ also dictates that when $\omega$ is taken to be a small parameter later, it must meet the requirement $\omega\ll\nu+\gamma$ as those are the variables $\omega$ is seen to be compared to in $M$. This notation allows Eq.\ \ref{FLJuxtc1} to be written as
\begin{equation}
\sum_{l}M_{jl}\left(\omega\right)\tilde{\delta m}_{l} = \beta\tilde{\delta r}_{j}+\tilde{\eta}_{mj} \implies \tilde{\delta m}_{j} = \sum_{l}M_{jl}^{-1}\left(\omega\right)\left(\beta\tilde{\delta r}_{l}+\tilde{\eta}_{ml}\right)
\label{FLJuxtc2}
\end{equation}
Utilizing Eq.\ \ref{FLJuxtc2} yields the cross spectrum of $m_{j}\left(t\right)$ and $m_{l}\left(t\right)$ to be
\begin{align}
\left\langle\tilde{\delta m}_{l}^{*}\left(\omega'\right)\tilde{\delta m}_{j}\left(\omega\right)\right\rangle &= \left\langle\left(\sum_{u}M_{lu}^{-1}\left(\omega'\right)\left(\beta\tilde{\delta r}_{u}\left(\omega'\right)+\tilde{\eta}_{mu}\left(\omega'\right)\right)\right)^{*}\right. \nonumber\\
&\quad \left.\cdot\left(\sum_{s}M_{js}^{-1}\left(\omega\right)\left(\beta\tilde{\delta r}_{s}\left(\omega\right)+\tilde{\eta}_{ms}\left(\omega\right)\right)\right)\right\rangle.
\label{mjspec1}
\end{align}
At this point it is necessary to determine the properties of $\eta_{mj}$. Just as for $\eta_{rj}$, Eq.\ \ref{Juxtsysdefc} is in the form of a birth-death process, which allows the power spectrum of $\eta_{mj}$ to simply be written as the sum of the mean propensities. However, due to the exchange term, $\eta_{mj}$ and $\eta_{ml}$ cannot be independent if $l\in\mathcal{N}_{j}$. The cross spectrum, in this case, must be negative due to the fact that exchange means one cell is losing $m$ molecules when the other is gaining them and must also be the sum of propensities of the exchange reaction. Thus, the power spectrum of $\eta_{mj}$ takes the form
\begin{align}
\left\langle\tilde{\eta}_{ml}^{*}\left(\omega'\right)\tilde{\eta}_{mj}\left(\omega\right)\right\rangle &= \left(\beta\bar{r}_{j}+\nu\bar{m}_{j}+\sum_{s\in\mathcal{N}_{j}}\gamma\left(\bar{m}_{s}+\bar{m}_{j}\right)\right)\delta_{jl}\left(2\pi\delta\left(\omega-\omega'\right)\right) \nonumber\\
&\quad -\gamma\left(\bar{m}_{l}+\bar{m}_{j}\right)\delta_{l\in\mathcal{N}_{j}}\left(2\pi\delta\left(\omega-\omega'\right)\right),
\label{etamspec1}
\end{align}
which is the Fourier transform of the $\eta_{mj}$ correlator in the main text. Utilizing Eqs.\ \ref{rmbarrel2} and \ref{Mdef} allows Eq.\ \ref{etamspec1} to be simplified to
\begin{equation}
\left\langle\tilde{\eta}_{ml}^{*}\left(\omega'\right)\tilde{\eta}_{mj}\left(\omega\right)\right\rangle = 2\bar{m}\text{Re}\left(M_{jl}\left(\omega\right)\right)\left(2\pi\delta\left(\omega-\omega'\right)\right)
\label{etamspec2}
\end{equation}
Since the ligand binding-unbinding process is independent of the noise in the $m$ molecule production, any cross terms between $\delta r_{j}$ and $\eta_{ml}$ must vanish. This allows Eq.\ \ref{mjspec1} to be written as
\begin{align}
&\left\langle\tilde{\delta m}_{l}^{*}\left(\omega'\right)\tilde{\delta m}_{j}\left(\omega\right)\right\rangle = \sum_{s,u}M_{js}^{-1}\left(\omega\right)\left(M_{lu}^{-1}\left(\omega'\right)\right)^{*}\left(\beta^{2}\left\langle\tilde{\delta r}_{u}^{*}\left(\omega'\right)\tilde{\delta r}_{s}\left(\omega\right)\right\rangle+\left\langle\tilde{\eta}_{ms}^{*}\left(\omega'\right)\tilde{\eta}_{mu}\left(\omega\right)\right\rangle\right) \nonumber\\
&\quad =2\left(2\pi\delta\left(\omega-\omega'\right)\right)\sum_{s,u}M_{js}^{-1}\left(\omega\right)\left(M_{lu}^{-1}\left(\omega\right)\right)^{*}\left(\beta^{2}\frac{\alpha\bar{c}}{\omega}\text{Im}\left(R_{su}^{-1}\left(\omega\right)\right)+\bar{m}\text{Re}\left(\left(M_{su}\left(\omega\right)\right)^{*}\right)\right),
\label{mjspec2}
\end{align}
where Eq.\ \ref{rjspec7} has been used, $\text{Re}\left(M_{su}\left(\omega\right)\right)$ from Eq.\ \ref{etamspec2} has been freely changed to $\text{Re}\left(\left(M_{su}\left(\omega\right)\right)^{*}\right)$, and all instances of $\omega'$ outside the $\delta$ function were freely replaced with $\omega$ due to the $\delta$ function being a global factor.

In order to simplify Eq.\ \ref{mjspec2}, first let $a$ and $b$ be two arbitrary complex numbers. The product $a\text{Re}\left(b\right)$ can be reordered as
\begin{align}
a\text{Re}\left(b\right) &= \text{Re}\left(a\right)\text{Re}\left(b\right)+i\text{Im}\left(a\right)\text{Re}\left(b\right) \nonumber\\
&= \left(\text{Re}\left(a\right)\text{Re}\left(b\right)-\text{Im}\left(a\right)\text{Im}\left(b\right)\right)+\text{Im}\left(a\right)\left(\text{Im}\left(b\right)+i\text{Re}\left(b\right)\right) \nonumber\\
&= \text{Re}\left(ab\right)+ib^{*}\text{Im}\left(a\right) = \text{Re}\left(ab\right)-ib^{*}\text{Im}\left(a^{*}\right)
\label{abprodrule2}
\end{align}
Temporarily ignorning the $R_{su}^{-1}$ term and applying Eq.\ \ref{abprodrule2} to the $\left(M_{lu}^{-1}\left(\omega\right)\right)^{*}\text{Re}\left(\left(M_{su}\left(\omega\right)\right)^{*}\right)$ term in Eq.\ \ref{mjspec2} yields
\begin{align}
&\sum_{s,u}M_{js}^{-1}\left(\omega\right)\left(M_{lu}^{-1}\left(\omega\right)\right)^{*}\text{Re}\left(\left(M_{su}\left(\omega\right)\right)^{*}\right) \nonumber\\
&\quad = \sum_{s,u}M_{js}^{-1}\left(\omega\right)\left(\text{Re}\left(\left(M_{lu}^{-1}\left(\omega\right)M_{su}\left(\omega\right)\right)^{*}\right)-iM_{su}\left(\omega\right)\text{Im}\left(M_{lu}^{-1}\left(\omega\right)\right)\right).
\label{mjspecsum1}
\end{align}
Separating out the sums and freely changing $\text{Re}\left(\left(M_{lu}^{-1}\left(\omega\right)M_{su}\left(\omega\right)\right)^{*}\right)$ to $\text{Re}\left(M_{lu}^{-1}\left(\omega\right)M_{su}\left(\omega\right)\right)$ then yields
\begin{align}
&\sum_{s,u}M_{js}^{-1}\left(\omega\right)\left(M_{lu}^{-1}\left(\omega\right)\right)^{*}\text{Re}\left(\left(M_{su}\left(\omega\right)\right)^{*}\right) \nonumber\\
&\quad = \sum_{s}M_{js}^{-1}\left(\omega\right)\sum_{u}\text{Re}\left(M_{lu}^{-1}\left(\omega\right)M_{su}\left(\omega\right)\right)-i\sum_{u}\text{Im}\left(M_{lu}^{-1}\left(\omega\right)\right)\sum_{s}M_{js}^{-1}\left(\omega\right)M_{su}\left(\omega\right).
\label{mjspecsum2}
\end{align}
Since $M$ is symmetric, $M_{su}\left(\omega\right)$ in the first term can be freely changed to $M_{us}\left(\omega\right)$. This implies that when the summation over $u$ is brought inside the Re operator, $M_{lu}^{-1}\left(\omega\right)M_{su}\left(\omega\right)$ will collapse to $\delta_{ls}$, which has no imaginary part. Thus, the entire summation simplifies to $\delta_{ls}$. Similarly, the summation over $s$ in the second term will collapse into $\delta_{ju}$. These along with once again using the symmetry of $M$ to freely change $M_{lu}^{-1}\left(\omega\right)$ to $M_{ul}^{-1}\left(\omega\right)$ in the second term then yields
\begin{align}
&\sum_{s,u}M_{js}^{-1}\left(\omega\right)\left(M_{lu}^{-1}\left(\omega\right)\right)^{*}\text{Re}\left(\left(M_{su}\left(\omega\right)\right)^{*}\right) = \sum_{s}M_{js}^{-1}\left(\omega\right)\delta_{ls}-i\sum_{u}\text{Im}\left(M_{ul}^{-1}\left(\omega\right)\right)\delta_{ju} \nonumber\\
&\quad = M_{jl}^{-1}\left(\omega\right)-i\text{Im}\left(M_{jl}^{-1}\left(\omega\right)\right) = \text{Re}\left(M_{jl}^{-1}\left(\omega\right)\right).
\label{mjspecsum3}
\end{align}
Applying Eq.\ \ref{mjspecsum3} to Eq.\ \ref{mjspec2} then yields
\begin{align}
&\left\langle\tilde{\delta m}_{l}^{*}\left(\omega'\right)\tilde{\delta m}_{j}\left(\omega\right)\right\rangle \nonumber\\
&\quad =2\left(2\pi\delta\left(\omega-\omega'\right)\right)\left(\bar{m}\text{Re}\left(M_{jl}^{-1}\left(\omega\right)\right)+\frac{\alpha\beta^{2}\bar{c}}{\omega}\sum_{s,u}M_{js}^{-1}\left(\omega\right)\left(M_{lu}^{-1}\left(\omega\right)\right)^{*}\text{Im}\left(R_{su}^{-1}\left(\omega\right)\right)\right).
\label{mjspec3}
\end{align}
This allows the power spectrum of $m_{j}$ to take the form
\begin{align}
S_{m}\left(\omega\right) &= \int\frac{d\omega'}{2\pi}\left\langle\tilde{\delta m}_{j}^{*}\left(\omega'\right)\tilde{\delta m}_{j}\left(\omega\right)\right\rangle \nonumber\\
&= 2\bar{m}\text{Re}\left(M_{jj}^{-1}\left(\omega\right)\right)+\frac{2\alpha\beta^{2}\bar{c}}{\omega}\sum_{s,u}M_{js}^{-1}\left(\omega\right)\left(M_{ju}^{-1}\left(\omega\right)\right)^{*}\text{Im}\left(R_{su}^{-1}\left(\omega\right)\right),
\label{mjspec4}
\end{align}
which by Eq.\ \ref{notcorr0T1} yields a noise-to-signal ratio of
\begin{align}
\frac{\left(\delta m\right)^{2}}{\bar{m}^{2}} &= \frac{C_{m,T}\left(0\right)}{\bar{m}^{2}} \approx \frac{S_{m}\left(0\right)}{\bar{m}^{2}T} \nonumber\\
&= \frac{2}{\bar{m}T}\text{Re}\left(M_{jj}^{-1}\left(0\right)\right)+\frac{2\mu\nu^{2}}{\bar{r}T}\sum_{s,u}\underbrace{M_{js}^{-1}\left(0\right)\left(M_{ju}^{-1}\left(0\right)\right)^{*}}_\text{integration}\underbrace{\left(\lim_{\omega\to 0}\frac{1}{\omega}\text{Im}\left(R_{su}^{-1}\left(\omega\right)\right)\right)}_\text{correlation},
\label{mjNSR}
\end{align}
where Eqs.\ \ref{crbarrel} and \ref{rmbarrel2} were used to alter the form of the prefactor in the second term.

\begin{figure}
\centering
\includegraphics[width=.5\textwidth]{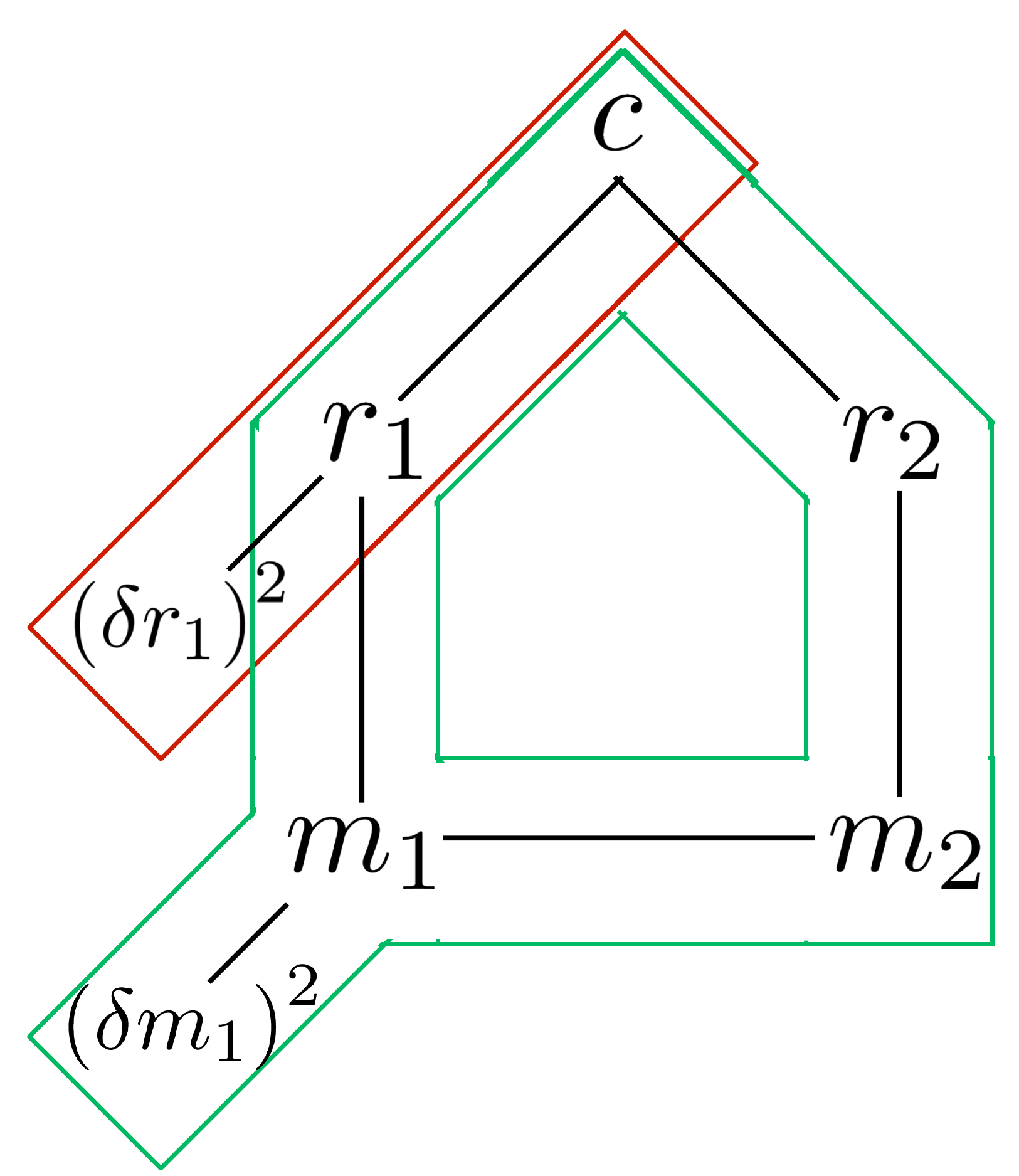}
\caption{The variables $r$ and $m$ and their respective variances are shown with black lines connecting the variables which have direct connections. The variances of each variable depend on particular subsets of the variables. These subsets are shown bounded by the various colors. $\left(\delta r_{1}\right)^{2}$, bounded by red, can be seen through Eq.\ \ref{rjNSR2} to have two distinct terms, one which depends on $c$ and the other on $r$. It is also known from Eq.\ \ref{Rinvdiag} that $r_{2}$ has no influence on the variance of $r_{1}$. $\left(\delta m_{1}\right)^{2}$, bounded by green, can be seen through Eq.\ \ref{mjNSR} (and Eq.\ \ref{ParamjNSR} in the autocrine case) to depend on $c$, $r_{1}$, and $r_{2}$ through the correlation factor and on $m_{1}$ and $m_{2}$ through the intrinsic noise terms as well as the integration factor. Physically, this means that the variance in any individual cell's bound receptor number does not depend on the presence of other cells, since the receptors are not communicated. Nonetheless, the variance in any individual cell's messenger molecule number does depend on the other cells, since the messenger molecules are communicated. Specifically, a given cell's messenger molecules depend on the bound receptor cross-spectrum, which inherits the ligand cross-correlations mediated by diffusion.}
\label{suppfig}
\end{figure}

The second term in Eq.\ \ref{mjNSR} consists of the noise that the messenger molecules inherit from the dynamics of the bound receptor numbers and ligand field. The summation can be seen to be composed of two distinct factors. The first factor, $M_{js}^{-1}\left(0\right)\left(M_{ju}^{-1}\left(0\right)\right)^{*}$, comes from the matrix $M$ which dictates which cells can communicate directly. Thus, this factor is highly dependent on the geometry of the cluster and can be interpreted as accounting for the spatial integration of the messenger molecule. The second factor, $\lim_{\omega\to 0}\frac{1}{\omega}\text{Im}\left(R_{su}^{-1}\left(\omega\right)\right)$, comes from Eq.\ \ref{rjspec7} and thus directly accounts for the cross correlations between the bound receptor numbers in each cell. This interplay between the ligand field, bound receptor number, and messenger molecule count is visualized for a two cell system in Fig. \ref{suppfig}.

Eq.\ \ref{mjNSR} can be further separated into a more intuitive form by acknowledging the explicit forms of the $R$ and $M$ matrices. As seen in Eq.\ \ref{Mdef}, $M$ has the same units as $\nu$, so clustering $M$ or $M^{-1}$ with $\nu^{-1}$ or $\nu$ respectively will produce a unitless factor dependent only on $\gamma/\nu$. Additionally, Eqs.\ \ref{Rinvoffd} and \ref{Rinvdiag} can be used to rewrite the correlation factor as
\begin{equation}
\lim_{\omega\to 0}\frac{1}{\omega}\text{Im}\left(R_{su}^{-1}\left(\omega\right)\right) = \frac{1}{\mu^{2}}\left(\delta_{su}+\text{Re}\left(\Sigma\left(\vec{x}_{s}-\vec{x}_{u},0\right)\right)\right) = \frac{1}{\mu^{2}}\left(\delta_{su}+\frac{\alpha}{4\pi aD_{c}}\Theta\left(\frac{\abs{\vec{x}_{s}-\vec{x}_{u}}}{a}\right)\right),
\label{Rrewrite}
\end{equation}
where
\begin{equation}
\Theta\left(x\right) = \begin{cases}
1 & x<1 \\
\frac{1}{x} & x\ge 1 \end{cases}
\label{Thetadef}
\end{equation}
is a formalization of the approximation made in Eq.\ \ref{sig0eval}. These, along with further simplifications from Eqs.\ \ref{crbarrel} and \ref{rmbarrel2} allow Eq.\ \ref{mjNSR} to be rewritten as
\begin{align}
\frac{\left(\delta m\right)^{2}}{\bar{m}^{2}} = &\frac{2}{\nu\bar{m}T}\text{Re}\left(\nu M_{jj}^{-1}\left(0\right)\right)+\frac{2}{\mu\bar{r}T}\sum_{s}\abs{\nu M_{js}^{-1}\left(0\right)}^{2} \nonumber\\
&+\frac{1}{2}\frac{1}{\pi a\bar{c}D_{c}T}\sum_{s,u}\left(\nu M_{js}^{-1}\left(0\right)\right)\left(\nu M_{ju}^{-1}\left(0\right)\right)^{*}\Theta\left(\frac{\abs{\vec{x}_{s}-\vec{x}_{u}}}{a}\right).
\label{mjNSRrewrite}
\end{align}
Eq.\ \ref{mjNSRrewrite} is the general expression for the error in the case of juxtacrine signaling and, when only the extrinsic term is considered, is utilized in calculating the phase boundaries in Fig. 3 of the main text.

From here, a few properties of $M$ can be used to simplify Eq.\ \ref{mjNSR} in particular limits of $\gamma$ or $N$. First, for $\gamma\ll\nu$, $M\left(0\right)$ approximately becomes $\nu I_{N}$, where $I_{N}$ is the identity matrix of rank $N$. This makes inverting $M$ trivial and reduces Eq.\ \ref{mjNSR} to
\begin{align}
\frac{\left(\delta m\right)^{2}}{\bar{m}^{2}} &= \frac{2}{\nu\bar{m}T}+\frac{2\nu^{2}}{\mu\bar{r}T}\sum_{s,u}\frac{1}{\nu^{2}}\delta_{js}\delta_{ju}\left(\delta_{su}+\text{Re}\left(\Sigma\left(\vec{x}_{s}-\vec{x}_{u},0\right)\right)\right) = \frac{2}{\nu\bar{m}T}+\frac{2}{\mu\bar{r}T}\left(1+\text{Re}\left(\Sigma\left(0,0\right)\right)\right) \nonumber\\
&= \frac{1}{2}\frac{1}{\pi a\bar{c}D_{c}T}+\frac{2}{\mu\bar{r}T}+\frac{2}{\nu\bar{m}T},
\label{mjspec0g}
\end{align}
as seen in Eq.\ 4 of the main text. Since setting $\gamma\ll\nu$ is equivalent to eliminating any communication between cells and the extrinsic noises from the ligand diffusion and binding and unbinding process are unaffected by the presence of other cells as determined in the previous section, the cell is effectively sensing no effects from the presence of any other cell. Thus, this result must also be valid for a single isolated cell.

Second, for $\gamma\gg\nu$ the cells are effectively communicating at infinite speed. Physically, this implies that every cell in the cluster communicates with every other cell equally, which mathematically translates to the dictation that $M_{jl}^{-1}$ must be independent of $j$ and $l$. Let $\vec{X}$ be a column vector with unit entries. By Eq.\ \ref{Mdef} it is easy to see that
\begin{equation}
M\vec{X} = \left(\nu-i\omega\right)\vec{X}.
\label{Meig}
\end{equation}
Thus, $\vec{X}$ is an eigenvector of $M$, which in turn implies it must also be an eigenvector of $M^{-1}$ satisfying
\begin{equation}
M^{-1}\vec{X} = \left(\nu-i\omega\right)^{-1}\vec{X}.
\label{Minveig}
\end{equation}
Since $M_{jl}^{-1}$ must be independent of $j$ and $l$, the only way Eq.\ \ref{Minveig} can be true is if
\begin{equation}
\lim_{\gamma\to\infty}M_{jl}^{-1} = \frac{1}{N\left(\nu-i\omega\right)}.
\label{Minvlim}
\end{equation}
This allows Eq.\ \ref{mjNSRrewrite} to reduce to
\begin{align}
\frac{\left(\delta m\right)^{2}}{\bar{m}^{2}} &= \frac{2}{\nu\bar{m}TN}+\frac{2}{\mu\bar{r}T}\sum_{s}\frac{1}{N^{2}}+\frac {1}{2}\frac {1}{\pi a\bar{c}D_{c}T}\sum_{s,u}\frac{1}{N^{2}}\Theta\left(\frac{\abs{\vec{x}_{s}-\vec{x}_{u}}}{a}\right) \nonumber\\
&= \frac{2}{\nu\bar{m}TN}+\frac {2}{\mu\bar{r}TN}+\frac{1}{2}\frac{1}{\pi a\bar{c}D_{c}TN^{2}}\sum_{s,u}\Theta\left(\frac{\abs{\vec{x}_{s}-\vec{x}_{u}}}{a}\right)
\label{mjNSRinfg}
\end{align}
Once again, when only the extrinsic term proportional to $\text{Re}\left(\Sigma\left(\vec{x}_{s}-\vec{x}_{u},0\right)\right)$ is considered, Eq.\ \ref{mjNSRinfg} is utilized in calculating the data depicted in Fig. 2 of the main text.

Finally, Eq.\ \ref{mjNSR} can be easily represented under no assumptions about $\gamma$ but for the limiting case of $N=2$ cells. When the cells are adjacent to each other, $M$ and $M^{-1}$ take the form
\begin{subequations}
\begin{equation}
M = \begin{bmatrix}
\nu+\gamma-i\omega & -\gamma \\
-\gamma & \nu+\gamma-i\omega \end{bmatrix}
\label{2cellJuxtMa}
\end{equation}
\begin{equation}
M^{-1} = \frac{1}{\left(\nu-i\omega\right)\left(\nu-i\omega+2\gamma\right)}\begin{bmatrix}
\nu+\gamma-i\omega & \gamma \\
\gamma & \nu+\gamma-i\omega \end{bmatrix}.
\label{2cellJuxtMb}
\end{equation}
\label{2cellJuxtM}%
\end{subequations}
Let $\vec{\ell}=\vec{x}_{1}-\vec{x}_{2}$. With these and Eqs.\ \ref{Rinvoffd} and \ref{Rinvdiag}, Eq.\ \ref{mjNSR} can be calculated to yield
\begin{align}
\frac{(\delta m)^{2}}{\bar{m}^{2}} &= \frac{2}{\bar{m}T}\text{Re}\left(M_{11}^{-1}\left(0\right)\right)+\frac{2\nu^{2}}{\mu\bar{r}T}\sum_{s,u}M_{js}^{-1}\left(0\right)\left(M_{lu}^{-1}\left(0\right)\right)^{*}\left(\delta_{su}+\text{Re}\left(\Sigma\left(\vec{x}_{s}-\vec{x}_{u},0\right)\right)\right) \nonumber\\
&= \frac{2}{\bar{m}T}\frac{\nu+\gamma}{\nu\left(\nu+2\gamma\right)}+\frac{2\nu^{2}}{\mu\bar{r}T}\left(2\frac{\nu+\gamma}{\nu\left(\nu+2\gamma\right)}\frac{\gamma}{\nu\left(\nu+2\gamma\right)}\text{Re}\left(\Sigma\left(\vec{\ell},0\right)\right)\vphantom{\left(\left(\frac{\nu+\gamma}{\nu\left(\nu+2\gamma\right)}\right)^{2}\right)}\right. \nonumber\\
&\quad \left.+\left(\left(\frac{\nu+\gamma}{\nu\left(\nu+2\gamma\right)}\right)^{2}+\left(\frac{\gamma}{\nu\left(\nu+2\gamma\right)}\right)^{2}\right)\left(1+\text{Re}\left(\Sigma\left(0,0\right)\right)\right)\right) \nonumber\\
&= \frac{1}{\pi a\bar{c}D_{c}T}\frac{\nu^{2}+3\nu\gamma+3\gamma^{2}}{2\left(\nu+2\gamma\right)^{2}}+\frac{1}{\mu\bar{r}T}\frac{2\left(\nu^{2}+2\nu\gamma+2\gamma^{2}\right)}{\left(\nu+2\gamma\right)^{2}}+\frac{1}{\nu\bar{m}T}\frac{2\left(\nu+\gamma\right)}{\nu+2\gamma},
\label{m1SJuxt1}
\end{align}
where $\abs{\vec{\ell}}$ has been set to be exactly $2a$ to reflect the necessity of the cells to be adjacent to each other in order to exchange $m$ molecules. Eq.\ \ref{m1SJuxt1} can be seen to be easily separable into the three distinct terms which reflect the noise inherited from the ligand diffusion, binding-unbinding process, and $m$ birth-death and exchange processes. Of important note is that under the $\gamma\ll\nu$ limit Eq.\ \ref{m1SJuxt1} reduces to Eq.\ \ref{rjNSR2} plus the term $\frac{2}{\nu\bar{m}T}$. Conversely, under the $\gamma\gg\nu$ limit (which is equivalent to $\lambda=2a\sqrt{\frac{\gamma}{\nu}}\gg a$) the second fraction of each term go to $\frac{3}{8}$, 1, and 1 respectively. The first term, which is the extrinsic noise, is reproduced in Eq.\ 5 of the main text.

\section{Autocrine signaling}

We here assume that the cells produce a messenger molecule with the same production and degradation rate as in the previous section, but instead of being exchanged between neighboring cells they are secreted into the environment and diffuse with a diffusion constant of $D_{\rho}$. The purpose of this section is to calculate the statistics of the long-time average of the number of messenger molecules within the volume of a particular cell.

This system can be modeled via
\begin{subequations}
\begin{equation}
\frac{\partial c}{\partial t} = D_{c}\nabla^{2}c-\sum_{j}\delta^{3}\left(\vec{x}-\vec{x}_{j}\right)\frac{\partial r_{j}}{\partial t}+\eta_{c}
\label{Parasysdefa}
\end{equation}
\begin{equation}
\frac{\partial r_{j}}{\partial t} = \alpha c\left(\vec{x}_{j},t\right)-\mu r_{j}+\eta_{rj}
\label{Parasysdefb}
\end{equation}
\begin{equation}
\frac{\partial \rho}{\partial t} = D_{\rho}\nabla^{2}\rho-\nu\rho+\sum_{j}\delta^{3}\left(\vec{x}-\vec{x}_{j}\right)\left(\beta r_{j}+\eta_{pj}\right)+\eta_{d},
\label{Parasysdefc}
\end{equation}
\label{Parasysdef}%
\end{subequations}
where $\rho\left(\vec{x},t\right)$ is the density field of the diffusing messenger molecule and $\eta_{pj}$ and $\eta_{d}$ are the production from the $j$th cell, and the degradation and diffusive noise terms, respectively, as in the main text. Again, let $\rho\left(\vec{x},t\right)=\bar{\rho}\left(\vec{x}\right)+\delta\rho\left(\vec{x},t\right)$, where $\bar{\rho}\left(\vec{x}\right)$ is the mean value of $\rho\left(\vec{x},t\right)$ as a function of space. Since $\rho$ is being produced at each cell and allowed to diffuse, $\bar{\rho}\left(\vec{x}\right)$ cannot be constant in space and by Eq.\ \ref{Parasysdefc} must obey
\begin{equation}
0 = D_{\rho}\nabla^{2}\bar{\rho}-\nu\bar{\rho}+\sum_{j}\delta^{3}\left(\vec{x}-\vec{x}_{j}\right)\beta\bar{r}_{j},
\label{rhobarrel1}
\end{equation}
which is solved by
\begin{equation}
\bar{\rho}\left(\vec{x}\right) = \frac{1}{4\pi D_{\rho}}\sum_{j}\frac{\beta\bar{r}_{j}}{\abs{\vec{x}-\vec{x}_{j}}}e^{-\abs{\vec{x}-\vec{x}_{j}}\sqrt{\frac{\nu}{D_{\rho}}}},
\label{rhobarrel2}
\end{equation}
as in Eq.\ 8 of the main text. Additionally, Eq.\ \ref{Parasysdefc} also dictates
\begin{equation}
\frac{\partial\delta\rho}{\partial t} = D_{\rho}\nabla^{2}\delta\rho-\nu\delta\rho+\sum_{j}\delta^{3}\left(\vec{x}-\vec{x}_{j}\right)\left(\beta\delta r_{j}+\eta_{pj}\right)+\eta_{d},
\label{LPara}
\end{equation}
which can be Fourier transformed into
\begin{align}
-i\omega\tilde{\delta\rho} &= -D_{\rho}k^{2}\tilde{\delta\rho}-\nu\tilde{\delta\rho}+\sum_{j}e^{i\vec{k}\cdot\left(\vec{x}-\vec{x}_{j}\right)}\left(\beta\tilde{\delta r}_{j}+\tilde{\eta}_{pj}\right)+\tilde{\eta}_{d} \nonumber\\
&\implies \tilde{\delta\rho} = \frac{\sum_{j}e^{i\vec{k}\cdot\vec{x}_{j}}\left(\beta\tilde{\delta r}_{j}+\tilde{\eta}_{pj}\right)+\tilde{\eta}_{d}}{\nu+D_{\rho}k^{2}-i\omega}.
\label{FTLPara}
\end{align}
Since the noise in each $r_{j}$ is independent of the noise in the production and degradation/diffusion of $\rho$, and the production and diffusion noises must be independent of each other, the cross spectrum of $\rho\left(\vec{x},t\right)$ can be written as
\begin{align}
&\left\langle\tilde{\delta\rho}^{*}\left(\vec{k}',\omega'\right)\tilde{\delta\rho}\left(\vec{k},\omega\right)\right\rangle = \left\langle\left(\frac{\sum_{l}e^{i\vec{k'}\cdot\vec{x}_{l}}\left(\beta\tilde{\delta r}_{l}\left(\omega'\right)+\tilde{\eta}_{pl}\left(\omega'\right)\right)+\tilde{\eta}_{d}\left(\vec{k}',\omega'\right)}{\nu+D_{\rho}{k'}^{2}-i\omega'}\right)^{*}\right. \nonumber\\
&\quad\quad \cdot\left.\left(\frac{\sum_{j}e^{i\vec{k}\cdot\vec{x}_{j}}\left(\beta\tilde{\delta r}_{j}\left(\omega\right)+\tilde{\eta}_{pj}\left(\omega\right)\right)+\tilde{\eta}_{d}\left(\vec{k},\omega\right)}{\nu+D_{\rho}k^{2}-i\omega}\right)\right\rangle \nonumber\\
&\quad = \frac{\sum_{j,l}e^{i\left(\vec{k}\cdot\vec{x}_{j}-\vec{k}'\cdot\vec{x}_{l}\right)}\left(\beta^{2}\left\langle\tilde{\delta r}_{l}^{*}\left(\omega'\right)\tilde{\delta r}_{j}\left(\omega\right)\right\rangle+\left\langle\tilde{\eta}_{pl}^{*}\left(\omega'\right)\tilde{\eta}_{pj}\left(\omega\right)\right\rangle\right)+ \left\langle\tilde{\eta}_{d}^{*}\left(\vec{k}',\omega'\right)\tilde{\eta}_{d}\left(\vec{k},\omega\right)\right\rangle}{\left(\nu+D_{\rho}k^{2}-i\omega\right)\left(\nu+D_{\rho}{k'}^{2}+i\omega'\right)}.
\label{rhospec1}
\end{align}

From here the spectra of both $\eta_{pj}$ and $\eta_{d}$ are needed. Since $\eta_{d}$ is also a diffusive noise term, it must follow the same formalism used in Eq.\ \ref{etacorr}. However, unlike $c$, $\rho$ can degrade, meaning a degradation term must be added to the noise. This yields
\begin{equation}
\left\langle\eta_{d}\left(\vec{x}',t'\right)\eta_{d}\left(\vec{x},t\right)\right\rangle = 2D_{\rho}\delta\left(t-t'\right)\vec{\nabla}_{x}\cdot\vec{\nabla}_{x'}\left(\bar{\rho}\left(\vec{x}\right)\delta^{3}\left(\vec{x}-\vec{x}'\right)\right)+\nu\bar{\rho}\left(\vec{x}\right)\delta\left(t-t'\right)\delta^{3}\left(\vec{x}-\vec{x}'\right),
\label{etarhocorr}
\end{equation}
as in the mian text. Eq.\ \ref{etarhocorr} can then be Fourier transformed to yield
\begin{align}
&\left\langle\tilde{\eta}_{d}^{*}\left(\vec{k}',\omega'\right)\tilde{\eta}_{d}\left(\vec{k},\omega\right)\right\rangle = \int d^{3}xd^{3}x'dtdt'\left\langle\eta_{d}\left(\vec{x}',t'\right)\eta_{d}\left(\vec{x},t\right)\right\rangle\left(e^{i\vec{k}\cdot\vec{x}}e^{i\omega t}\right)\left(e^{i\vec{k}'\cdot\vec{x}'}e^{i\omega' t'}\right)^{*} \nonumber\\
&\quad = 2D_{\rho}\int d^{3}xd^{3}x'dtdt'e^{i\left(\vec{k}\cdot\vec{x}-\vec{k}'\cdot\vec{x}'\right)}e^{i\left(\omega t-\omega' t'\right)}\delta\left(t-t'\right)\vec{\nabla}_{x}\cdot\vec{\nabla}_{x'}\left(\bar{\rho}\left(\vec{x}\right)\delta^{3}\left(\vec{x}-\vec{x}'\right)\right) \nonumber\\
&\quad\quad +\nu\int d^{3}xd^{3}x'dtdt'e^{i\left(\vec{k}\cdot\vec{x}-\vec{k}'\cdot\vec{x}'\right)}e^{i\left(\omega t-\omega' t'\right)}\bar{\rho}\left(\vec{x}\right)\delta\left(t-t'\right)\delta^{3}\left(\vec{x}-\vec{x}'\right).
\label{etarhospec1}
\end{align}
The $\delta$ function makes the $t'$ integrals trivial, which leaves the only time dependent term in the integrands as $e^{it\left(\omega-\omega'\right)}$. Eq.\ \ref{ndelta} can then be used to solve the $t$ integral and transform $\delta^{3}\left(\vec{x}-\vec{x}'\right)$ into an integral form in the first integral. In the second integral, the $\delta^{3}$ function makes the $x'$ integral trivial. Combining these with Eq.\ \ref{rhobarrel2} allows Eq.\ \ref{etarhospec1} to be written as
\begin{align}
&\left\langle\tilde{\eta}_{d}^{*}\left(\vec{k}',\omega'\right)\tilde{\eta}_{d}\left(\vec{k},\omega\right)\right\rangle = \frac{1}{2\pi}\left(2\pi\delta\left(\omega-\omega'\right)\right)\int d^{3}xd^{3}x'e^{i\left(\vec{k}\cdot\vec{x}-\vec{k}'\cdot\vec{x}'\right)} \nonumber\\
&\quad\quad \cdot\vec{\nabla}_{x}\cdot\vec{\nabla}_{x'}\left(\left(\sum_{j}\frac{\beta\bar{r}_{j}}{\abs{\vec{x}-\vec{x}_{j}}}e^{-\abs{\vec{x}-\vec{x}_{j}}\sqrt{\frac{\nu}{D_{\rho}}}}\right)\left(\int\frac{d^{3}\kappa}{\left(2\pi\right)^{3}}e^{i\vec{\kappa}\cdot\left(\vec{x}-\vec{x}'\right)}\right)\right) \nonumber\\
&\quad\quad +\frac{\nu}{4\pi D_{\rho}}\left(2\pi\delta\left(\omega-\omega'\right)\right)\int d^{3}xe^{i\vec{x}\cdot\left(\vec{k}-\vec{k}'\right)}\sum_{j}\frac{\beta\bar{r}_{j}}{\abs{\vec{x}-\vec{x}_{j}}}e^{-\abs{\vec{x}-\vec{x}_{j}}\sqrt{\frac{\nu}{D_{\rho}}}}.
\label{etarhospec2}
\end{align}
Focusing on the first integral, moving the $\kappa$ integral outside the gradient operators before applying them then yields
\begin{align}
&\left\langle\tilde{\eta}_{d}^{*}\left(\vec{k}',\omega'\right)\tilde{\eta}_{d}\left(\vec{k},\omega\right)\right\rangle = \frac{1}{\left(2\pi\right)^{4}}\left(2\pi\delta\left(\omega-\omega'\right)\right)\int d^{3}xd^{3}x'd^{3}\kappa e^{i\left(\vec{k}\cdot\vec{x}-\vec{k}'\cdot\vec{x}'\right)}\left(-i\vec{\kappa}\right) \nonumber\\
&\quad\quad \cdot\left(i\vec{\kappa}e^{i\vec{\kappa}\cdot\left(\vec{x}-\vec{x}'\right)}\sum_{j}\frac{\beta\bar{r}_{j}}{\abs{\vec{x}-\vec{x}_{j}}}e^{-\abs{\vec{x}-\vec{x}_{j}}\sqrt{\frac{\nu}{D_{\rho}}}}+e^{i\vec{\kappa}\cdot\left(\vec{x}-\vec{x}'\right)}\sum_{j}\vec{\nabla}_{x}\frac{\beta\bar{r}_{j}}{\abs{\vec{x}-\vec{x}_{j}}}e^{-\abs{\vec{x}-\vec{x}_{j}}\sqrt{\frac{\nu}{D_{\rho}}}}\right) \nonumber\\
&\quad\quad +\frac{\nu}{4\pi D_{\rho}}\left(2\pi\delta\left(\omega-\omega'\right)\right)\int d^{3}xe^{i\vec{x}\cdot\left(\vec{k}-\vec{k}'\right)}\sum_{j}\frac{\beta\bar{r}_{j}}{\abs{\vec{x}-\vec{x}_{j}}}e^{-\abs{\vec{x}-\vec{x}_{j}}\sqrt{\frac{\nu}{D_{\rho}}}}\nonumber\\
&\quad = \frac{1}{\left(2\pi\right)^{4}}\left(2\pi\delta\left(\omega-\omega'\right)\right)\int d^{3}xd^{3}x'd^{3}\kappa e^{i\vec{x}\cdot\left(\vec{k}+\vec{\kappa}\right)}e^{-i\vec{x}'\cdot\left(\vec{k}'+\vec{\kappa}\right)} \nonumber\\
&\quad\quad \cdot\sum_{j}\left(\kappa^{2}\frac{\beta\bar{r}_{j}}{\abs{\vec{x}-\vec{x}_{j}}}e^{-\abs{\vec{x}-\vec{x}_{j}}\sqrt{\frac{\nu}{D_{\rho}}}}-i\vec{\kappa}\cdot\vec{\nabla}_{x}\frac{\beta\bar{r}_{j}}{\abs{\vec{x}-\vec{x}_{j}}}e^{-\abs{\vec{x}-\vec{x}_{j}}\sqrt{\frac{\nu}{D_{\rho}}}}\right) \nonumber\\
&\quad\quad +\frac{\nu}{4\pi D_{\rho}}\left(2\pi\delta\left(\omega-\omega'\right)\right)\int d^{3}xe^{i\vec{x}\cdot\left(\vec{k}-\vec{k}'\right)}\sum_{j}\frac{\beta\bar{r}_{j}}{\abs{\vec{x}-\vec{x}_{j}}}e^{-\abs{\vec{x}-\vec{x}_{j}}\sqrt{\frac{\nu}{D_{\rho}}}}.
\label{etarhospec3}
\end{align}
Since $x'$ only appears in the term $e^{-i\vec{x}'\cdot\left(\vec{k}'+\vec{\kappa}\right)}$, Eq.\ \ref{ndelta} can again be used to solve the $x'$ integral, which will then make the $\kappa$ integral trivial due to the resultant $\delta$ function. This causes Eq.\ \ref{etarhospec3} to simplify to
\begin{align}
&\left\langle\tilde{\eta}_{d}^{*}\left(\vec{k}',\omega'\right)\tilde{\eta}_{d}\left(\vec{k},\omega\right)\right\rangle = \frac{1}{2\pi}\left(2\pi\delta\left(\omega-\omega'\right)\right)\int d^{3}xd^{3}\kappa e^{i\vec{x}\cdot\left(\vec{k}+\vec{\kappa}\right)}\delta^{3}\left(\vec{k}'+\vec{\kappa}\right) \nonumber\\
&\quad\quad \cdot\sum_{j}\left(\kappa^{2}\frac{\beta\bar{r}_{j}}{\abs{\vec{x}-\vec{x}_{j}}}e^{-\abs{\vec{x}-\vec{x}_{j}}\sqrt{\frac{\nu}{D_{\rho}}}}-i\vec{\kappa}\cdot\vec{\nabla}_{x}\frac{\beta\bar{r}_{j}}{\abs{\vec{x}-\vec{x}_{j}}}e^{-\abs{\vec{x}-\vec{x}_{j}}\sqrt{\frac{\nu}{D_{\rho}}}}\right) \nonumber\\
&\quad\quad +\frac{\nu}{4\pi D_{\rho}}\left(2\pi\delta\left(\omega-\omega'\right)\right)\int d^{3}xe^{i\vec{x}\cdot\left(\vec{k}-\vec{k}'\right)}\sum_{j}\frac{\beta\bar{r}_{j}}{\abs{\vec{x}-\vec{x}_{j}}}e^{-\abs{\vec{x}-\vec{x}_{j}}\sqrt{\frac{\nu}{D_{\rho}}}}\nonumber\\
&\quad = \frac{1}{2\pi}\left(2\pi\delta\left(\omega-\omega'\right)\right)\int d^{3}x e^{i\vec{x}\cdot\left(\vec{k}-\vec{k}'\right)} \nonumber\\
&\quad\quad \cdot\sum_{j}\beta\bar{r}_{j}\left({k'}^{2}\frac{1}{\abs{\vec{x}-\vec{x}_{j}}}e^{-\abs{\vec{x}-\vec{x}_{j}}\sqrt{\frac{\nu}{D_{\rho}}}}+i\vec{k}'\cdot\vec{\nabla}_{x}\frac{1}{\abs{\vec{x}-\vec{x}_{j}}}e^{-\abs{\vec{x}-\vec{x}_{j}}\sqrt{\frac{\nu}{D_{\rho}}}}\right) \nonumber\\
&\quad\quad +\frac{\nu}{4\pi D_{\rho}}\left(2\pi\delta\left(\omega-\omega'\right)\right)\int d^{3}xe^{i\vec{x}\cdot\left(\vec{k}-\vec{k}'\right)}\sum_{j}\frac{\beta\bar{r}_{j}}{\abs{\vec{x}-\vec{x}_{j}}}e^{-\abs{\vec{x}-\vec{x}_{j}}\sqrt{\frac{\nu}{D_{\rho}}}}.
\label{etarhospec4}
\end{align}
Let $\vec{v}_{j}=\vec{x}-\vec{x}_{j}$ in both integrals. Since the $x$ integral is over all of $x$-space, this transformation does not change the limits of integration. Additionally, $d^{3}x=d^{3}v_{j}$ and $\vec{\nabla}_{x}=\vec{\nabla}_{v_{j}}$ due to $x$ and $v_{j}$ being related by a simple translation. Utilizing this and moving the summations in Eq.\ \ref{etarhospec4} outside the integrals allows it to be written as
\begin{align}
&\left\langle\tilde{\eta}_{d}^{*}\left(\vec{k}',\omega'\right)\tilde{\eta}_{d}\left(\vec{k},\omega\right)\right\rangle = \frac{1}{2\pi}\left(2\pi\delta\left(\omega-\omega'\right)\right)\sum_{j}\beta\bar{r}_{j}e^{i\vec{x}_{j}\cdot\left(\vec{k}-\vec{k}'\right)}\int d^{3}v_{j}e^{i\vec{v}_{j}\cdot\left(\vec{k}-\vec{k}'\right)} \nonumber\\
&\quad\quad \cdot\left({k'}^{2}\frac{1}{v_{j}}e^{-v_{j}\sqrt{\frac{\nu}{D_{\rho}}}}+i\vec{k}'\cdot\vec{\nabla}_{v_{j}}\frac{1}{v_{j}}e^{-v_{j}\sqrt{\frac{\nu}{D_{\rho}}}}\right) \nonumber\\
&\quad\quad +\frac{\nu}{4\pi D_{\rho}}\left(2\pi\delta\left(\omega-\omega'\right)\right)\sum_{j}\beta\bar{r}_{j}e^{i\vec{x}_{j}\cdot\left(\vec{k}-\vec{k}'\right)}\int d^{3}v_{j}\frac{1}{v_{j}}e^{-v_{j}\sqrt{\frac{\nu}{D_{\rho}}}}e^{i\vec{v}_{j}\cdot\left(\vec{k}-\vec{k}'\right)}.
\label{etarhospec5}
\end{align}
Since $\frac{1}{\abs{\vec{v}_{j}}}e^{-\abs{\vec{v}_{j}}\sqrt{\frac{\nu}{D_{\rho}}}}$ goes to 0 exponentially as $\abs{\vec{v}_{j}}\to\infty$, the second term in the first integral of Eq.\ \ref{etarhospec5} can be integrated by parts with the net result of simply adding a factor of $-1$ and moving the gradient to apply to $e^{i\vec{v}_{j}\cdot\left(\vec{k}-\vec{k}'\right)}$. This causes Eq.\ \ref{etarhospec5} to simplify to
\begin{align}
&\left\langle\tilde{\eta}_{d}^{*}\left(\vec{k}',\omega'\right)\tilde{\eta}_{d}\left(\vec{k},\omega\right)\right\rangle = \frac{1}{2\pi}\left(2\pi\delta\left(\omega-\omega'\right)\right)\sum_{j}\beta\bar{r}_{j}e^{i\vec{x}_{j}\cdot\left(\vec{k}-\vec{k}'\right)} \nonumber\\
&\quad\quad \cdot\int d^{3}v_{j}\frac{1}{v_{j}}e^{-v_{j}\sqrt{\frac{\nu}{D_{\rho}}}}\left({k'}^{2}e^{i\vec{v}_{j}\cdot\left(\vec{k}-\vec{k}'\right)}-i\vec{k}'\cdot\vec{\nabla}_{v_{j}}e^{i\vec{v}_{j}\cdot\left(\vec{k}-\vec{k}'\right)}\right) \nonumber\\
&\quad\quad +\frac{\nu}{4\pi D_{\rho}}\left(2\pi\delta\left(\omega-\omega'\right)\right)\sum_{j}\beta\bar{r}_{j}e^{i\vec{x}_{j}\cdot\left(\vec{k}-\vec{k}'\right)}\int d^{3}v_{j}\frac{1}{v_{j}}e^{-v_{j}\sqrt{\frac{\nu}{D_{\rho}}}}e^{i\vec{v}_{j}\cdot\left(\vec{k}-\vec{k}'\right)}\nonumber\\
&\quad = \frac{1}{2\pi}\left(2\pi\delta\left(\omega-\omega'\right)\right)\sum_{j}\beta\bar{r}_{j}e^{i\vec{x}_{j}\cdot\left(\vec{k}-\vec{k}'\right)}\int d^{3}v_{j}\frac{\vec{k}\cdot\vec{k}'}{v_{j}}e^{-v_{j}\sqrt{\frac{\nu}{D_{\rho}}}}e^{i\vec{v}_{j}\cdot\left(\vec{k}-\vec{k}'\right)} \nonumber\\
&\quad\quad +\frac{\nu}{4\pi D_{\rho}}\left(2\pi\delta\left(\omega-\omega'\right)\right)\sum_{j}\beta\bar{r}_{j}e^{i\vec{x}_{j}\cdot\left(\vec{k}-\vec{k}'\right)}\int d^{3}v_{j}\frac{1}{v_{j}}e^{-v_{j}\sqrt{\frac{\nu}{D_{\rho}}}}e^{i\vec{v}_{j}\cdot\left(\vec{k}-\vec{k}'\right)}.
\label{etarhospec6}
\end{align}
The integrals in Eq.\ \ref{etarhospec6} can be solved via the known Fourier transformation
\begin{equation}
\int d^{3}z\frac{1}{z}e^{-zl}e^{i\vec{z}\cdot\vec{\kappa}} = \frac{4\pi}{l^{2}+\kappa^{2}}.
\label{KFT1}
\end{equation}
Letting $\vec{z}=\vec{v}_{j}$, $l=\sqrt{\frac{\nu}{D_{\rho}}}$ and $\vec{\kappa}=\vec{k}-\vec{k}'$, substituting Eq.\ \ref{KFT1} into Eq.\ \ref{etarhospec6} yields
\begin{equation}
\left\langle\tilde{\eta}_{d}^{*}\left(\vec{k}',\omega'\right)\tilde{\eta}_{d}\left(\vec{k},\omega\right)\right\rangle = \left(2\pi\delta\left(\omega-\omega'\right)\right)\frac{2D_{\rho}\vec{k}\cdot\vec{k}'+\nu}{\nu+D_{\rho}\abs{\vec{k}-\vec{k}'}^{2}}\sum_{j}\beta\bar{r}_{j}e^{i\vec{x}_{j}\cdot\left(\vec{k}-\vec{k}'\right)}.
\label{etarhospec7}
\end{equation}

Returning to $\eta_{pj}$, since each cell produces $\rho$ independently of each other cell, the production noises must be independent. Additionally, since the production is a birth only process, its power spectrum must simply be the mean propensity of the production, which in turn yields
\begin{equation}
\left\langle\tilde{\eta}_{pl}^{*}\left(\omega'\right)\tilde{\eta}_{pj}\left(\omega\right)\right\rangle = \beta\bar{r}_{j}\delta_{jl}\left(2\pi\delta\left(\omega-\omega'\right)\right),
\label{etapspec}
\end{equation}
as in the main text. Substituting Eqs.\ \ref{rjspec7}, \ref{etarhospec7}, and \ref{etapspec} into Eq.\ \ref{rhospec1} then yields
\begin{align}
&\left\langle\tilde{\delta\rho}^{*}\left(\vec{k}',\omega'\right)\tilde{\delta\rho}\left(\vec{k},\omega\right)\right\rangle = \frac{2\pi\delta\left(\omega-\omega'\right)}{\left(\nu+D_{\rho}k^{2}-i\omega\right)\left(\nu+D_{\rho}{k'}^{2}+i\omega'\right)} \nonumber\\
&\quad\quad \cdot\left(\sum_{j,l}e^{i\left(\vec{k}\cdot\vec{x}_{j}-\vec{k}'\cdot\vec{x}_{l}\right)}\left(\beta^{2}\frac{2\alpha\bar{c}}{\omega}\text{Im}\left(R_{jl}^{-1}\left(\omega\right)\right)+\beta\bar{r}_{j}\delta_{jl}\right)+\frac{2D_{\rho}\vec{k}\cdot\vec{k}'+\nu}{\nu+D_{\rho}\abs{\vec{k}-\vec{k}'}^{2}}\sum_{j}\beta\bar{r}_{j}e^{i\vec{x}_{j}\cdot\left(\vec{k}-\vec{k}'\right)}\right) \nonumber\\
&\quad = \frac{2\pi\delta\left(\omega-\omega'\right)}{\left(\nu+D_{\rho}k^{2}-i\omega\right)\left(\nu+D_{\rho}{k'}^{2}+i\omega\right)}\left(\sum_{j,l}\frac{2\alpha\beta^{2}\bar{c}}{\omega}e^{i\left(\vec{k}\cdot\vec{x}_{j}-\vec{k}'\cdot\vec{x}_{l}\right)}\text{Im}\left(R_{jl}^{-1}\left(\omega\right)\right)\right. \nonumber\\
&\quad\quad \left.+\left(1+\frac{2D_{\rho}\vec{k}\cdot\vec{k}'+\nu}{\nu+D_{\rho}\abs{\vec{k}-\vec{k}'}^{2}}\right)\sum_{j}\beta\bar{r}_{j}e^{i\vec{x}_{j}\cdot\left(\vec{k}-\vec{k}'\right)}\right),
\label{rhospec2}
\end{align}
where all instances of $\omega'$ outside the $\delta$ function have been freely replaced with $\omega$ due to the $\delta$ function being a global factor.

Now, let $m_{j}\left(t\right)$ be the number of $\rho$ molecules in the $j$th cell, which has volume $V_{j}$ and radius $a$. $m_{j}\left(t\right)$ can be calculated from $\rho\left(\vec{x},t\right)$ via
\begin{equation}
m_{j}\left(t\right) = \int_{V_{j}}d^{3}x\rho\left(\vec{x},t\right).
\label{Paramjdef}
\end{equation}
Once again, let $m_{j}\left(t\right)=\bar{m}_{j}+\delta m_{j}\left(t\right)$, where $\bar{m}_{j}$ is the mean value of $m_{j}\left(t\right)$. Since $\bar{\rho}\left(\vec{x}\right)$ is the mean value of $\rho\left(\vec{x},t\right)$, this implies
\begin{equation}
\bar{m}_{j} = \int_{V_{j}}d^{3}x\bar{\rho}\left(\vec{x}\right) \implies \delta m_{j}\left(t\right) = \int_{V_{j}}d^{3}x\delta\rho\left(\vec{x},t\right).
\label{Paramjbar}
\end{equation}
Fourier transforming the second part of Eq.\ \ref{Paramjbar} then yields
\begin{equation}
\tilde{\delta m}_{j}\left(\omega\right) = \int_{V_{j}}d^{3}x\int\frac{d^{3}k}{\left(2\pi\right)^{3}}\tilde{\delta\rho}\left(\vec{k},\omega\right)e^{-i\vec{k}\cdot\vec{x}}.
\label{FTLParamj}
\end{equation}
With this, the cross spectrum of $m_{j}\left(t\right)$ can be calculated to be
\begin{align}
&\left\langle\tilde{\delta m}_{j}^{*}\left(\omega'\right)\tilde{\delta m}_{j}\left(\omega\right)\right\rangle = \left\langle\left(\int_{V_{j}}d^{3}x'\int\frac{d^{3}k'}{\left(2\pi\right)^{3}}\tilde{\delta\rho}^{*}\left(\vec{k}',\omega'\right)e^{i\vec{k}'\cdot\vec{x}'}\right)\left(\int_{V_{j}}d^{3}x\int\frac{d^{3}k}{\left(2\pi\right)^{3}}\tilde{\delta\rho}\left(\vec{k},\omega\right)e^{-i\vec{k}\cdot\vec{x}}\right)\right\rangle \nonumber\\
&\quad = \frac{1}{\left(2\pi\right)^{6}}\int_{V_{j}}d^{3}xd^{3}x'\int d^{3}kd^{3}k'\left\langle\tilde{\delta\rho}^{*}\left(\vec{k}',\omega'\right)\tilde{\delta\rho}\left(\vec{k},\omega\right)\right\rangle e^{i\left(\vec{k}'\cdot\vec{x}'-\vec{k}\cdot\vec{x}\right)}.
\label{mjPspec1}
\end{align}
Again, let $\vec{v}_{j}=\vec{x}-\vec{x}_{j}$ and $\vec{v}'_{j}=\vec{x}'-\vec{x}_{j}$ and $V$ be the volume of the cell centered at the origin. This, along with Eq.\ \ref{rhospec2}, transforms Eq.\ \ref{mjPspec1} into
\begin{align}
&\left\langle\tilde{\delta m}_{j}^{*}\left(\omega'\right)\tilde{\delta m}_{j}\left(\omega\right)\right\rangle \nonumber\\
&\quad = \frac{1}{\left(2\pi\right)^{6}}\int_{V}d^{3}v_{j}d^{3}v'_{j}\int d^{3}kd^{3}k'\left\langle\tilde{\delta\rho}^{*}\left(\vec{k}',\omega'\right)\tilde{\delta\rho}\left(\vec{k},\omega\right)\right\rangle e^{i\left(\vec{k}'\cdot\vec{v}'_{j}-\vec{k}\cdot\vec{v}_{j}\right)}e^{i\vec{x}_{j}\cdot\left(\vec{k}'-\vec{k}\right)} \nonumber\\
&\quad = 2\pi\delta\left(\omega-\omega'\right)\left(I_{1}\left(\omega\right)+I_{2}\left(\omega\right)+I_{3}\left(\omega\right)+I_{4}\left(\omega\right)\right),
\label{mjPspec2}
\end{align}
where
\begin{subequations}
\begin{align}
I_{1}\left(\omega\right) &= \frac{1}{\left(2\pi\right)^{6}}\int_{V}d^{3}v_{j}d^{3}v'_{j}\int d^{3}kd^{3}k'\frac{1}{\left(\nu+D_{\rho}k^{2}-i\omega\right)\left(\nu+D_{\rho}{k'}^{2}+i\omega\right)} \nonumber\\
&\quad \cdot e^{i\left(\vec{k}'\cdot\vec{v}'_{j}-\vec{k}\cdot\vec{v}_{j}\right)}e^{i\vec{x}_{j}\cdot\left(\vec{k}'-\vec{k}\right)}\sum_{s,u}\frac{2\alpha\beta^{2}\bar{c}}{\omega}e^{i\left(\vec{k}\cdot\vec{x}_{s}-\vec{k}'\cdot\vec{x}_{u}\right)}\text{Im}\left(R_{su}^{-1}\left(\omega\right)\right)
\label{Idefa}
\end{align}
\begin{align}
I_{2}\left(\omega\right) &= \frac{1}{\left(2\pi\right)^{6}}\int_{V}d^{3}v_{j}d^{3}v'_{j}\int d^{3}kd^{3}k'\frac{1}{\left(\nu+D_{\rho}k^{2}-i\omega\right)\left(\nu+D_{\rho}{k'}^{2}+i\omega\right)} \nonumber\\
&\quad \cdot e^{i\left(\vec{k}'\cdot\vec{v}'_{j}-\vec{k}\cdot\vec{v}_{j}\right)}e^{i\vec{x}_{j}\cdot\left(\vec{k}'-\vec{k}\right)} \sum_{s}\beta\bar{r}_{s}e^{i\vec{x}_{s}\cdot\left(\vec{k}-\vec{k}'\right)}
\label{Idefb}
\end{align}
\begin{align}
I_{3}\left(\omega\right) &= \frac{1}{\left(2\pi\right)^{6}}\int_{V}d^{3}v_{j}d^{3}v'_{j}\int d^{3}kd^{3}k'\frac{1}{\left(\nu+D_{\rho}k^{2}-i\omega\right)\left(\nu+D_{\rho}{k'}^{2}+i\omega\right)} \nonumber\\
&\quad \cdot e^{i\left(\vec{k}'\cdot\vec{v}'_{j}-\vec{k}\cdot\vec{v}_{j}\right)}e^{i\vec{x}_{j}\cdot\left(\vec{k}'-\vec{k}\right)}\frac{2D_{\rho}\vec{k}\cdot\vec{k}'}{\nu+D_{\rho}\abs{\vec{k}-\vec{k}'}^{2}}\sum_{s}\beta\bar{r}_{s}e^{i\vec{x}_{s}\cdot\left(\vec{k}-\vec{k}'\right)}.
\label{Idefc}
\end{align}
\begin{align}
I_{4}\left(\omega\right) &= \frac{1}{\left(2\pi\right)^{6}}\int_{V}d^{3}v_{j}d^{3}v'_{j}\int d^{3}kd^{3}k'\frac{1}{\left(\nu+D_{\rho}k^{2}-i\omega\right)\left(\nu+D_{\rho}{k'}^{2}+i\omega\right)} \nonumber\\
&\quad \cdot e^{i\left(\vec{k}'\cdot\vec{v}'_{j}-\vec{k}\cdot\vec{v}_{j}\right)}e^{i\vec{x}_{j}\cdot\left(\vec{k}'-\vec{k}\right)}\frac{\nu}{\nu+D_{\rho}\abs{\vec{k}-\vec{k}'}^{2}}\sum_{s}\beta\bar{r}_{s}e^{i\vec{x}_{s}\cdot\left(\vec{k}-\vec{k}'\right)}.
\label{Idefd}
\end{align}
\label{Idef}%
\end{subequations}

Beginning with $I_{1}\left(\omega\right)$, moving the summation outside the integral and collecting terms exponential in $\vec{k}$ and $\vec{k}'$ yields
\begin{align}
I_{1}\left(\omega\right) &= \frac{2\alpha\beta^{2}\bar{c}}{\left(2\pi\right)^{6}D_{\rho}^{2}\omega}\sum_{s,u}\text{Im}\left(R_{su}^{-1}\left(\omega\right)\right)\int_{V}d^{3}v_{j}d^{3}v'_{j}\int d^{3}kd^{3}k' \nonumber\\
&\quad \cdot \frac{1}{\left(\frac{\nu-i\omega}{D_{\rho}}+k^{2}\right)\left(\frac{\nu+i\omega}{D_{\rho}}+{k'}^{2}\right)}e^{i\vec{k}\cdot\left(\vec{x}_{s}-\vec{x}_{j}-\vec{v}_{j}\right)}e^{-i\vec{k}'\cdot\left(\vec{x}_{u}-\vec{x}_{j}-\vec{v}'_{j}\right)}.
\label{I11}
\end{align}
Inverting Eq.\ \ref{KFT1} allows for the $k$ and $k'$ integrals to be easily solved, simplifying $I_{1}\left(\omega\right)$ to
\begin{align}
I_{1}\left(\omega\right) &= \frac{\alpha\beta^{2}\bar{c}}{2\left(2\pi D_{\rho}\right)^{2}\omega}\sum_{s,u}\text{Im}\left(R_{su}^{-1}\left(\omega\right)\right)\int_{V}d^{3}v_{j}d^{3}v'_{j} \nonumber\\
&\quad \cdot\left(\frac{1}{\abs{\vec{x}_{p}-\vec{x}_{j}-\vec{v}_{j}}}e^{-\abs{\vec{x}_{s}-\vec{x}_{j}-\vec{v}_{j}}\sqrt{\frac{\nu-i\omega}{D_{\rho}}}}\right)\left(\frac{1}{\abs{\vec{x}_{u}-\vec{x}_{j}-\vec{v}'_{j}}}e^{-\abs{\vec{x}_{q}-\vec{x}_{j}-\vec{v}'_{j}}\sqrt{\frac{\nu+i\omega}{D_{\rho}}}}\right) \nonumber\\
&= \frac{\alpha\beta^{2}\bar{c}a^{4}}{2D_{\rho}^{2}\omega}\sum_{s,u}\text{Im}\left(R_{su}^{-1}\left(\omega\right)\right)\Lambda\left(\abs{\vec{x}_{s}-\vec{x}_{j}},a,\lambda\left(\omega\right)\right) \nonumber\\
&\quad \cdot\Lambda\left(\abs{\vec{x}_{u}-\vec{x}_{j}},a,\lambda\left(-\omega\right)\right),
\label{I12}
\end{align}
where $a$ is the radius of the volume $V$,
\begin{equation}
\lambda\left(\omega\right)\equiv\sqrt{\frac{D_{\rho}}{\nu-i\omega}} = \sqrt[4]{\frac{D_{\rho}^{2}}{4\left(\nu^{2}+\omega^{2}\right)}}\left(\sqrt{1+\frac{\nu}{\sqrt{\nu^{2}+\omega^{2}}}}+i\;\text{sgn}\left(\omega\right)\sqrt{1-\frac{\nu}{\sqrt{\nu^{2}+\omega^{2}}}}\right),
\label{lambdadef}
\end{equation}
and
\begin{equation}
\Lambda\left(x,y,z\right)\equiv\begin{cases}
\frac{2z^{3}}{xy^{2}}\left(\frac{x}{z}-\left(1+\frac{y}{z}\right)e^{-\frac{y}{z}}\sinh\left(\frac{x}{z}\right)\right) & x<y \\
\frac{2z^{3}}{xy^{2}}e^{-\frac{x}{z}}\left(\frac{y}{z}\cosh\left(\frac{y}{z}\right)-\sinh\left(\frac{y}{z}\right)\right) & x>y \end{cases}
\label{Lambdadef}
\end{equation}
comes from the relation
\begin{equation}
\int_{V}d^{3}z\frac{1}{\abs{\vec{\kappa}-\vec{z}}}e^{-\frac{\abs{\vec{\kappa}-\vec{z}}}{l}} = 2\pi a^{2}\Lambda\left(\kappa,a,l\right),
\label{Lambdarel}
\end{equation}
which can be shown by writing $\vec{z}$ in spherical coordinates and evaluating.

Moving to $I_{2}\left(\omega\right)$, following the exact same procedure as was done for $I_{1}\left(\omega\right)$ yields
\begin{align}
I_{2}\left(\omega\right) &= \frac{\beta}{\left(2\pi\right)^{6}D_{\rho}^{2}}\sum_{s}\bar{r}_{s}\int_{V}d^{3}v_{j}d^{3}v'_{j}\int d^{3}kd^{3}k' \nonumber\\
&\quad \cdot \frac{1}{\left(\frac{\nu-i\omega}{D_{\rho}}+k^{2}\right)\left(\frac{\nu+i\omega}{D_{\rho}}+{k'}^{2}\right)}e^{i\vec{k}\cdot\left(\vec{x}_{s}-\vec{x}_{j}-\vec{v}_{j}\right)}e^{-i\vec{k}'\cdot\left(\vec{x}_{s}-\vec{x}_{j}-\vec{v}'_{j}\right)} \nonumber\\
&= \frac{\beta}{\left(4\pi D_{\rho}\right)^{2}}\sum_{s}\bar{r}_{s}\int_{V}d^{3}v_{j}d^{3}v'_{j} \nonumber\\
&\quad \cdot\left(\frac{1}{\abs{\vec{x}_{s}-\vec{x}_{j}-\vec{v}_{j}}}e^{-\abs{\vec{x}_{s}-\vec{x}_{j}-\vec{v}_{j}}\sqrt{\frac{\nu-i\omega}{D_{\rho}}}}\right)\left(\frac{1}{\abs{\vec{x}_{s}-\vec{x}_{j}-\vec{v}'_{j}}}e^{-\abs{\vec{x}_{s}-\vec{x}_{j}-\vec{v}'_{j}}\sqrt{\frac{\nu+i\omega}{D_{\rho}}}}\right) \nonumber\\
&= \frac{\beta a^{4}}{4D_{\rho}^{2}}\sum_{s}\bar{r}_{s}\abs{\Lambda\left(\abs{\vec{x}_{s}-\vec{x}_{j}},a,\lambda\left(\omega\right)\right)}^{2}.
\label{I2}
\end{align}

Unfortunately, $I_{3}\left(\omega\right)$ cannot be solved by the same procedure as $I_{1}\left(\omega\right)$ and $I_{2}\left(\omega\right)$, but it can be solved. First, utilizing Eq.\ \ref{KFT1} again and letting $l = \sqrt{\frac{D_{\rho}}{\nu}}$ and $\vec{\kappa}=\vec{k}-\vec{k}'$ allows the factor of $\frac{2D_{\rho}}{\nu+D_{\rho}\abs{\vec{k}-\vec{k}'}^{2}}$ to be transformed into another integral, yielding
\begin{align}
I_{3}\left(\omega\right) &= \frac{1}{\left(2\pi\right)^{7}D_{\rho}^{2}}\int_{V}d^{3}v_{j}d^{3}v'_{j}\int d^{3}kd^{3}k'd^{3}z\frac{1}{\left(\frac{\nu-i\omega}{D_{\rho}}+k^{2}\right)\left(\frac{\nu+i\omega}{D_{\rho}}+{k'}^{2}\right)} \nonumber\\
&\quad \cdot e^{i\left(\vec{k}'\cdot\vec{v}'_{j}-\vec{k}\cdot\vec{v}_{j}\right)}e^{i\vec{x}_{j}\cdot\left(\vec{k}'-\vec{k}\right)}\frac{\vec{k}\cdot\vec{k}'}{z}e^{-z\sqrt{\frac{\nu}{D_{\rho}}}}e^{i\vec{z}\cdot\left(\vec{k}-\vec{k}'\right)}\sum_{s}\beta\bar{r}_{s}e^{i\vec{x}_{s}\cdot\left(\vec{k}-\vec{k}'\right)}.
\label{I31}
\end{align}
Since $v_{j}$ and $v_{j}'$ only appear in a single exponential within the integrand, the factor of $\vec{k}\cdot\vec{k}'$ can be replaced by $\vec{\nabla}_{v_{j}}\cdot\vec{\nabla}_{v'_{j}}$ acting on the exponential. The gradient operators can then be moved outside the $k$, $k'$, and $z$ integrals while the summation is moved outside of all the integrals to produce
\begin{align}
I_{3}\left(\omega\right) &= \frac{\beta}{\left(2\pi\right)^{7}D_{\rho}^{2}}\sum_{s}\bar{r}_{s}\int_{V}d^{3}v_{j}d^{3}v'_{j}\vec{\nabla}_{v_{j}}\cdot\vec{\nabla}_{v'_{j}}\int d^{3}kd^{3}k'd^{3}z\frac{1}{\left(\frac{\nu-i\omega}{D_{\rho}}+k^{2}\right)\left(\frac{\nu+i\omega}{D_{\rho}}+{k'}^{2}\right)} \nonumber\\
&\quad \cdot \frac{1}{z}e^{-z\sqrt{\frac{\nu}{D_{\rho}}}}e^{i\left(\vec{k}'\cdot\vec{v}'_{j}-\vec{k}\cdot\vec{v}_{j}\right)}e^{i\vec{x}_{j}\cdot\left(\vec{k}'-\vec{k}\right)} e^{i\vec{z}\cdot\left(\vec{k}-\vec{k}'\right)}e^{i\vec{x}_{s}\cdot\left(\vec{k}-\vec{k}'\right)}.
\label{I32}
\end{align}
Utilizing the inverse of Eq.\ \ref{KFT1} to solve the $k$ and $k'$ integrals then yields
\begin{align}
I_{3}\left(\omega\right) &= \frac{\beta}{4\left(2\pi\right)^{3}D_{\rho}^{2}}\sum_{s}\bar{r}_{s}\int_{V}d^{3}v_{j}d^{3}v'_{j}\vec{\nabla}_{v_{j}}\cdot\vec{\nabla}_{v'_{j}}\int d^{3}z\frac{1}{z}e^{-z\sqrt{\frac{\nu}{D_{\rho}}}} \nonumber\\
&\quad \cdot\left(\frac{1}{\abs{\vec{z}+\vec{x}_{s}-\vec{x}_{j}-\vec{v}_{j}}}e^{-\abs{\vec{z}+\vec{x}_{s}-\vec{x}_{j}-\vec{v}_{j}}\sqrt{\frac{\nu-i\omega}{D_{\rho}}}}\right)\left(\frac{1}{\abs{\vec{z}+\vec{x}_{s}-\vec{x}_{j}-\vec{v}'_{j}}}e^{-\abs{\vec{z}+\vec{x}_{s}-\vec{x}_{j}-\vec{v}'_{j}}\sqrt{\frac{\nu+i\omega}{D_{\rho}}}}\right).
\label{I33}
\end{align}
From here the $z$ integral can be moved outside the $v_{j}$ and $v'_{j}$ integrals, which can in turn be separated into the product of two independent integrals to produce
\begin{align}
I_{3}\left(\omega\right) &= \frac{\beta}{4\left(2\pi\right)^{3}D_{\rho}^{2}}\sum_{s}\bar{r}_{s}\int d^{3}z\frac{1}{z}e^{-z\sqrt{\frac{\nu}{D_{\rho}}}}\left(\int_{V}d^{3}v_{j}\vec{\nabla}_{v_{j}}\frac{1}{\abs{\vec{z}+\vec{x}_{s}-\vec{x}_{j}-\vec{v}_{j}}}e^{-\abs{\vec{z}+\vec{x}_{s}-\vec{x}_{j}-\vec{v}_{j}}\sqrt{\frac{\nu-i\omega}{D_{\rho}}}}\right) \nonumber\\
&\quad \cdot\left(\int_{V}d^{3}v'_{j}\vec{\nabla}_{v'_{j}}\frac{1}{\abs{\vec{z}+\vec{x}_{s}-\vec{x}_{j}-\vec{v}'_{j}}}e^{-\abs{\vec{z}+\vec{x}_{s}-\vec{x}_{j}-\vec{v}'_{j}}\sqrt{\frac{\nu+i\omega}{D_{\rho}}}}\right).
\label{I34}
\end{align}
Due to the fact that the $v_{j}$ and $v'_{j}$ integrands in Eq.\ \ref{I34} depend only on $\abs{\vec{z}+\vec{x}_{s}-\vec{x}_{j}-\vec{v}_{j}}$ and $\abs{\vec{z}+\vec{x}_{s}-\vec{x}_{j}-\vec{v}'_{j}}$ respectively, taking the gradient with respect to $v_{j}$ and $v'_{j}$ is identical to taking the gradient with respect to $z$ and multiplying by a factor of $-1$ in both cases. The extra factors of $-1$ can be ignored, however, as they will multiply to unity. This allows the gradients to be moved outside of the $v_{j}$ and $v'_{j}$ integrals, which in turn allows them to be solved via Eq.\ \ref{Lambdarel} to produce
\begin{equation}
I_{3}\left(\omega\right) = \frac{\beta a^{4}}{8\pi D_{\rho}^{2}}\sum_{s}\bar{r}_{s}\int d^{3}z\frac{1}{z}e^{-z\sqrt{\frac{\nu}{D_{\rho}}}}\abs{\vec{\nabla}_{z}\Lambda\left(\abs{\vec{z}+\vec{x}_{s}-\vec{x}_{j}},a,\lambda\left(\omega\right)\right)}^{2}.
\label{I35}
\end{equation}
Let $\vec{y}=\vec{z}+\vec{x}_{s}-\vec{x}_{j}$. Since the $z$ integral is over all of $z$-space, this transformation does not change the limits of integration. Additionally, $d^{3}y=d^{3}z$ and $\vec{\nabla}_{y}=\vec{\nabla}_{z}$ since $y$ and $z$ are related by a simple translation. This transformation allows Eq.\ \ref{I35} to be written as
\begin{equation}
I_{3}\left(\omega\right) = \frac{\beta a^{4}}{8\pi D_{\rho}^{2}}\sum_{s}\bar{r}_{s}\int d^{3}y\frac{1}{\abs{\vec{y}+\vec{x}_{j}-\vec{x}_{s}}}e^{-\abs{\vec{y}+\vec{x}_{j}-\vec{x}_{s}}\sqrt{\frac{\nu}{D_{\rho}}}}\abs{\vec{\nabla}_{y}\Lambda\left(y,a,\lambda\left(\omega\right)\right)}^{2}.
\label{I36}
\end{equation}
Let $V_{y}$ be the spherical volume in $y$-space centered at the origin with radius $a$ and $V'_{y}$ be all of $y$-space excluding $V_{y}$. These along with Eq.\ \ref{Lambdadef} allow the integral in Eq.\ \ref{I36} to be broken into two separate pieces along the piecewise boundary of $\Lambda\left(y,a,\lambda\left(\omega\right)\right)$ to produce
\begin{align}
I_{3}\left(\omega\right) &= \frac{\beta a^{4}}{8\pi D_{\rho}^{2}}\sum_{s}\bar{r}_{s}\int_{V_{y}}d^{3}y\frac{1}{\abs{\vec{y}+\vec{x}_{j}-\vec{x}_{s}}}e^{-\abs{\vec{y}+\vec{x}_{j}-\vec{x}_{s}}\sqrt{\frac{\nu}{D_{\rho}}}} \nonumber\\
&\quad \cdot\abs{\vec{\nabla}_{y}\frac{2}{ya^{2}}\left(\frac{D_{\rho}}{\nu-i\omega}\right)^{\frac{3}{2}}\left(y\sqrt{\frac{\nu-i\omega}{D_{\rho}}}-\left(1+a\sqrt{\frac{\nu-i\omega}{D_{\rho}}}\right)e^{-a\sqrt{\frac{\nu-i\omega}{D_{\rho}}}}\sinh\left(y\sqrt{\frac{\nu-i\omega}{D_{\rho}}}\right)\right)}^{2} \nonumber\\
&\quad +\int_{V'_{y}}d^{3}y\frac{1}{\abs{\vec{y}+\vec{x}_{j}-\vec{x}_{s}}}e^{-\abs{\vec{y}+\vec{x}_{j}-\vec{x}_{s}}\sqrt{\frac{\nu}{D_{\rho}}}} \nonumber\\
&\quad \cdot\abs{\vec{\nabla}_{y}\frac{2}{ya^{2}}\left(\frac{D_{\rho}}{\nu-i\omega}\right)^{\frac{3}{2}}e^{-y\sqrt{\frac{\nu-i\omega}{D_{\rho}}}}\left(a\sqrt{\frac{\nu-i\omega}{D_{\rho}}}\cosh\left(a\sqrt{\frac{\nu-i\omega}{D_{\rho}}}\right)-\sinh\left(a\sqrt{\frac{\nu-i\omega}{D_{\rho}}}\right)\right)}^{2}.
\label{I37}
\end{align}
Performing the gradient operators then yields
\begin{align}
I_{3}\left(\omega\right) &= \frac{\beta a^{4}}{8\pi D_{\rho}^{2}}\sum_{s}\bar{r}_{s}\int_{V_{y}}d^{3}y\frac{1}{\abs{\vec{y}+\vec{x}_{j}-\vec{x}_{s}}}e^{-\frac{\abs{\vec{y}+\vec{x}_{j}-\vec{x}_{s}}}{\lambda\left(0\right)}} \nonumber\\
&\quad \cdot\left|\frac{\vec{y}}{y}\frac{2\left(\lambda\left(\omega\right)\right)^{3}}{y^{2}a^{2}}\left(1+\frac{a}{\lambda\left(\omega\right)}\right)e^{-\frac{a}{\lambda\left(\omega\right)}}\left(\sinh\left(\frac{y}{\lambda\left(\omega\right)}\right)\right.\right. \nonumber\\
&\quad \left.\vphantom{\frac{2\left(\lambda\left(\omega\right)\right)^{3}}{y^{2}a^{2}}}\left.-\frac{y}{\lambda\left(\omega\right)}\cosh\left(\frac{y}{\lambda\left(\omega\right)}\right)\right)\right|^{2}+\int_{V'_{y}}d^{3}y\frac{1}{\abs{\vec{y}+\vec{x}_{j}-\vec{x}_{s}}}e^{-\frac{\abs{\vec{y}+\vec{x}_{j}-\vec{x}_{s}}}{\lambda\left(0\right)}} \nonumber\\
&\quad \cdot\left|\frac{\vec{y}}{y}\frac{2\left(\lambda\left(\omega\right)\right)^{3}}{y^{2}a^{2}}\left(1+\frac{y}{\lambda\left(\omega\right)}\right)e^{-\frac{y}{\lambda\left(\omega\right)}}\left(\sinh\left(\frac{a}{\lambda\left(\omega\right)}\right)\right.\right. \nonumber\\
&\quad \left.\vphantom{\frac{2\left(\lambda\left(\omega\right)\right)^{3}}{y^{2}a^{2}}}\left.-\frac{a}{\lambda\left(\omega\right)}\cosh\left(\frac{a}{\lambda\left(\omega\right)}\right)\right)\right|^{2}.
\label{I38}
\end{align}
Once the magnitude squared of each vector is taken, the only term in either integral that depends on the angle of $\vec{y}$ will be $\abs{\vec{y}+\vec{x}_{j}-\vec{x}_{s}}$. Thus, the angular portion of each integral can be performed to yield
\begin{align}
I_{3}\left(\omega\right) &= \frac{\beta\abs{\lambda\left(\omega\right)}^{6}}{D_{\rho}^{2}}\sum_{s}\bar{r}_{s}\int_{0}^{a}dy\frac{\lambda\left(0\right)}{y^{3}\abs{\vec{x}_{s}-\vec{x}_{j}}}\left(e^{-\frac{\abs{y-\abs{\vec{x}_{s}-\vec{x}_{j}}}}{\lambda\left(0\right)}}-e^{-\frac{y+\abs{\vec{x}_{s}-\vec{x}_{j}}}{\lambda\left(0\right)}}\right) \nonumber\\
&\quad \cdot\abs{\left(1+\frac{a}{\lambda\left(\omega\right)}\right)e^{-\frac{a}{\lambda\left(\omega\right)}}\left(\sinh\left(\frac{y}{\lambda\left(\omega\right)}\right)-\frac{y}{\lambda\left(\omega\right)}\cosh\left(\frac{y}{\lambda\left(\omega\right)}\right)\right)}^{2} \nonumber\\
&\quad +\int_{a}^{\infty}dy\frac{\lambda\left(0\right)}{y^{3}\abs{\vec{x}_{s}-\vec{x}_{j}}}\left(e^{-\frac{\abs{y-\abs{\vec{x}_{s}-\vec{x}_{j}}}}{\lambda\left(0\right)}}-e^{-\frac{y+\abs{\vec{x}_{s}-\vec{x}_{j}}}{\lambda\left(0\right)}}\right) \nonumber\\
&\quad \cdot\abs{\left(1+\frac{y}{\lambda\left(\omega\right)}\right)e^{-\frac{y}{\lambda\left(\omega\right)}}\left(\sinh\left(\frac{a}{\lambda\left(\omega\right)}\right)-\frac{a}{\lambda\left(\omega\right)}\cosh\left(\frac{a}{\lambda\left(\omega\right)}\right)\right)}^{2}.
\label{I39}
\end{align}
The integrals in Eq.\ \ref{I39} are well defined and very involved. Nonetheless, they can be performed piece-by-piece with the aid of integral tables or symbolic computational solvers. The result is
\begin{align}
I_{3}\left(\omega\right) &= \frac{\beta a^{4}}{D_{\rho}^{2}}\sum_{s}\bar{r}_{s}\left(\frac{\abs{\left(1+\frac{a}{\lambda\left(\omega\right)}\right)e^{-\frac{a}{\lambda\left(\omega\right)}}}^{2}}{18}\Upsilon\left(\abs{\vec{x}_{s}-\vec{x}_{j}},a,\lambda\left(0\right),\lambda\left(\omega\right)\right)\right. \nonumber\\
&\quad \left.\vphantom{\frac{\abs{\left(1+\frac{a}{\lambda\left(\omega\right)}\right)e^{-\frac{a}{\lambda\left(\omega\right)}}}^{2}}{18}}+\abs{\left(\frac{\lambda\left(\omega\right)}{a}\right)^{3}\left(\sinh\left(\frac{a}{\lambda\left(\omega\right)}\right)-\frac{a}{\lambda\left(\omega\right)}\cosh\left(\frac{a}{\lambda\left(\omega\right)}\right)\right)}^{2}\Xi\left(\abs{\vec{x}_{s}-\vec{x}_{j}},a,\lambda\left(0\right),\lambda\left(\omega\right)\right)\right),
\label{I310}
\end{align}
where
\begin{equation}
\Upsilon\left(x,y,z,w\right)\equiv\begin{cases}
\frac{9z\abs{w}^{6}\sinh\left(\frac{x}{z}\right)}{xy^{6}}e^{-\frac{y}{z}}\left(2y\text{Re}\left(\frac{1}{w}\right)\sinh\left(2y\text{Re}\left(\frac{1}{w}\right)\right) +2y\text{Im}\left(\frac{1}{w}\right)\sin\left(2y\text{Im}\left(\frac{1}{w}\right)\right)\vphantom{\left(\cosh\left(2y\text{Re}\left(\frac{1}{w}\right)\right)\right)}\right. & \\
\quad \left.-\left(1-\frac{y}{z}\right)\left(\cosh\left(2y\text{Re}\left(\frac{1}{w}\right)\right)-\cos\left(2y\text{Im}\left(\frac{1}{w}\right)\right)\right)\right) & \\
\quad -\frac{9\abs{w}^{6}}{x^{2}y^{4}}\left(\cosh\left(2x\text{Re}\left(\frac{1}{w}\right)\right)-\cos\left(2x\text{Im}\left(\frac{1}{w}\right)\right)\right) & \\
\quad -\frac{9\abs{w}^{6}}{2xy^{4}z}e^{-\frac{x}{z}}\left(\text{Shi}\left(x\left(\frac{1}{z}+2\text{Re}\left(\frac{1}{w}\right)\right)\right)+\text{Shi}\left(x\left(\frac{1}{z}-2\text{Re}\left(\frac{1}{w}\right)\right)\right)\right. & \\
\quad \left.-\text{Shi}\left(x\left(\frac{1}{z}+2i\text{Im}\left(\frac{1}{w}\right)\right)\right)-\text{Shi}\left(x\left(\frac{1}{z}-2i\text{Im}\left(\frac{1}{w}\right)\right)\right)\right) & \\
\quad -\frac{9\abs{w}^{6}\sinh\left(\frac{x}{z}\right)}{2xy^{4}z}\left(\text{Ei}\left(x\left(\frac{1}{z}+2\text{Re}\left(\frac{1}{w}\right)\right)\right)+\text{Ei}\left(x\left(\frac{1}{z}-2\text{Re}\left(\frac{1}{w}\right)\right)\right)\right. & \\
\quad -\text{Ei}\left(x\left(\frac{1}{z}+2i\text{Im}\left(\frac{1}{w}\right)\right)\right)-\text{Ei}\left(x\left(\frac{1}{z}-2i\text{Im}\left(\frac{1}{w}\right)\right)\right) & \\
\quad -\text{Ei}\left(y\left(\frac{1}{z}+2\text{Re}\left(\frac{1}{w}\right)\right)\right)-\text{Ei}\left(y\left(\frac{1}{z}-2\text{Re}\left(\frac{1}{w}\right)\right)\right) & \\
\quad \left.+\text{Ei}\left(y\left(\frac{1}{z}+2i\text{Im}\left(\frac{1}{w}\right)\right)\right)+\text{Ei}\left(y\left(\frac{1}{z}-2i\text{Im}\left(\frac{1}{w}\right)\right)\right)\right) & x<y \\
\frac{9z\abs{w}^{6}\sinh\left(\frac{y}{z}\right)}{xy^{6}}e^{-\frac{x}{z}}\left(2y\text{Re}\left(\frac{1}{w}\right)\sinh\left(2y\text{Re}\left(\frac{1}{w}\right)\right)+2y\text{Im}\left(\frac{1}{w}\right)\sin\left(2y\text{Im}\left(\frac{1}{w}\right)\right)\right) & \\
\quad -\frac{9z\abs{w}^{6}\left(\sinh\left(\frac{y}{z}\right)+\frac{y}{z}\cosh\left(\frac{y}{z}\right)\right)}{xy^{6}}e^{-\frac{x}{z}}\left(\cosh\left(2y\text{Re}\left(\frac{1}{w}\right)\right)-\cos\left(2y\text{Im}\left(\frac{1}{w}\right)\right)\right) & \\
\quad -\frac{9\abs{w}^{6}}{2xy^{4}z}e^{-\frac{x}{z}}\left(\text{Shi}\left(y\left(\frac{1}{z}+2\text{Re}\left(\frac{1}{w}\right)\right)\right)+\text{Shi}\left(y\left(\frac{1}{z}-2\text{Re}\left(\frac{1}{w}\right)\right)\right)\right. & \\
\quad \left.-\text{Shi}\left(y\left(\frac{1}{z}+2i\text{Im}\left(\frac{1}{w}\right)\right)\right)-\text{Shi}\left(y\left(\frac{1}{z}-2i\text{Im}\left(\frac{1}{w}\right)\right)\right)\right) & x>y, \end{cases}
\label{Updef}
\end{equation}
\begin{equation}
\Xi\left(x,y,z,w\right)\equiv\begin{cases}
\frac{z\sinh\left(\frac{x}{z}\right)}{x}\left(\left(1+y\left(2\text{Re}\left(\frac{1}{w}\right)-\frac{1}{z}\right)\right)e^{-y\left(2\text{Re}\left(\frac{1}{w}\right)+\frac{1}{z}\right)}-\frac{y^{2}}{z^{2}}\text{Ei}\left(y\left(2\text{Re}\left(\frac{1}{w}\right)+\frac{1}{z}\right)\right)\right) & x<y \\
\frac{z}{x}e^{-\frac{x}{z}}\left(\vphantom{\left(\left(\left(\text{Re}\left(\frac{1}{w}\right)\right)\right)\right)} e^{-2y\text{Re}\left(\frac{1}{w}\right)}\left(\left(1+2y\text{Re}\left(\frac{1}{w}\right)\right)\sinh\left(\frac{y}{z}\right)+\frac{y}{z}\cosh\left(\frac{y}{z}\right)\right)\right. & \\
\quad -\frac{y^{2}}{2z^{2}}\left(e^{\frac{2x}{z}}\text{Ei}\left(x\left(2\text{Re}\left(\frac{1}{w}\right)+\frac{1}{z}\right)\right)-\text{Ei}\left(x\left(2\text{Re}\left(\frac{1}{w}\right)-\frac{1}{z}\right)\right)\right. & \\
\quad \left.\left.-\text{Ei}\left(y\left(2\text{Re}\left(\frac{1}{w}\right)+\frac{1}{z}\right)\right)+\text{Ei}\left(y\left(2\text{Re}\left(\frac{1}{w}\right)-\frac{1}{z}\right)\right)\right)-\frac{y^{2}}{xz}e^{-x\left(2\text{Re}\left(\frac{1}{w}\right)-\frac{1}{z}\right)}\right) & x>y, \end{cases}
\label{Xidef}
\end{equation}
\begin{equation}
\text{Shi}\left(x\right)\equiv\int_{0}^{x}dt\frac{\sinh\left(t\right)}{t},
\label{Shidef}
\end{equation}
and
\begin{equation}
\text{Ei}\left(x\right)\equiv\int_{x}^{\infty}dt\frac{e^{-t}}{t}.
\label{Eidef}
\end{equation}

Lastly, $I_{4}\left(\omega\right)$ must be solved. Similarly utilizing Eq.\ \ref{KFT1} allows Eq.\ \ref{Idefd} to be transformed into
\begin{align}
I_{4}\left(\omega\right) &= \frac{\nu}{2\left(2\pi\right)^{7}D_{\rho}^{3}}\int_{V}d^{3}v_{j}d^{3}v'_{j}\int d^{3}kd^{3}k'd^{3}z\frac{1}{\left(\frac{\nu-i\omega}{D_{\rho}}+k^{2}\right)\left(\frac{\nu+i\omega}{D_{\rho}}+{k'}^{2}\right)} \nonumber\\
&\quad \cdot e^{i\left(\vec{k}'\cdot\vec{v}'_{j}-\vec{k}\cdot\vec{v}_{j}\right)}e^{i\vec{x}_{j}\cdot\left(\vec{k}'-\vec{k}\right)}\frac{1}{z}e^{-z\sqrt{\frac{\nu}{D_{\rho}}}}e^{i\vec{z}\cdot\left(\vec{k}-\vec{k}'\right)}\sum_{s}\beta\bar{r}_{s}e^{i\vec{x}_{s}\cdot\left(\vec{k}-\vec{k}'\right)}.
\label{I41}
\end{align}
Utilizing the inverse of Eq.\ \ref{KFT1} to solve the $k$ and $k'$ integrals then yields
\begin{align}
I_{4}\left(\omega\right) &= \frac{\beta\nu}{8\left(2\pi\right)^{3}D_{\rho}^{3}}\sum_{s}\bar{r}_{s}\int_{V}d^{3}v_{j}d^{3}v'_{j}\int d^{3}z\frac{1}{z}e^{-z\sqrt{\frac{\nu}{D_{\rho}}}} \nonumber\\
&\quad \cdot\left(\frac{1}{\abs{\vec{z}+\vec{x}_{s}-\vec{x}_{j}-\vec{v}_{j}}}e^{-\abs{\vec{z}+\vec{x}_{s}-\vec{x}_{j}-\vec{v}_{j}}\sqrt{\frac{\nu-i\omega}{D_{\rho}}}}\right)\left(\frac{1}{\abs{\vec{z}+\vec{x}_{s}-\vec{x}_{j}-\vec{v}'_{j}}}e^{-\abs{\vec{z}+\vec{x}_{s}-\vec{x}_{j}-\vec{v}'_{j}}\sqrt{\frac{\nu+i\omega}{D_{\rho}}}}\right).
\label{I42}
\end{align}
From here the $z$ integral can be moved outside the $v_{j}$ and $v'_{j}$ integrals, which can in turn be separated into the product of two independent integrals to produce
\begin{align}
I_{4}\left(\omega\right) &= \frac{\beta\nu}{8\left(2\pi\right)^{3}D_{\rho}^{3}}\sum_{s}\bar{r}_{s}\int d^{3}z\frac{1}{z}e^{-z\sqrt{\frac{\nu}{D_{\rho}}}}\left(\int_{V}d^{3}v_{j}\frac{1}{\abs{\vec{z}+\vec{x}_{s}-\vec{x}_{j}-\vec{v}_{j}}}e^{-\abs{\vec{z}+\vec{x}_{s}-\vec{x}_{j}-\vec{v}_{j}}\sqrt{\frac{\nu-i\omega}{D_{\rho}}}}\right) \nonumber\\
&\quad \cdot\left(\int_{V}d^{3}v'_{j}\frac{1}{\abs{\vec{z}+\vec{x}_{s}-\vec{x}_{j}-\vec{v}'_{j}}}e^{-\abs{\vec{z}+\vec{x}_{s}-\vec{x}_{j}-\vec{v}'_{j}}\sqrt{\frac{\nu+i\omega}{D_{\rho}}}}\right).
\label{I43}
\end{align}
The $v_{j}$ and $v'_{j}$ integrals can then be solved via Eq.\ \ref{Lambdarel} to produce
\begin{equation}
I_{4}\left(\omega\right) = \frac{\beta\nu a^{4}}{16\pi D_{\rho}^{3}}\sum_{s}\bar{r}_{s}\int d^{3}z\frac{1}{z}e^{-z\sqrt{\frac{\nu}{D_{\rho}}}}\abs{\Lambda\left(\abs{\vec{z}+\vec{x}_{s}-\vec{x}_{j}},a,\lambda\left(\omega\right)\right)}^{2}.
\label{I44}
\end{equation}
Again, let $\vec{y}=\vec{z}+\vec{x}_{s}-\vec{x}_{j}$ as well as $V_{y}$ be the spherical volume in $y$-space centered at the origin with radius $a$ and $V'_{y}$ be all of $y$-space excluding $V_{y}$. These transformations allow the $\Lambda$ function to be split along its piecewise boundary again and Eq.\ \ref{I44} to be written as
\begin{align}
I_{4}\left(\omega\right) &= \frac{\beta\nu a^{4}}{16\pi D_{\rho}^{3}}\sum_{s}\bar{r}_{s}\int_{V_{y}}d^{3}y\frac{1}{\abs{\vec{y}+\vec{x}_{j}-\vec{x}_{s}}}e^{-\abs{\vec{y}+\vec{x}_{j}-\vec{x}_{s}}\sqrt{\frac{\nu}{D_{\rho}}}} \nonumber\\
&\quad \cdot\abs{\frac{2\left(\lambda\left(\omega\right)\right)^{3}}{ya^{2}}\left(\frac{y}{\lambda\left(\omega\right)}-\left(1+\frac{a}{\lambda\left(\omega\right)}\right)e^{-\frac{a}{\lambda\left(\omega\right)}}\sinh\left(\frac{y}{\lambda\left(\omega\right)}\right)\right)}^{2} \nonumber\\
&\quad +\int_{V'_{y}}d^{3}y\frac{1}{\abs{\vec{y}+\vec{x}_{j}-\vec{x}_{s}}}e^{-\abs{\vec{y}+\vec{x}_{j}-\vec{x}_{s}}\sqrt{\frac{\nu}{D_{\rho}}}} \nonumber\\
&\quad \cdot\abs{\frac{2\left(\lambda\left(\omega\right)\right)^{3}}{ya^{2}}e^{-\frac{y}{\lambda\left(\omega\right)}}\left(\frac{a}{\lambda\left(\omega\right)}\cosh\left(\frac{a}{\lambda\left(\omega\right)}\right)-\sinh\left(\frac{a}{\lambda\left(\omega\right)}\right)\right)}^{2}.
\label{I45}
\end{align}
Once again, the only term in either integral that depends on the angle of $\vec{y}$ will be $\abs{\vec{y}+\vec{x}_{j}-\vec{x}_{s}}$. Thus, the angular portion of each integral can be performed to yield
\begin{align}
I_{4}\left(\omega\right) &= \frac{\beta\nu\abs{\lambda\left(\omega\right)}^{6}}{2D_{\rho}^{3}}\sum_{s}\bar{r}_{s}\int_{0}^{a}dy\frac{\lambda\left(0\right)}{y\abs{\vec{x}_{s}-\vec{x}_{j}}}\left(e^{-\frac{\abs{y-\abs{\vec{x}_{s}-\vec{x}_{j}}}}{\lambda\left(0\right)}}-e^{-\frac{y+\abs{\vec{x}_{s}-\vec{x}_{j}}}{\lambda\left(0\right)}}\right) \nonumber\\
&\quad \cdot\abs{\frac{y}{\lambda\left(\omega\right)}-\left(1+\frac{a}{\lambda\left(\omega\right)}\right)e^{-\frac{a}{\lambda\left(\omega\right)}}\sinh\left(\frac{y}{\lambda\left(\omega\right)}\right)}^{2} \nonumber\\
&\quad +\int_{a}^{\infty}dy\frac{\lambda\left(0\right)}{y\abs{\vec{x}_{s}-\vec{x}_{j}}}\left(e^{-\frac{\abs{y-\abs{\vec{x}_{s}-\vec{x}_{j}}}}{\lambda\left(0\right)}}-e^{-\frac{y+\abs{\vec{x}_{s}-\vec{x}_{j}}}{\lambda\left(0\right)}}\right) \nonumber\\
&\quad \cdot\abs{e^{-\frac{y}{\lambda\left(\omega\right)}}\left(\frac{a}{\lambda\left(\omega\right)}\cosh\left(\frac{a}{\lambda\left(\omega\right)}\right)-\sinh\left(\frac{a}{\lambda\left(\omega\right)}\right)\right)}^{2}.
\label{I46}
\end{align}
Relying again on integral tables or computational solvers, these integrals can also be performed and yield
\begin{align}
I_{4}\left(\omega\right) &= \frac{\beta\nu\abs{\lambda\left(\omega\right)}^{6}}{2D_{\rho}^{3}}\sum_{s}\bar{r}_{s}\left(\Psi\left(\abs{\vec{x}_{s}-\vec{x}_{j}},a,\lambda\left(0\right),\lambda\left(\omega\right)\right)+\Omega\left(\abs{\vec{x}_{s}-\vec{x}_{j}},a,\lambda\left(0\right),\lambda\left(\omega\right)\right)\right)
\label{I47}
\end{align}
where
\begin{equation}
\Psi\left(x,y,z,w\right) = \begin{cases}
\frac{2z^3}{x\abs{w}^{2}}\left(\frac{x}{z}-\left(1+\frac{y}{z}\right)e^{-\frac{y}{z}}\sinh\left(\frac{x}{z}\right)\right) & \\
\quad +\text{Re}\left(\frac{4z^{2}w}{xw^{*}}\frac{1+\frac{y}{w}}{w^{2}-z^{2}}\left(ze^{-\frac{y}{z}}\sinh\left(\frac{x}{z}\right)+we^{-\frac{y}{w}}\sinh\left(\frac{x}{w}\right)\right)\right) & \\
\quad +\frac{z}{2x}\abs{\left(1+\frac{y}{w}\right)e^{-\frac{y}{w}}}^{2}\left(e^{-\frac{x}{z}}\left(\text{Shi}\left(x\left(\frac{1}{z}+2\text{Re}\left(\frac{1}{w}\right)\right)\right)\right.\right. & \\
\quad +\text{Shi}\left(x\left(\frac{1}{z}-2\text{Re}\left(\frac{1}{w}\right)\right)\right)-\text{Shi}\left(x\left(\frac{1}{z}+2i\text{Im}\left(\frac{1}{w}\right)\right)\right) & \\
\quad \left.-\text{Shi}\left(x\left(\frac{1}{z}-2i\text{Im}\left(\frac{1}{w}\right)\right)\right)\right)+\sinh\left(\frac{x}{z}\right)\left(\text{Ei}\left(x\left(\frac{1}{z}+2\text{Re}\left(\frac{1}{w}\right)\right)\right)\right. & \\
\quad+\text{Ei}\left(x\left(\frac{1}{z}-2\text{Re}\left(\frac{1}{w}\right)\right)\right)-\text{Ei}\left(x\left(\frac{1}{z}+2i\text{Im}\left(\frac{1}{w}\right)\right)\right) & \\
\quad -\text{Ei}\left(x\left(\frac{1}{z}-2i\text{Im}\left(\frac{1}{w}\right)\right)\right)-\text{Ei}\left(y\left(\frac{1}{z}+2\text{Re}\left(\frac{1}{w}\right)\right)\right) & \\
\quad -\text{Ei}\left(y\left(\frac{1}{z}-2\text{Re}\left(\frac{1}{w}\right)\right)\right)+\text{Ei}\left(y\left(\frac{1}{z}+2i\text{Im}\left(\frac{1}{w}\right)\right)\right) & \\
\quad \left.\left.+\text{Ei}\left(y\left(\frac{1}{z}-2i\text{Im}\left(\frac{1}{w}\right)\right)\right)\right)\right) & x<y \\
\frac{2z^{3}}{x\abs{w}^{2}}e^{-\frac{x}{z}}\left(\frac{y}{z}\cosh\left(\frac{y}{z}\right)-\sinh\left(\frac{y}{z}\right)\right) & \\
\quad +\text{Re}\left(\frac{4z^{2}w}{xw^{*}}\frac{1+\frac{y}{w}}{w^{2}-z^{2}}e^{-\frac{x}{z}}\left(\left(w+z\right)\sinh\left(y\left(\frac{1}{z}-\frac{1}{w}\right)\right)\right.\right. & \\
\quad \left.\left.-\left(w-z\right)\sinh\left(y\left(\frac{1}{z}+\frac{1}{w}\right)\right)\right)\right) & \\
\quad +\frac{z}{2x}\abs{\left(1+\frac{y}{w}\right)e^{-\frac{y}{w}}}^{2}e^{-\frac{x}{z}}\left(\text{Shi}\left(a\left(\frac{1}{z}+2\text{Re}\left(\frac{1}{w}\right)\right)\right)\right. & \\
\quad +\text{Shi}\left(a\left(\frac{1}{z}-2\text{Re}\left(\frac{1}{w}\right)\right)\right)-\text{Shi}\left(a\left(\frac{1}{z}+2i\text{Im}\left(\frac{1}{w}\right)\right)\right) & \\
\quad \left.-\text{Shi}\left(a\left(\frac{1}{z}-2i\text{Im}\left(\frac{1}{w}\right)\right)\right)\right) & x>y, \end{cases}
\label{Psidef}
\end{equation}
and
\begin{equation}
\Omega\left(x,y,z,w\right) = \begin{cases}
\frac{2z}{x}\sinh\left(\frac{x}{z}\right)\abs{\frac{y}{w}\cosh\left(\frac{y}{w}\right)-\sinh\left(\frac{y}{w}\right)}^{2}\text{Ei}\left(y\left(\frac{1}{z}+2\text{Re}\left(\frac{1}{w}\right)\right)\right) & x<y \\
\frac{2z}{x}\sinh\left(\frac{x}{z}\right)\abs{\frac{y}{w}\cosh\left(\frac{y}{w}\right)-\sinh\left(\frac{y}{w}\right)}^{2}\text{Ei}\left(x\left(\frac{1}{z}+2\text{Re}\left(\frac{1}{w}\right)\right)\right) & \\
\quad +\frac{z}{x}e^{-\frac{x}{z}}\abs{\frac{y}{w}\cosh\left(\frac{y}{w}\right)-\sinh\left(\frac{y}{w}\right)}^{2}\left(\text{Ei}\left(y\left(2\text{Re}\left(\frac{1}{w}\right)-\frac{1}{z}\right)\right)\right. & \\
\quad -\text{Ei}\left(y\left(2\text{Re}\left(\frac{1}{w}\right)+\frac{1}{z}\right)\right)-\text{Ei}\left(x\left(2\text{Re}\left(\frac{1}{w}\right)-\frac{1}{z}\right)\right) & \\
\quad \left.+\text{Ei}\left(x\left(2\text{Re}\left(\frac{1}{w}\right)+\frac{1}{z}\right)\right)\right) & x>y. \end{cases}
\label{Omegadef}
\end{equation}

With the power spectrum of $m_{j}\left(t\right)$ solved, the mean $\bar{m}_{j}$ needs to now be calculated in order to obtain the noise-to-signal ratio. Combining Eqs.\ \ref{rhobarrel2} and \ref{Paramjbar} yields
\begin{equation}
\bar{m}_{j} = \int_{V_{j}}d^{3}x\frac{1}{4\pi D_{\rho}}\sum_{l}\frac{\beta\bar{r}_{l}}{\abs{\vec{x}-\vec{x}_{l}}}e^{-\abs{\vec{x}-\vec{x}_{l}}\sqrt{\frac{\nu}{D_{\rho}}}}.
\label{Parambarsol1}
\end{equation}
Again, let $\vec{v}_{j} = \vec{x}-\vec{x}_{j}$. Utilizing this substitution and Eq.\ \ref{Lambdarel} to solve Eq.\ \ref{Parambarsol1} yields
\begin{align}
\bar{m}_{j} &= \frac{\beta}{4\pi D_{\rho}}\sum_{l}\bar{r}_{l}\int_{V_{j}}d^{3}v_{j}\frac{1}{\abs{\vec{v_{j}}-\left(\vec{x}_{l}-\vec{x}_{j}\right)}}e^{-\abs{\vec{v_{j}}-\left(\vec{x}_{l}-\vec{x}_{j}\right)}\sqrt{\frac{\nu}{D_{\rho}}}} \nonumber\\
&= \frac{\beta a^{2}}{2D_{\rho}}\sum_{l}\bar{r}_{l}\Lambda\left(\abs{\vec{x}_{l}-\vec{x}_{j}},a,\lambda\left(0\right)\right)
\label{Parambarsol2}
\end{align}

Finally, combining Eqs.\ \ref{mjPspec2}, \ref{I12}, \ref{I2}, \ref{I310}, \ref{I47}, and \ref{Parambarsol2} yields the time averaged noise-to-signal ratio of $m_{j}\left(t\right)$. To determine the criteria for $T$ in this equation, it is important to note that $\omega$ only appears in $\lambda\left(\omega\right)$, which directly compares $\omega$ to $\nu$ in Eq.\ \ref{lambdadef}. However, $\nu$ can be taken to 0 without complication, thus leaving $\omega$ to be directly compared to $\frac{D_{\rho}}{a^{2}}$ as $\lambda\left(\omega\right)$ is always found in proportion to $a$ or $\abs{\vec{x}_{i}-\vec{x}_{j}}$, but $\abs{\vec{x}_{i}-\vec{x}_{j}}$ can be taken to $\infty$ without complication as well. Thus, $\omega\ll\nu$ must be true unless $\nu\ll\frac{D_{\rho}}{a^{2}}$, at which point $\omega\ll\frac{D_{\rho}}{a^{2}}$ must be true. This in turn implies $T\gg\tau_{4}=\left(\nu+\frac{D_{\rho}}{a^{2}}\right)^{-1}$ can be taken as the appropriate criterion for $T$. Once this is met, the time averaged noise-to-signal ratio of $m_{j}\left(t\right)$ can be calculated to be
\begin{align}
\frac{(\delta m_j)^{2}}{\bar{m}_{j}^{2}} &= \frac{S_{m}\left(0\right)}{\bar{m}_{j}^{2}T} = \frac{1}{\bar{m}_{j}^{2}T}\int\frac{d\omega'}{2\pi}\left\langle\tilde{\delta m}_{j}^{*}\left(\omega'\right)\tilde{\delta m}_{j}\left(0\right)\right\rangle \nonumber\\
&= \frac{I_{1}\left(0\right)+I_{2}\left(0\right)+I_{3}\left(0\right)+I_{4}\left(0\right)}{\left(\frac{\beta a^{2}}{2D_{\rho}}\sum_{l}\bar{r}_{l}\Lambda\left(\abs{\vec{x}_{l}-\vec{x}_{j}},a,\lambda\left(0\right)\right)\right)^{2}T} \nonumber\\
&= \frac{1}{\beta T}\left(\sum_{l}\bar{r}_{l}\Lambda\left(\abs{\vec{x}_{l}-\vec{x}_{j}},a,\lambda\left(0\right)\right)\right)^{-2}\left(\sum_{s}\bar{r}_{s}\abs{\Lambda\left(\abs{\vec{x}_{s}-\vec{x}_{j}},a,\lambda\left(0\right)\right)}^{2}\vphantom{\left(\frac{\abs{\left(1+\frac{a}{\lambda\left(\omega\right)}\right)e^{-\frac{a}{\lambda\left(\omega\right)}}}^{2}}{18}\right)}\right. \nonumber\\
&\quad +\sum_{s,u}2\alpha\beta\bar{c}\underbrace{\Lambda\left(\abs{\vec{x}_{s}-\vec{x}_{j}},a,\lambda\left(0\right)\right)\Lambda\left(\abs{\vec{x}_{u}-\vec{x}_{j}},a,\lambda\left(0\right)\right)}_\text{integration}\underbrace{\left(\lim_{\omega\to 0}\frac{1}{\omega}\text{Im}\left(R_{su}^{-1}\left(\omega\right)\right)\right)}_\text{correlation} \nonumber\\
&\quad +\sum_{s}4\bar{r}_{s}\left(\frac{\abs{\left(1+\frac{a}{\lambda\left(0\right)}\right)e^{-\frac{a}{\lambda\left(0\right)}}}^{2}}{18}\Upsilon\left(\abs{\vec{x}_{s}-\vec{x}_{j}},a,\lambda\left(0\right),\lambda\left(0\right)\right)\right. \nonumber\\
&\quad \left.\vphantom{\frac{\abs{\left(1+\frac{a}{\lambda\left(0\right)}\right)e^{-\frac{a}{\lambda\left(0\right)}}}^{2}}{18}}+\abs{\left(\frac{\lambda\left(0\right)}{a}\right)^{3}\left(\sinh\left(\frac{a}{\lambda\left(0\right)}\right)-\frac{a}{\lambda\left(0\right)}\cosh\left(\frac{a}{\lambda\left(0\right)}\right)\right)}^{2}\Xi\left(\abs{\vec{x}_{s}-\vec{x}_{j}},a,\lambda\left(0\right),\lambda\left(0\right)\right)\right) \nonumber\\
&\quad \left.\vphantom{\left(\frac{\abs{\left(1+\frac{a}{\lambda\left(0\right)}\right)e^{-\frac{a}{\lambda\left(0\right)}}}^{2}}{18}\right)}+\sum_{s}2\bar{r}_{s}\left(\frac{\lambda\left(0\right)}{a}\right)^{4}\left(\Psi\left(\abs{\vec{x}_{s}-\vec{x}_{j}},a,\lambda\left(0\right),\lambda\left(0\right)\right)+\Omega\left(\abs{\vec{x}_{s}-\vec{x}_{j}},a,\lambda\left(0\right),\lambda\left(0\right)\right)\right)\right).
\label{ParamjNSR}
\end{align}

Eq.\ \ref{Rrewrite} along with simplifications from Eq.\ \ref{crbarrel} can once again be used to separate out Eq.\ \ref{ParamjNSR} into its separate, more intuitive terms.

\begin{align}
\frac{\left(\delta m_{j}\right)^{2}}{\bar{m}_{j}^{2}} = &\frac{1}{2}\frac{1}{\pi a\bar{c}D_{c}T}\frac{\sum_{s,u}\Lambda\left(\abs{\vec{x}_{s}-\vec{x}_{j}},a,\lambda\left(0\right)\right)\Lambda\left(\abs{\vec{x}_{u}-\vec{x}_{j}},a,\lambda\left(0\right)\right)\Theta\left(\frac{\abs{\vec{x}_{s}-\vec{x}_{u}}}{a}\right)}{\left(\sum_{l}\Lambda\left(\abs{\vec{x}_{l}-\vec{x}_{j}},a,\lambda\left(0\right)\right)\right)^{2}}+\frac{2}{\mu\bar{r}T}\frac{\sum_{s}\left(\Lambda\left(\abs{\vec{x}_{s}-\vec{x}_{j}},a,\lambda\left(0\right)\right)\right)^{2}}{\left(\sum_{l}\Lambda\left(\abs{\vec{x}_{l}-\vec{x}_{j}},a,\lambda\left(0\right)\right)\right)^{2}} \nonumber\\
&+\frac{1}{\beta\bar{r} T}\left(\sum_{l}\Lambda\left(\abs{\vec{x}_{l}-\vec{x}_{j}},a,\lambda\left(0\right)\right)\right)^{-2}\left(\sum_{s}\abs{\Lambda\left(\abs{\vec{x}_{s}-\vec{x}_{j}},a,\lambda\left(0\right)\right)}^{2}\vphantom{\left(\frac{\abs{\left(1+\frac{a}{\lambda\left(\omega\right)}\right)e^{-\frac{a}{\lambda\left(\omega\right)}}}^{2}}{18}\right)}\right. \nonumber\\
&\quad +\sum_{s}4\left(\frac{\abs{\left(1+\frac{a}{\lambda\left(0\right)}\right)e^{-\frac{a}{\lambda\left(0\right)}}}^{2}}{18}\Upsilon\left(\abs{\vec{x}_{s}-\vec{x}_{j}},a,\lambda\left(0\right),\lambda\left(0\right)\right)\right. \nonumber\\
&\quad \left.\vphantom{\frac{\abs{\left(1+\frac{a}{\lambda\left(0\right)}\right)e^{-\frac{a}{\lambda\left(0\right)}}}^{2}}{18}}+\abs{\left(\frac{\lambda\left(0\right)}{a}\right)^{3}\left(\sinh\left(\frac{a}{\lambda\left(0\right)}\right)-\frac{a}{\lambda\left(0\right)}\cosh\left(\frac{a}{\lambda\left(0\right)}\right)\right)}^{2}\Xi\left(\abs{\vec{x}_{s}-\vec{x}_{j}},a,\lambda\left(0\right),\lambda\left(0\right)\right)\right) \nonumber\\
&\quad \left.\vphantom{\left(\frac{\abs{\left(1+\frac{a}{\lambda\left(0\right)}\right)e^{-\frac{a}{\lambda\left(0\right)}}}^{2}}{18}\right)}+\sum_{s}2\left(\frac{\lambda\left(0\right)}{a}\right)^{4}\left(\Psi\left(\abs{\vec{x}_{s}-\vec{x}_{j}},a,\lambda\left(0\right),\lambda\left(0\right)\right)+\Omega\left(\abs{\vec{x}_{s}-\vec{x}_{j}},a,\lambda\left(0\right),\lambda\left(0\right)\right)\right)\right).
\label{ParamjNSRrewrite}
\end{align}
Eq.\ \ref{ParamjNSRrewrite} is the general expression for the error in the case of autocrine signaling and, when only the extrinsic term is considered, is utilized in calculating the phase boundaries in Fig. 3 of the main text.

As with Eq.\ \ref{mjNSRrewrite}, the term responsible for holding the inheritted noise from the bound receptor numbers and ligand field has a summation that is comprised of two clear factors. The first factor, $\Lambda\left(\abs{\vec{x}_{s}-\vec{x}_{j}},a,\lambda\left(0\right)\right)\Lambda\left(\abs{\vec{x}_{u}-\vec{x}_{j}},a,\lambda\left(0\right)\right)$, comes from the integration of the ligand field over the cell volume and thus accounts for the spatial integration of the messenger molecule. The second term, $\lim_{\omega\to 0}\frac{1}{\omega}\text{Im}\left(R_{su}^{-1}\left(\omega\right)\right)$, is the same as the correlation term in Eq.\ \ref{mjNSR} and thus is known to account for the cross correlation between bound receptor numbers. Once again, the interplay between the ligand field, bound receptor number, and messenger molecule count can be visualized for a two cell system in Fig. \ref{suppfig}.

Eq.\ \ref{ParamjNSRrewrite} can be greatly simplified in form under the limit $\nu\ll\frac{D_{\rho}}{a^{2}}$, which by Eq.\ \ref{lambdadef} implies $\lambda\left(0\right)\gg a$. When this limit is taken, the $\Psi$ and $\Omega$ functions vanish due to their original multiplication by $\nu$ in Eq.\ \ref{Idefd} while the $\Lambda$, $\Upsilon$, and $\Xi$ functions simplify to
\begin{subequations}
\begin{equation}
\lim_{z\to\infty}\Lambda\left(x,y,z\right)\equiv\Lambda_{\infty}\left(\frac{x}{y}\right)=\begin{cases}
1-\frac{1}{3}\left(\frac{x}{y}\right)^{2} & \frac{x}{y}<1 \\
\frac{2y}{3x} & \frac{x}{y}>1, \end{cases}
\label{nuto0a}
\end{equation}
\begin{equation}
\lim_{z\to\infty}\Upsilon\left(x,y,z,z\right)\equiv\Upsilon_{\infty}\left(\frac{x}{y}\right)=\begin{cases}
1-\frac{1}{5}\left(\frac{x}{y}\right)^{4} & \frac{x}{y}<1 \\
\frac{4y}{5x} & \frac{x}{y}>1, \end{cases}
\label{nuto0b}
\end{equation}
\begin{equation}
\lim_{z\to\infty}\Xi\left(x,y,z,z\right)\equiv\Xi_{\infty}\left(\frac{x}{y}\right)=\begin{cases}
1 & \frac{x}{y}<1 \\
\frac{2y}{x}-\left(\frac{y}{x}\right)^{2} & \frac{x}{y}>1, \end{cases}
\label{nuto0c}
\end{equation}
\label{nuto0}%
\end{subequations}
which can be shown by Taylor expanding all the functions in Eqs.\ \ref{Lambdadef}, \ref{Updef}, and \ref{Xidef} for small $\frac{1}{z}$ and evaluating. Utilizing the same method for the other instances of $\lambda\left(0\right)$ in Eq.\ \ref{ParamjNSRrewrite} allows it to simplify to
\begin{align}
\lim_{\lambda\left(0\right)\to\infty}\frac{(\delta m_j)^{2}}{\bar{m}_{j}^{2}} = &\frac{1}{2}\frac{1}{\pi a\bar{c}D_{c}T}\frac{\sum_{s,u}\Lambda_{\infty}\left(\frac{\abs{\vec{x}_{s}-\vec{x}_{j}}}{a}\right)\Lambda_{\infty}\left(\frac{\abs{\vec{x}_{u}-\vec{x}_{j}}}{a}\right)\Theta\left(\frac{\abs{\vec{x}_{s}-\vec{x}_{u}}}{a}\right)}{\left(\sum_{l}\Lambda_{\infty}\left(\frac{\abs{\vec{x}_{l}-\vec{x}_{j}}}{a}\right)\right)^{2}}+\frac{2}{\mu\bar{r}T}\frac{\sum_{s}\left(\Lambda_{\infty}\left(\frac{\abs{\vec{x}_{s}-\vec{x}_{j}}}{a}\right)\right)^{2}}{\left(\sum_{l}\Lambda_{\infty}\left(\frac{\abs{\vec{x}_{l}-\vec{x}_{j}}}{a}\right)\right)^{2}} \nonumber\\
&+\frac{1}{\beta\bar{r} T}\left(\sum_{l}\Lambda_{\infty}\left(\frac{\abs{\vec{x}_{l}-\vec{x}_{j}}}{a}\right)\right)^{-2}\left(\sum_{s}\left(\Lambda_{\infty}\left(\frac{\abs{\vec{x}_{s}-\vec{x}_{j}}}{a}\right)\right)^{2}\right. \nonumber\\
&\quad +\left.\sum_{s}4\left(\frac{1}{18}\Upsilon_{\infty}\left(\frac{\abs{\vec{x}_{s}-\vec{x}_{j}}}{a}\right)+\frac{1}{9}\Xi_{\infty}\left(\frac{\abs{\vec{x}_{s}-\vec{x}_{j}}}{a}\right)\right)\right)
\label{ParamjNSRlim}
\end{align}
The data presented in Fig. 2 of the main text is similarly obtained by separating Eq.\ \ref{ParamjNSRlim} into its extrinsic and intrinsic terms and considering only those extrinsic terms caused by the ligand diffusion.

For the two cell case, Eq.\ \ref{ParamjNSRlim} can be further evaluated to yield for either cell
\begin{align}
\frac{(\delta m)^{2}}{\bar{m}^{2}} &= \frac{1}{\pi a\bar{c}D_{c}T}\frac{1+\left(\Lambda_{\infty}\left(\frac{\ell}{a}\right)\right)^{2}+2\Lambda_{\infty}\left(\frac{\ell}{a}\right)\Theta\left(\frac{\ell}{a}\right)}{2\left(1+\Lambda_{\infty}\left(\frac{\ell}{a}\right)\right)^{2}}+\frac{1}{\mu\bar{r}T}\frac{2\left(1+\left(\Lambda_{\infty}\left(\frac{\ell}{a}\right)\right)^{2}\right)}{\left(1+\Lambda_{\infty}\left(\frac{\ell}{a}\right)\right)^{2}} \nonumber\\
&\quad +\frac{1}{\beta\bar{r}T}\frac{\frac{5}{3}+\left(\Lambda_{\infty}\left(\frac{\ell}{a}\right)\right)^{2}+\frac{2}{9}\Upsilon_{\infty}\left(\frac{\ell}{a}\right)+\frac{4}{9}\Xi_{\infty}\left(\frac{\ell}{a}\right)}{\left(1+\Lambda_{\infty}\left(\frac{\ell}{a}\right)\right)^{2}}
\label{ParamjNSR2cell2}
\end{align}
The first term is the extrinsic noise, and when the assumption $\ell>a$ is made so as to replace $\Theta\left(\frac{\ell}{a}\right)$ with $\frac{a}{\ell}$, it is seen to reduce to Eq.\ 9 of the main text. Let $\bar{r}$ and $\beta$ be large enough such that the second two terms in Eq.\ \ref{ParamjNSR2cell2} can be neglected. Additionally, extend the $\ell>a$ assumption to the $\Lambda_{\infty}$ functions. Under these, Eq.\ \ref{nuto0a} can be used to reduce Eq.\ \ref{ParamjNSR2cell2} to
\begin{equation}
\frac{(\delta m)^{2}}{\bar{m}^{2}} = \frac{1}{\pi a\bar{c}D_{c}T}\frac{1+\left(\frac{2a}{3\ell}\right)^{2}+2\frac{a}{\ell}\frac{2a}{3\ell}}{2\left(1+\frac{2a}{3\ell}\right)^{2}} = \frac{1}{\pi a\bar{c}D_{c}T}\frac{1+\frac{16a^{2}}{9\ell^{2}}}{2\left(1+\frac{2a}{3\ell}\right)^{2}}
\label{ParamjNSR2cellBP}
\end{equation}
The coefficient of $\frac{1}{\pi a\bar{c}D_{c}T}$ in Eq.\ \ref{ParamjNSR2cellBP} achieves its minimum value of $\frac{2}{5}$ at $\ell^*=\frac{8}{3}a$, which is within the bounds of the $\ell>a$ assumption and is also presented in Eq.\ 10 of the main text.

\section{Summary of Mechanisms of Improved Sensing}

Our findings for how cell-cell communication reduces sensory error can be summarized as arising from two main effects. First, we find very generally that the error in a cell's ability to sense a ligand concentration decreases as the number of other cells with which it communicates increases. This effect is demonstrated by Fig.\ 2A-C in the main text: for both juxtacrine and autocrine communication, the error decreases with the cell number. The reason for this effect is that when cells share information via diffusive spatial averaging, a single cell gains more information than what it can measure alone, which reduces its sensory error. The idea that spatial averaging among cells can reduce sensory error has been investigated experimentally in the context of boundary formation in the fruit fly embryo (references 11 and 12 in the main text), and this phenomenon has been studied in detail theoretically by Erdmann et al.\ (reference 13 in the main text). Indeed, Erdmann et al.\ showed that the noise in the number of molecules in a cell is reduced when that molecule is allowed to diffuse between cells.

One important difference between the result of Erdmann et al.\ and our results is that Erdmann et al.\ found that only the extrinsic, super-Poissonian noise term was decreased with increasing population size. Conversely, in this work we have shown that all noise terms decrease with increasing population size $N$, as seen in Eq.\ \ref{mjNSRinfg}, including the intrinsic noise from the messenger molecule itself (first term). The reason for this difference is that Erdmann et al.\ considered an instantaneous measurement while our work considers a time-averaged measurement. Time averaging reduces the intrinsic noise by the effective number of independent measurements, equal to the product of the number of molecular turnovers in a degradation time in a single cell ($\nu T$ in the first term of Eq.\ \ref{mjNSRinfg}) and the number of cells with which it is communicating in that time ($N$ in the first term of Eq.\ \ref{mjNSRinfg}). Time integration is necessary for spatial averaging to affect the intrinsic noise, which is why a reduction similar to this factor of $N$ is not present in the intrinsic noise term of Erdmann et al.

The second main effect that we discover is that autocrine signaling results in a lower error than juxtacrine signaling for large populations in two- or three-dimensional arrangements, as seen in Fig.\ 2B and C in the main text. The reason for this effect is that juxtacrine signaling requires cells to be adjacent to one another, while autocrine signaling does not. Surprisingly, the optimal cell-cell spacing for autocrine signaling can be large, and in fact it increases as population size increases, as seen in Fig.\ 2D of the main text. The optimal spacing arises due to a tradeoff between two effects: on the one hand, the mean messenger molecule number per cell decreases with cell spacing, since the messenger concentration profile decreases with distance from the source (Eq.\ 8 of the main text). This effect is evident from the denominator of the first term in Eq.\ \ref{ParamjNSRlim}, which originates from the mean: since $\Lambda_{\infty}\left(x\right)$ is a monotonically decreasing function of $x$ for $x>1$, the denominator decreases as the cells move further apart. On the other hand, the variance in the messenger molecule number per cell also decreases with cell spacing. This effect has two origins, as evidenced by the $\Theta(x)$ and $\Lambda_{\infty}(x)$ functions in the numerator of the first term in Eq.\ \ref{ParamjNSRlim}, which both decrease with $x$ for $x>1$. The $\Theta$ function originates from the $\left\langle\tilde{\delta r}_{l}^{*}\left(\omega'\right)\tilde{\delta r}_{j}\left(\omega\right)\right\rangle$ term of Eq.\ \ref{rhospec1} and thus accounts for the decorrelation of the ligand measurements made by distinct cells as those cells are spaced farther apart. The $\Lambda_{\infty}$ functions originate from the integration term in Eq.\ \ref{I12} and thus account for the decorrelation of the messenger molecule counts during their diffusion from one cell to another. This latter effect was also observed by Erdmann et al.: they found that super-Poissonian fluctuations in a messenger molecule concentration profile decrease with the distance to the source of the molecules, since diffusion ``washes them out.'' Together, these two noise-reduction effects trade off with the mean-reduction effect to give an optimal cell spacing, at which sparsely arranged cells communicating via autocrine signaling outperform tightly packed communicating cells in terms of sensory precision.


\begin{thebibliography}{47}
\bibitem[{foo({\natexlab{a}})}]{foot1}
\bibinfo{note}{Autocrine signaling occurs when a cell both secretes and detects
  a diffusible signaling molecule. When many cells are present, a given cell
  detects molecules from both itself and the other cells, resulting in
  potentially long-range cell-cell communication via the extracellular medium.
  Some authors call this type of communication paracrine signaling, although
  that term is usually reserved for communication between cells of different
  type.}
\end{thebibliography}
\end{document}